\documentclass{aa}  
\usepackage{graphicx}
\usepackage{lscape}
\usepackage{subfigure}

\usepackage{txfonts}
\begin{document}
\title{Rotational studies in the Orion Nebula Cluster:\\ from solar mass stars to brown dwarfs.}

 	\author{Mar\'{i}a V. Rodr\'{i}guez-Ledesma
          \inst{1,}\inst{2},         
          Reinhard Mundt\inst{1}
\and
          Jochen Eisl\"offel\inst{3}
	 }

   \offprints{Mar\'{i}a V. Rodr\'{i}guez-Ledesma}

   \institute{Max-Planck-Institut f\"ur Astronomie (MPIA), K\"onigstuhl 17, D-69117 Heidelberg, Germany\\
              \email{vicrodriguez@mpia.de}
	     \and
International Max Planck Research School for Astronomy \& Cosmic Physics at the University of Heidelberg
\and
             Th\"uringer Landessternwarte Tautenburg, Sternwarte 5, D-07778 Tautenburg, Germany
             }


 
\abstract
   {}
   {Rotational studies at a variety of ages and masses are important for constraining the angular momentum evolution of young stellar objects (YSO). Of particular interest are the very low mass (VLM) stars and brown dwarfs (BDs), because of the significant lack of known rotational periods in that mass range. We aim to extend previous studies well down into the substellar regime, providing for the first time information on rotational periods for a large sample of young VLM stars and BDs.}
   {This extensive rotational period study of YSOs in the 1 Myr old Orion Nebula Cluster (ONC) is based on a deep photometric monitoring campaign using the Wide Field Imager (WFI) camera on the ESO/MPG 2.2m telescope on La Silla, Chile. Time series data was obtained with about 95 data points spread over 19 nights. Accurate I-band photometry of 2908 stars was obtained within a magnitude range of 13 to 21 mag, i.e. extending three magnitudes deeper than previous studies in the ONC. Two different power spectral analysis techniques were used to search for periodic variability. In addition, the $\chi^{2}$ variability test was used for the detection of irregular variables.}
   {We found 487 periodic variables with estimated masses between 0.5\,$M_\odot$ and 0.015\,$M_\odot$, 124 of which are BD candidates.  This is by far the most extensive and complete rotational period data set for young VLM stars and BDs. In addition to the periodic variables, 808 objects show non-periodic brightness variations. We study the dependence of the period distribution on mass and variability level and compare this with known objects in the ONC with masses up to 1.5\,$M_\odot$ (Herbst et al. 2002) and with the $\sim$\,2 Myr old cluster NGC 2264 (Lamm et al., 2004). We find that substellar objects rotate on average faster than the VLM stars. In addition, our rotational data suggest a dependence of the rotational periods on position within the field, which can be explained by a possible age spread in the ONC with a somewhat younger central region. The results of a comparison between the period distributions of the ONC and NGC 2264 favours this hypothesis. In addition, periodic variables with larger peak-to-peak amplitudes rotate on average slower than those with small peak-to-peak amplitude variations, which can possibly be explained by different magnetic field topologies.\thanks{Tables 1 and 2 are only available in electronic form at the CDS via anonymous ftp to cdsarc.u-strasbg.fr (130.79.128.5)
or via http://cdsweb.u-strasbg.fr/cgi-bin/qcat?J/A+A/}}
   {}
\titlerunning{Rotational studies in the ONC}
\authorrunning{Rodr\'iguez-Ledesma et al.} 
\maketitle
%
\section{Introduction}
\begin{figure}
 \centering
 \includegraphics[width=8.8cm]{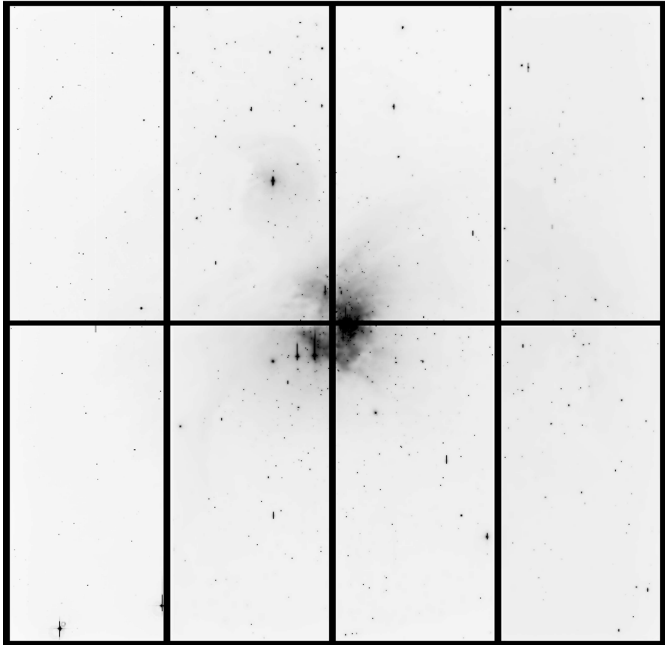}
 \caption{ The observed field of the Orion Nebula Cluster (ONC). North is up and east is left. The WFI 4k x 2k mosaic, covering a field of 34x33 arcmin$^2$, is easily seen. The ONC cluster is in the middle of the field.
          }
 \label{FigORI}
 \end{figure}

The basic angular momentum problem of star formation is that molecular cloud cores have five to six orders of magnitude higher specific angular momentum (j\,=\,J\,$\diagup$\,M  where J is the angular momentum and M the mass) than zero-age main sequence (ZAMS) stars (e.g. Bodenheimer \cite{Bodenheimer}). The various processes involved in angular momentum loss during the pre-main sequence (PMS) phase are not completely understood. As pointed out by Bodenheimer (\cite{Bodenheimer}) a combination of several phenomena acting during different phases of star formation is required to solve the angular momentum problem. Important processes include magnetic braking of the molecular cloud, due to coupling between the collapsing core and the surrounding medium, angular momentum transfer and distribution in disks and orbital motions, as well as angular momentum loss by magnetically driven outflows and jets.  Magnetic braking through stellar winds is also an important effect for angular momentum loss in late type main sequence (MS) stars, although it is most probably ineffective during the PMS phase.\\
In the last phase of star formation, when PMS stars become optically visible, we can study their angular momentum evolution by measuring rotational periods for objects at different ages and masses. A photometric monitoring technique can be used to derive the rotational periods of these young low mass objects from their brightness modulation (e.g. Rydgren \& Vrba, 1983; Vrba et al. 1986, 1989; Bouvier \& Bertout 1989; Bouvier et al., 1993; Herbst et al., 2000, 2001, 2002 from now on H2000, H2001, and H2002; Bailer-Jones \& Mundt 1999, 2001). This technique is applicable over a broad mass range, from $\sim$\,2\,$M_\odot$ down into the BD regime. The observed variations are largely due to the rotational brightness modulations by magnetically induced cool spots in the case of weak-line TTauri stars (WTTS), with the addition of an irregular component caused by accretion hot spots in the case of classical TTauri stars (CTTS) (e.g. Vrba et al. 1986, 1989; Bouvier \& Bertout, 1989; Bouvier et al., 1993; Herbst et al. 1994). The rotation period measurements from this technique can be derived to an accuracy of about 1\%, even for slow rotating stars. This accuracy, plus independence from inclination effects, provides a great advantage over $V_{rot}sin{i}$ measurements based on high resolution spectral observations.\\
Several authors have investigated the rotational properties of solar mass and low mass stars in clusters of various ages (e.g. Stauffer \& Hartmann, 1987; Prosser et al. 1993, 1995; Terndrup et al. 2000; Scholz \& Eisl\"offel \cite{Scholz04b}; Irwin et al. \cite{Irwin2008}). In young clusters and star forming regions, an enormous spread in rotational periods has been observed at all stellar masses (e.g. Hillenbrand, \cite{Hill97}; Hillenbrand \& Hartmann \cite{Hill98}; H2002; Lamm et al. \cite{Lamm04}; \cite{Lamm05} from now on L2004 and L2005; Scholz \& Eisl\"offel, \cite{Scholz04a}; \cite{Scholz05}). Particularly in very young star forming regions like the 1 Myr old Orion Nebula Cluster (ONC), the large spread observed has to be understood, at least in part, as being inherited from the protostellar phase (e.g. Hillenbrand 1997; H2000; H2002; Sicilia et al., 2005). The observed dependence of the measured rotational periods on mass and age reveals information about rotational braking due to magnetic coupling to the disk as well as due to magnetically driven outflows. Correlations of infrared excess (which results from a circumstellar-disk) and rotational periods have been performed for a broad mass range in several regions, suggesting, in some of them, that objects showing infrared excess rotate slower than their counterparts without infrared excess (e.g. Edwards et al. 1993, Rebull et al., 2002, 2006; H2002; L2004; L2005; Cieza \& Baliber, 2007). We will analyse this important correlation in a follow-up paper (Rodr\'iguez-Ledesma et al., 2009). The same conclusion arises from studies of the  H$\mathrm{\alpha}$ equivalent width and UV excess emission (e.g. Sicilia et al., 2005; Fallscheer \& Herbst, 2006). These results indicate that magnetic coupling to the disks and magnetically driven outflows are important ingredients in the angular momentum regulation during these PMS phases. However, most of these studies extend only down to masses of about 0.2\,$M_\odot$. Only few very low mass stars and BDs in different star forming regions have rotational periods measured so far (e.g. Bailer-Jones \& Mundt, 2001; Joergens et al., 2003; Scholz \& Eisl\"{o}ffel, 2004; Mohanty et al., 2004) making it hard to derive statistically significant information for this mass regime.\\
The ONC is an excellent target for this kind of study since it contains thousands of very young PMS objects with masses even below the deuterium burning limit, within 2pc of the central Trapezium stars. This means that a large sample of young stellar and substellar objects can be photometrically monitored simultaneously in order to derive rotational periods. Several rotational studies in the ONC have been performed in the past (e.g. Mandel \& Herbst, 1991; Attridge \& Herbst, 1992; Choi \& Herbst, 1996; H2000; H2002 ). H2002 confirmed that the period distribution found for stars with masses higher than 0.25\,M$_\odot$ (based on D'Antonna and Mazzitelli, \cite{Dantona04} models) is bimodal with peaks around 2 and 8\,days. The bimodal distribution was not seen in the period distribution of the lower mass stars (M\,$\leq$\,0.25\,M$_\odot$). Similar results have been found by L2005 for their rotational studies in the 2-3 times older cluster NGC2264, for which D'Antonna and Mazzitelli (\cite{Dantona07}) models have been used. They found that the peaks of the distributions are shifted towards the fast rotators with respect to the peaks of the distributions measured in the ONC. The results of these two studies reveal a clear dependence of the rotational periods with mass (down to $\sim$\,0.1\,M$_\odot$) and age. Lower mass objects were found to rotate on average faster and younger objects to rotate slower. These results have been interpreted in terms of disk-star interactions, which was first proposed as an efficient braking mechanism in young stellar objects during the 90s (Camenzind, 1990; K\"{o}nigl, 1991). In this scenario, objects rotating slowly are likely to be magnetically coupled to their disks and are prevented from spinning up by magnetic braking, while those objects showing short periods are more likely to be released from their disks or at least the magnetic coupling is much weaker (e.g. H2002; L2005).\\
The goal of this work is to fill the observational gap for the very low mass and, particularly, substellar regime. For this purpose we have carried out a deep and extensive photometric monitoring campaign on the ONC, which resulted in 487 rotational periods measured, 377 of which are new detections and 124 of which are BD candidates.\\
In Sect.\,2 we give details of the observations. In Sect.\,3 and 4 we describe the astrometric and photometric analysis of the data. The time series analysis is described in Sect.\,5 while Sect.\,6 is dedicated to the irregular variable objects. We present our results in Sect.\,7 and 8. We summarise our results in Sect.\,9, where we also discuss possible scenarios that could provide physical explanations for what is observed. 
  \begin{figure}
   \centering
   \includegraphics[width=9cm]{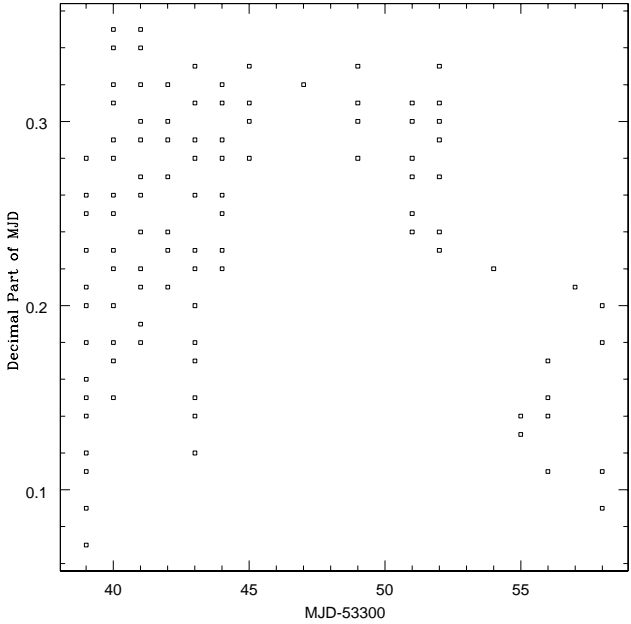}
      \caption{ Time distribution of the observations, showing the decimal part of the night on the ordinate and the modified Julian day on the abscissa. Please note the dense sampling in the first six nights.
              }
         \label{FigMJD}
   \end{figure}
\section{Observations and image reduction}
Observations were carried out over 19 nights from 30 Nov. 2004 to 19 Dec. 2004 with the Wide Field Imager (WFI) at the MPG/ESO 2.2m telescope on La Silla, Chile. The WFI consists of eight 2Kx4K CCDs in a 4x2 mosaic. The pixel scale of 0\farcs238 gives a field-of-view of 34\,x\,33 arcmin$^2$. All data were obtained through the ESO filter 851, with a central wavelength of 815.9\,nm and a FWHM of 20.1\,nm. In this wavelength range the contamination by nebular line emission is relatively small, while the signal from faint, red objects is relatively high. Fig. 1 shows the observed region; the bright Trapezium stars were placed in a gap near the center of the field.\\
One hundred images were obtained in total. As shown in detail in Fig.\,2, our time sampling
is not evenly spread over the observing period. During the first six nights, 10 to 15 exposures per
night were obtained, while during the rest of the campaign, about five exposures per night were acquired. The exposure time of each image was 1130 seconds, which provides a three magnitude deeper data set in I than the one obtained by H2002. The seeing during these observations varied between 0\farcs45 and 1\farcs4 although in most cases it oscilated between 0\farcs9 and 1\farcs0.\\
Image processing was done using standard tasks in IRAF (Image Reduction and Analysis Facility distributed by the U.S National Optical Astronomy Observatory). Each step in the data reduction, astrometry, photometry, and time series analysis was done separately for each of the eight CCDs, in
order to account for the differences between them. A one-dimensional bias was subtracted using the overscan region, and a two-dimensional residual was removed using the zero integration time (bias)
frames. For the flat-fielding correction, illumination-corrected dome flats were used, following
the method described by Bailer-Jones \& Mundt (\cite{Bailer-Jones01}).\\
A catalogue source image (or ``template'' image) was created by adding all the aligned science images, in order to have an image with about ten times higher signal-to-noise ratio (S/N) than the individual images. For the four inner CCDs, a median filtered
image was subtracted from the template images, in order to
remove large scale nebular structures which helps considerably to reduce large gradients in the sky
background. DAOFIND was used on the combined images to create the catalogue source list. In a visual inspection, some faint stellar sources were added and as many as possible non-stellar sources were removed. Finally about 3300 objects were taken into account for the photometric analysis.\\
\begin{figure}[ht]
\centering
\subfigure[]{\rotatebox{270}{\includegraphics[width=5cm,height=8.2cm]{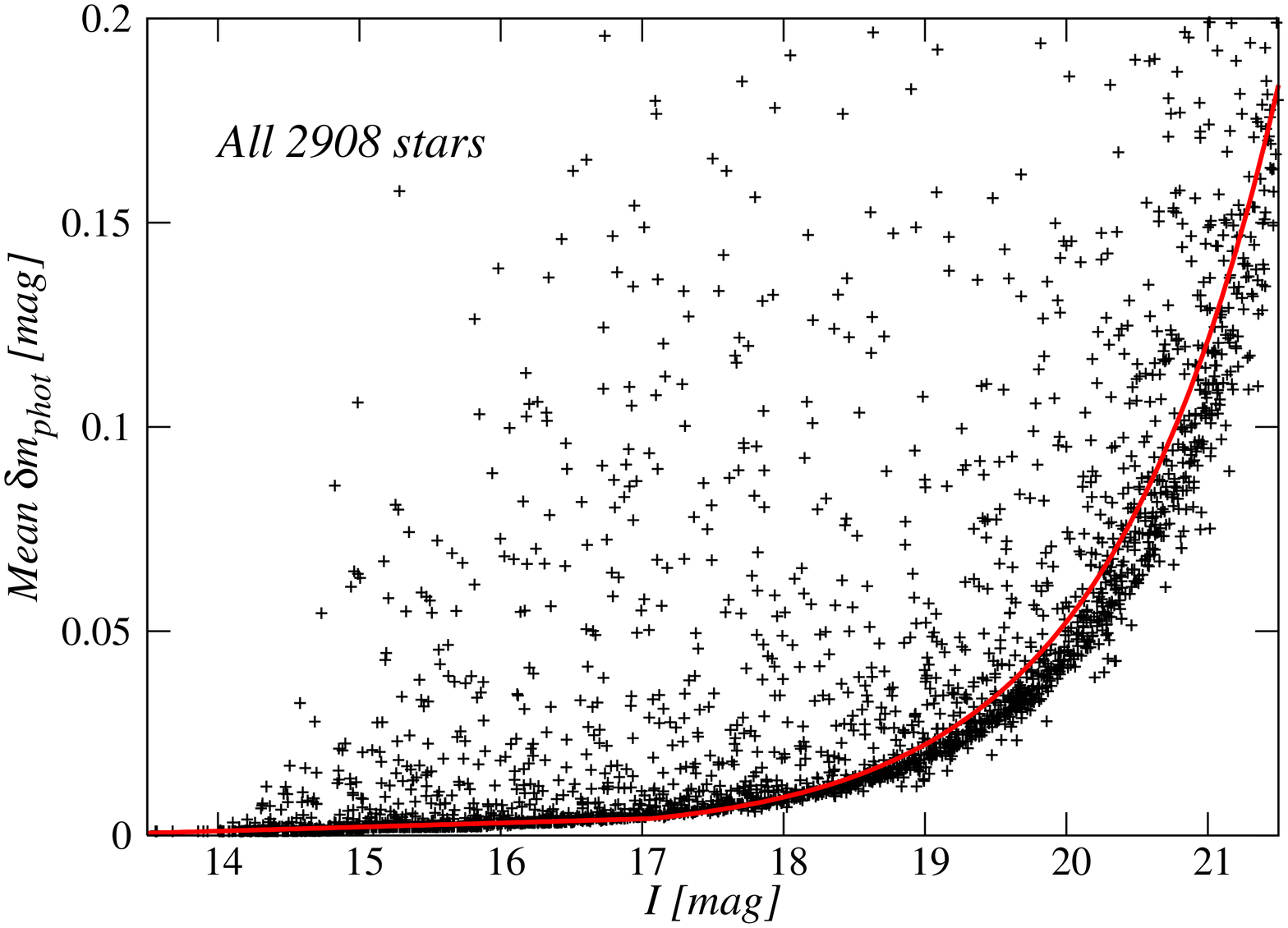}}}\\
\subfigure[]{\rotatebox{270}{\includegraphics[width=5cm,height=8.2cm]{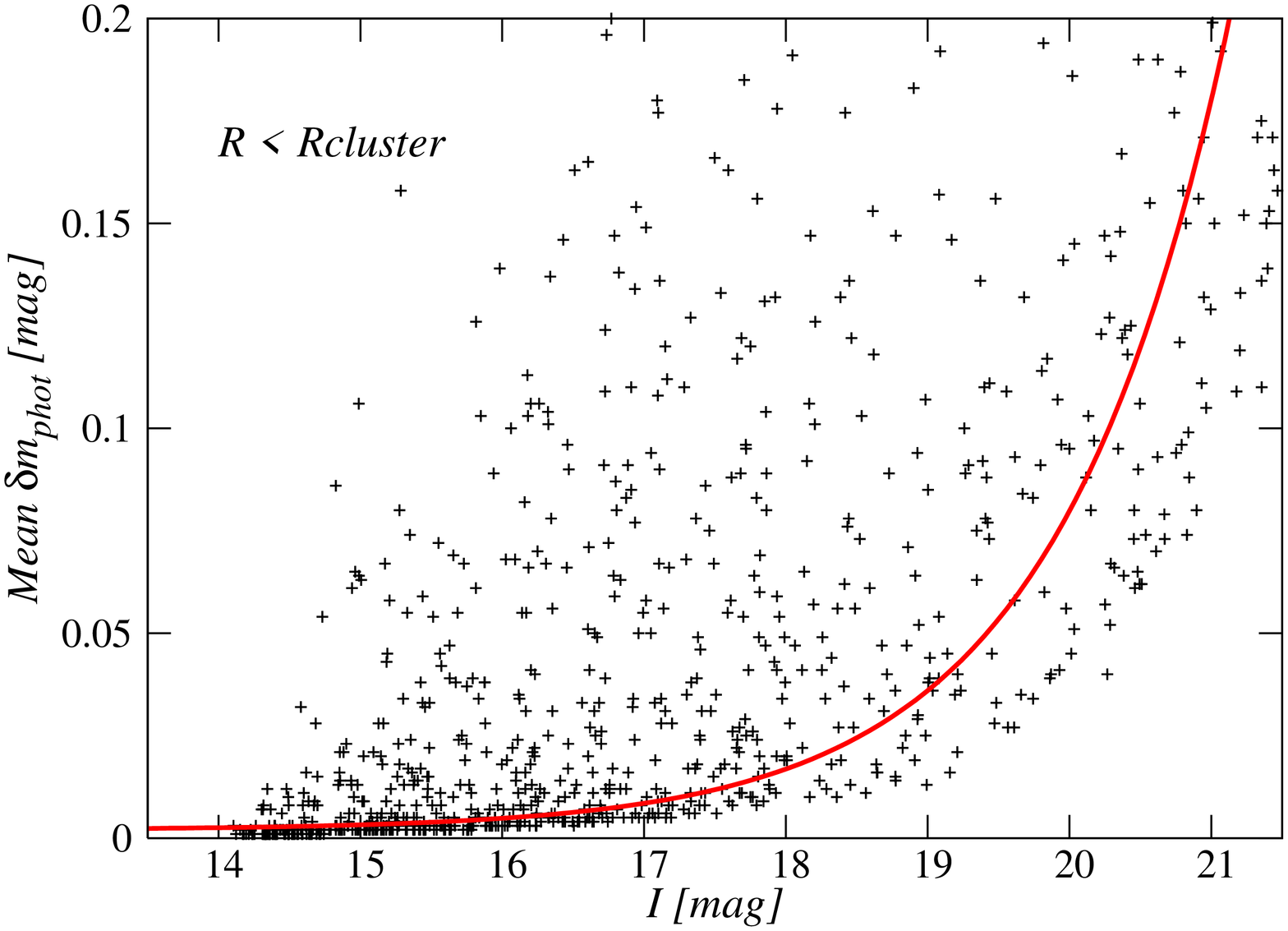}}}\\
\subfigure[]{\rotatebox{270}{\includegraphics[width=5cm,height=8.2cm]{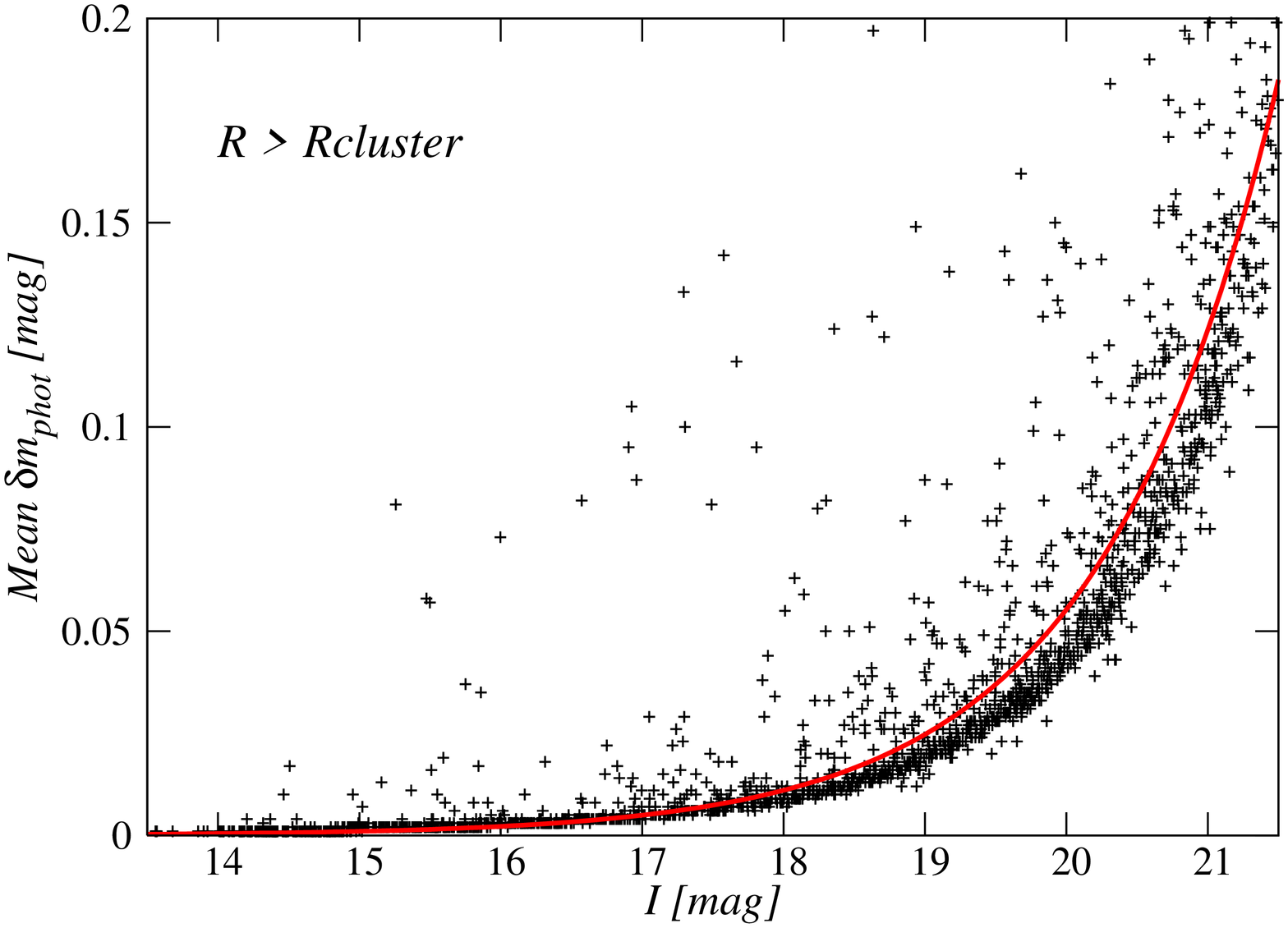}}}
\caption{\textit{a}) The mean photometric errors ($<\delta m_{phot}>$) of all the monitored stars as a function of I. The error curve, which was calculated as a two component fit function (linear for the bright regime, and exponential for the faint regime), is also shown. \textit{b}) As in \textit{a}) but for the stars located inside the so-called cluster radius (see text) while in \textit{c}) objects located outside the same radius are shown. In \textit{b}) and \textit{c}) the error curves correspond to exponential fit functions. The high dispersion in the photometric errors measured for objects located inside the cluster radius (where the nebular background is very strong) is evident from panel \textit{b}).
              }
         \label{FigPhoterrors}
   \end{figure}
\section{Astrometry}
The celestial coordinates of all the objects in the field were computed
using the Guided Star Catalogue 2.3.2 as the reference coordinate list. The average rms in the plate
solution fit was 0\farcs15. The coordinates were checked for stars in
different regions of each \@CCD. The highest positional differences reached
1\farcs1, although these cases were very rare and the differences were normally
not higher than 0\farcs4.
A cross-identification of our sources with H2002 and 2MASS was done. Since
our study extends to much fainter sources, we have a considerably larger number of sources (about 1000)
which are not included in any of these catalogues. In all tables given here we keep the numbering system by H2002 and extend
it for the new objects found in this study by using numbers beginning at 20000. In total our source list contains about 1600 new entries compared to H2002.\\
\section{Photometry}
Aperture photometry was used to obtain the instrumental
magnitudes. Even in the central region of the ONC the stellar density is sufficiently low for this technique to be
accurate enough. The aperture was chosen to be nine pixels (2\farcs1) in diameter in the four inner CCDs
and of ten pixels (2\farcs38) in the four corner ones. These values maximise the S/N
and the somewhat smaller apertures for the four inner CCDs provide lower contamination due to nebular background structures which are stronger in the center of our field. The sky was calculated as the median of an annulus with an inner diameter of 20 pixels
(4\farcs8) and a width of ten pixels (2\farcs4) centered on the source. Close
companions (closer than $\sim$\,40 pixels, i.e. 9\farcs5) were removed from the source list, because of their non-reliable aperture photometry magnitude determination. Objects located close to the edges of the CCDs as well as those with high photometric errors (i.e. $>$\,0.5 mag) were also removed from our sample. In the end, accurate photometry of
2908 stars was performed with photometric errors ranging from 0.001\,mag
to 0.2\,mag for the brighter and fainter limits, respectively. Our instrumental
magnitude system was placed roughly on the Cousins I filter system,
following H2002. Since we observed in only one filter, a proper transformation to this system including a colour term
was not possible. 
In Table 1 our photometric results for all measured (2908) objects are summarised (full version
is available electronically). The magnitude range extends from 13
to 21 mag in I, i.e. three magnitudes deeper than H2002. \\
The mean photometric errors of all monitored stars were calculated as:
\begin{equation}
     {\mathrm{<\delta{m{_{phot}}}>}} = \frac{1}{N_B}\displaystyle\sum^{N_B}_{j=1}\mathrm{\delta{m{_{phot}}}} ,
 \end{equation}

where \textit{N$\mathrm{_{B}}$} indicates the number of images and $\delta{m{_{phot}}}$ is the photometric error of a single measurement. Fig.\,3 shows $<\delta{m{_{phot}}}>$ as a function of magnitude and the error curve, which was calculated as a two component fit to the median values of $\delta{m{_{phot}}}$. The medians were calculated in magnitude bins of 0.5 magnitudes. For objects brighter than 17, a linear fit was applied, while for objects fainter than 17, an exponential fit was used. Above 16.5 the mean photometric errors start to increase slightly, while above I\,=\,18 the increase of the mean photometric errors is rather strong. This introduces a bias in our results, in the sense that it is more likely to measure somewhat higher amplitude modulations for the fainter objects in our sample. We will address this effect again in Sect. 7.
\subsection{Relative photometry}
Relative photometry of each individual chip was performed in order to obtain the light-curves of all the 2908 sources. This relative photometry is performed with respect to carefully selected non-variable flux reference stars. We followed the procedure adopted by Choi \& Herbst
(1996), in which the light-curves are calibrated using
an average light-curve constructed from a set of non-variable reference stars:
An initial sample of potential reference stars is formed using the following criteria:

\begin{enumerate}
 \item Presence in all the images.
      \item Photometric errors below 0.03\,mag.
      \item Sufficiently isolated from other sources.
\end{enumerate}
The quality of all images was checked before determining the final reference
star sample. For this purpose we used the approach described by Scholz \& Eisl\"offel (2004)
and is summarised as follows:

\begin{enumerate}
\item Compute the average instrumental magnitude $\bar{\mathrm{m_i}}$ for each potential reference star and subtract it from each value of the time series.
\begin{equation}
     \bar{m}_i = \frac{1}{N_B}\displaystyle\sum^{N_B}_{j=1} m_i(j),
 \end{equation}
and\\
\begin{equation}
\mathrm{m{^{0}_i}} = m_i(j)-\bar{m}_i,
\end{equation}
in which the \textit{i}=  1...\textit{N$\mathrm{_{R}}$} suffix indicates the number of the reference star while
\textit{j}=1...\textit{N$\mathrm{_{B}}$} indicates the image number. \\
\item Calculate the average and standard deviation of the previous differences for each individual image. 
\begin{equation}
     \bar{\mathrm{m{^{0}_j}}} = \frac{1}{N_R}\displaystyle\sum^{N_R}_{i=1} \mathrm{m{^{0}_i}} ,
 \end{equation}
and\\
\begin{equation}
     \sigma_{j} = \sqrt{\frac{1}{(N_R-1)}\displaystyle\sum^{N_R}_{i=1}\mathrm{(m{^{0}_i}-\bar{\mathrm{m{^{0}_j}}})^{2}}} ,
 \end{equation}
\item Reject the so-called ``bad'' images in the time series, which are the images with high standard deviation.
\end{enumerate}
Once a few bad images were removed (i.e. three to five depending on the CCD), potential reference stars showing indications for variability were rejected by examining the standard deviation of each light-curve with respect to the average light-curve of all other stars. This procedure
was iterated a few times, and about 30 stars relatively evenly distributed around the field were finally chosen on each CCD. Since most of the objects in the field are young and presumably variable, using many reference stars not only increases the precision in the relative photometry, but also strongly minimises the effect of very low level variations that any of the selected reference stars could still have. For this reference star sample the individual standard deviation $\sigma_{i}$ in the light-curves was typically below 0.007\,mag, where $\sigma_{i}$ is defined as:

\begin{equation}
     \sigma_{i} = \sqrt{\frac{1}{(N_B-1)}\displaystyle\sum^{N_B}_{j=1}\mathrm{(m{_j}-\bar{\mathrm{m{_i}}})^{2}}} ,
 \end{equation}
in which again the \textit{i}=  1...\textit{N$\mathrm{_{R}}$} suffix indicates the number of the reference star, while
\textit{j}=1...\textit{N$\mathrm{_{B}}$} indicates the image number. \\
For the final relative photometry, the averaged reference light-curve of the reference stars in each of the eight CCDs was subtracted from all 2908 stars. The light-curves of three periodic variables and a flux reference star are shown in Fig.\,4 to illustrate the sampling of the data, and the variability level of the periodic variables compared to a reference star. The light curves of these three periodic variables phased with the found period are also shown in Fig.\,4. 
\begin{table*}[t!]
\renewcommand{\arraystretch}{1.2}  
\begin{minipage}[]{2\columnwidth}
    \centering
\caption{Sample of the photometric catalogue. The complete catalogue is available electronically at the CDS.}
\label{PHOTtable}
\addtolength{\tabcolsep}{5pt}
\begin{tabular}{ccccccccc} 
\hline\hline             
ID\footnote{Identification numbers from Hillenbrand (1997) for numbers below 10000; from H2002 for numbers between 10000 and 11115, and from this study for identification numbers above 20000.} & RA(2000) & DEC(2000) & I\footnote{I magnitude.}[mag] & I$_{err}$\footnote{Photometric I magnitude error.}[mag] & N \footnote{Number of data points considered in the time series analysis.}& CCD\footnote{CCD identification. Each CCD of the mosaic is named from \textit{a} to \textit{h}, starting from the upper left corner and going clockwise to the bottom left corner.} & VarFlag\footnote{ The variability flag column catalogues all objects from their variation properties. The possibilities are: non-variable (NonV), irregular variable (IV), periodic variable (PV), possible periodic variable (PPV) and possible eclipsing system (PES).}& 2MASS-ID\footnote{2MASS identification number.}\\
\hline
1&	5:34:28.21&	-5:26:39.51&	15.535&	0.002&	80&    e&      PV&     05342821-0526397\\
10&	5:34:30.24&	-5:11:48.00&	13.859&	0.001&	87&    d&      IV&     05343024-0511481\\
89&	5:34:47.65&	-5:37:24.91&	20.885&	0.141&	53&    f&      IV&     05344765-0537250\\
3049&	5:35:36.49&	-5:20:8.96&	16.431&	0.004&	74&    b&      IV&     05353650-0520094\\
3051&	5:34:26.78&	-5:19:41.19&	16.642&	0.004&	87&    d&      PV&     05342678-0519414\\
3056&	5:35:10.18&	-5:21:0.29&	16.758&	0.013&	85&    c&      PV\\
5050&	5:35:10.07&	-5:17:6.52&	18.389&	0.019&	85&    c&      PV&     05351007-0517068\\
10667&	5:35:28.51&	-5:25:4.12&	16.253&	0.009&	90&    g&      IV\\
10668&	5:35:28.56&	-5:33:4.25&	18.611&	0.016&	90&    g&      PV&     05352856-0533043\\
10669&	5:35:28.65&	-5:22:24.89&	17.863&	0.080&	74&    b&      NonV&   05352865-0522253\\
10670&	5:35:28.88&	-5:29:14.40&	17.847&	0.017&	90&    g&      PV&     05352889-0529144\\
11102&	5:36:20.46&	-5:29:47.99&	19.894&	0.037&	77&    h&      NonV\\  
11111&	5:36:21.20&	-5:29:12.61&	17.889&	0.008&	81&    h&      NonV&   05362123-0529128\\
11112&	5:36:21.25&	-5:18:26.71&	18.168&	0.010&	82&    a&      NonV&   05362126-0518266\\
11114&	5:36:21.29&	-5:20:12.27&	15.399&	0.002&	82&    a&      IV\\    
11115&	5:36:21.58&	-5:36:37.37&	17.953&	0.008&	79&    h&      PV\\    
20000&	5:35:16.05&	-5:23:27.33&	15.474&	0.058&	51&    f&      IV\\    
20022&	5:35:14.82&	-5:23:5.47&	17.119&	0.337&	85&    c&      PV&     05351487-0523050\\
20245&	5:35:12.84&	-5:21:4.92&	18.535&	0.103&	85&    c&      PV&     05351285-0521051\\
20440&	5:35:16.60&	-5:17:23.26&	17.623&	0.008&	85&    c&      PV&     05351661-0517234\\
20441&	5:35:10.81&	-5:17:32.90&	18.932&	0.030&	85&    c&      NonV&   05351080-0517332\\
20442&	5:35:34.14&	-5:27:26.98&	19.078&	0.039&	90&    g&      NonV&   05353415-0527271\\
20443&	5:35:34.90&	-5:27:16.76&	20.833&	0.258&	90&    g&      NonV\\  
21231&	5:35:40.14&	-5:37:23.63&	20.074&	0.049&	89&    g&      PV\\    
21234&	5:34:42.10&	-5:35:56.59&	21.378&	0.174&	75&    e&      NonV\\
\hline
\end{tabular}
\renewcommand{\footnoterule}{} 
\end{minipage}
\end{table*}
\begin{figure*}
   \centering
   \subfigure{\includegraphics[width=13cm, height=11cm]{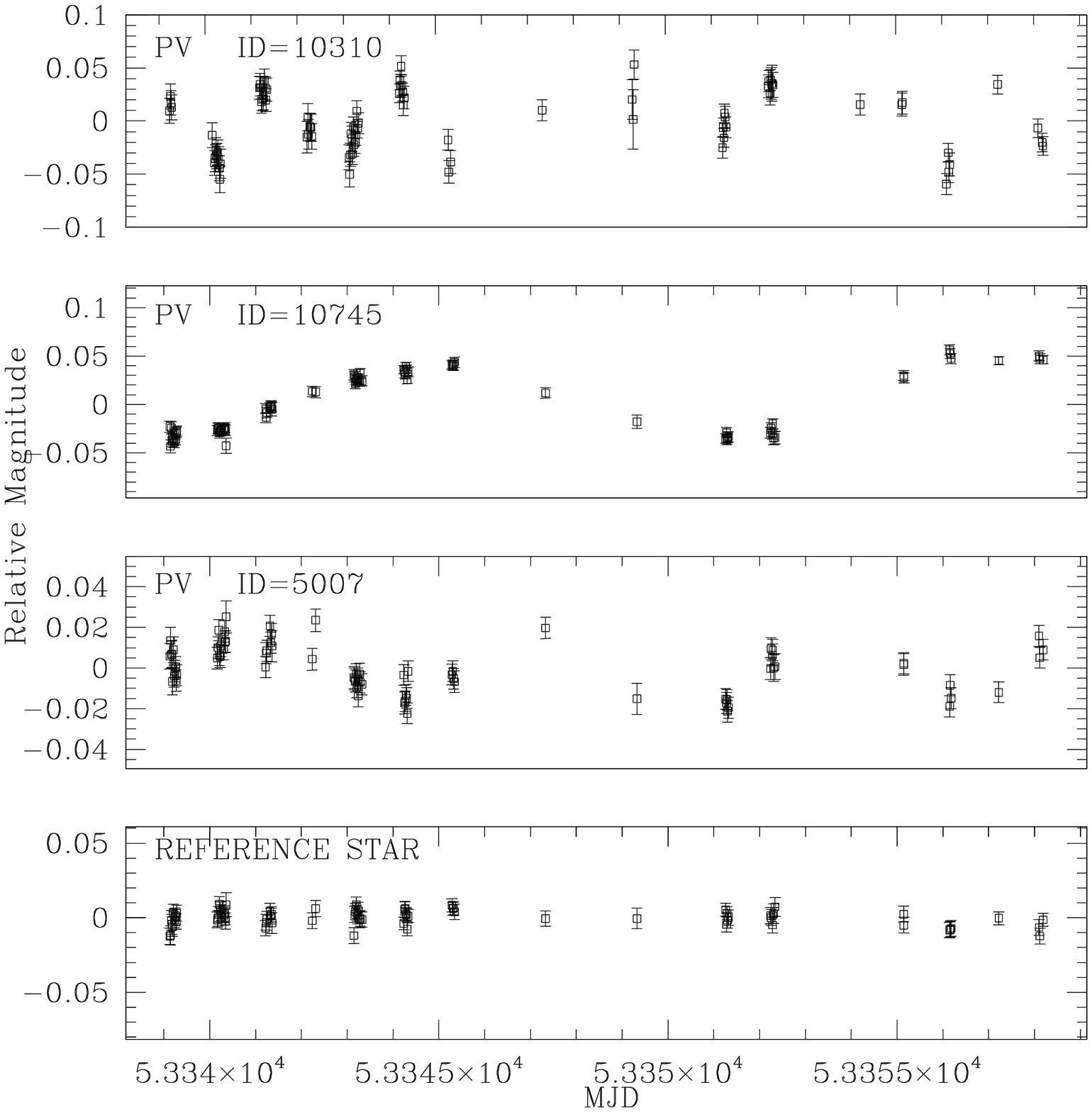}}\\
\subfigure{\includegraphics [width=6cm, height=4.5cm]{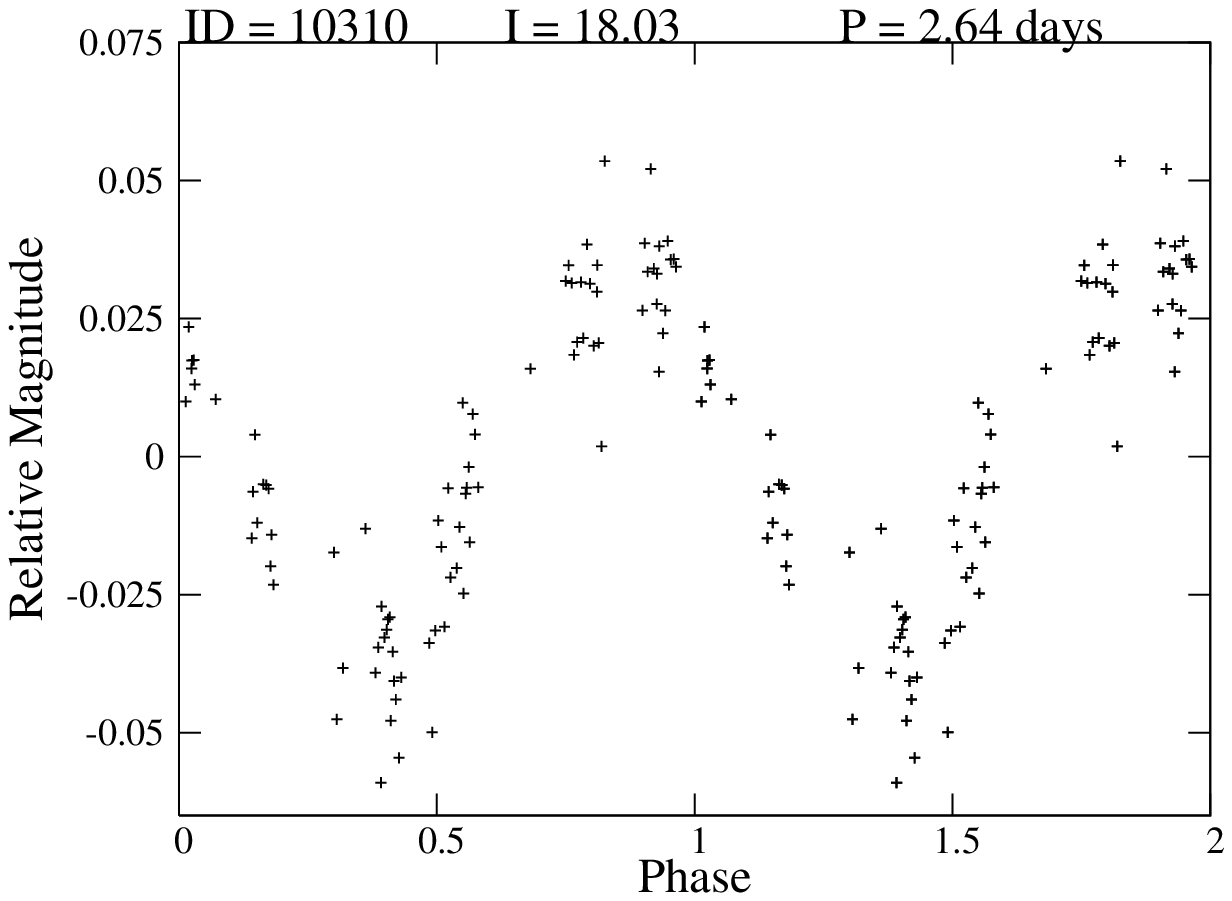}}
\subfigure{\includegraphics [width=6cm, height=4.5cm]{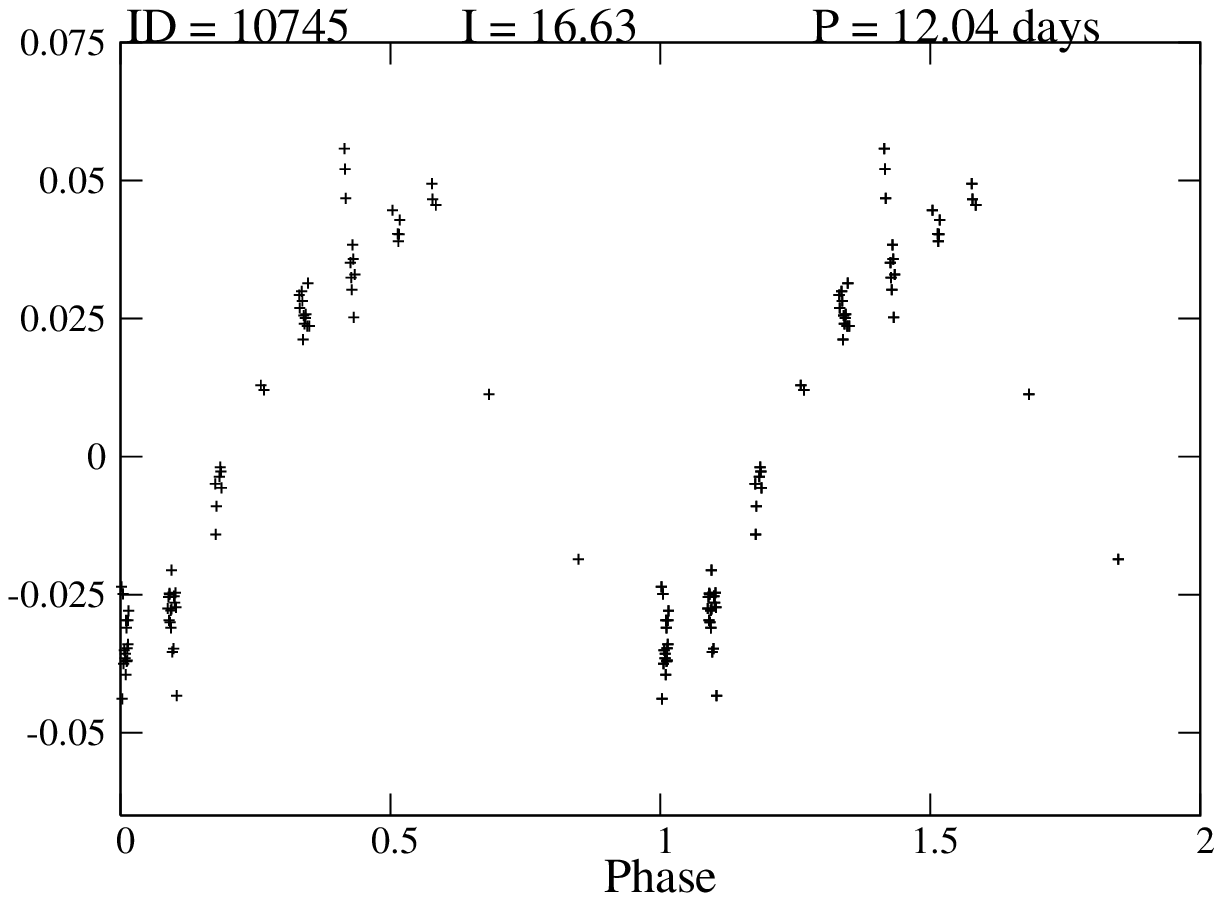}}
\subfigure{\includegraphics [width=6cm, height=4.5cm]{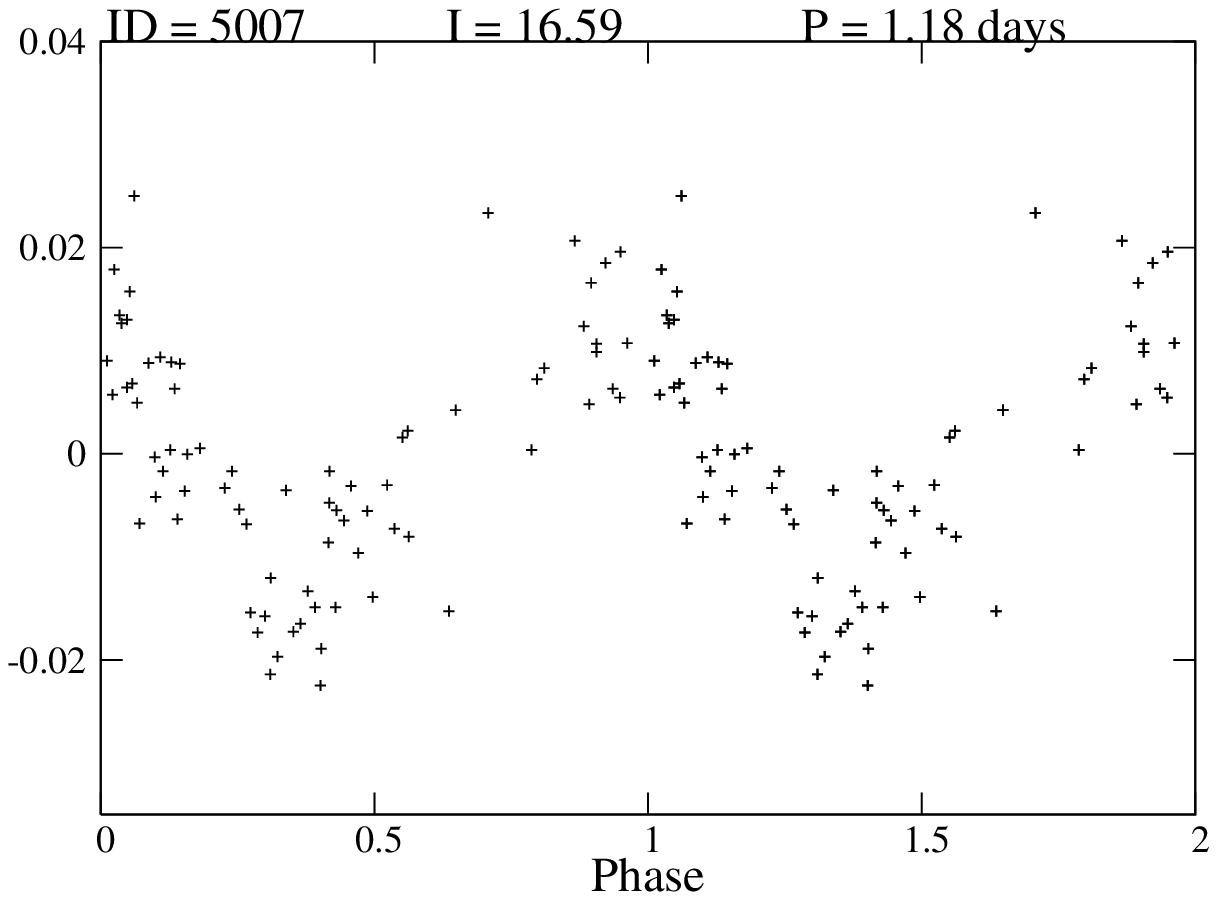}}
\caption{ Examples of light curves of three periodic variable stars and a reference star (top panels). For the three periodic variable stars the light curves phased with the found period are also shown (bottom panels). The I magnitude as well as the period in days is given for all three examples. Note the different amplitude scales which gives an idea of the variability level of the objects.              }
         \label{FigLightcurves}
   \end{figure*}
\begin{figure}
\includegraphics[width=8cm]{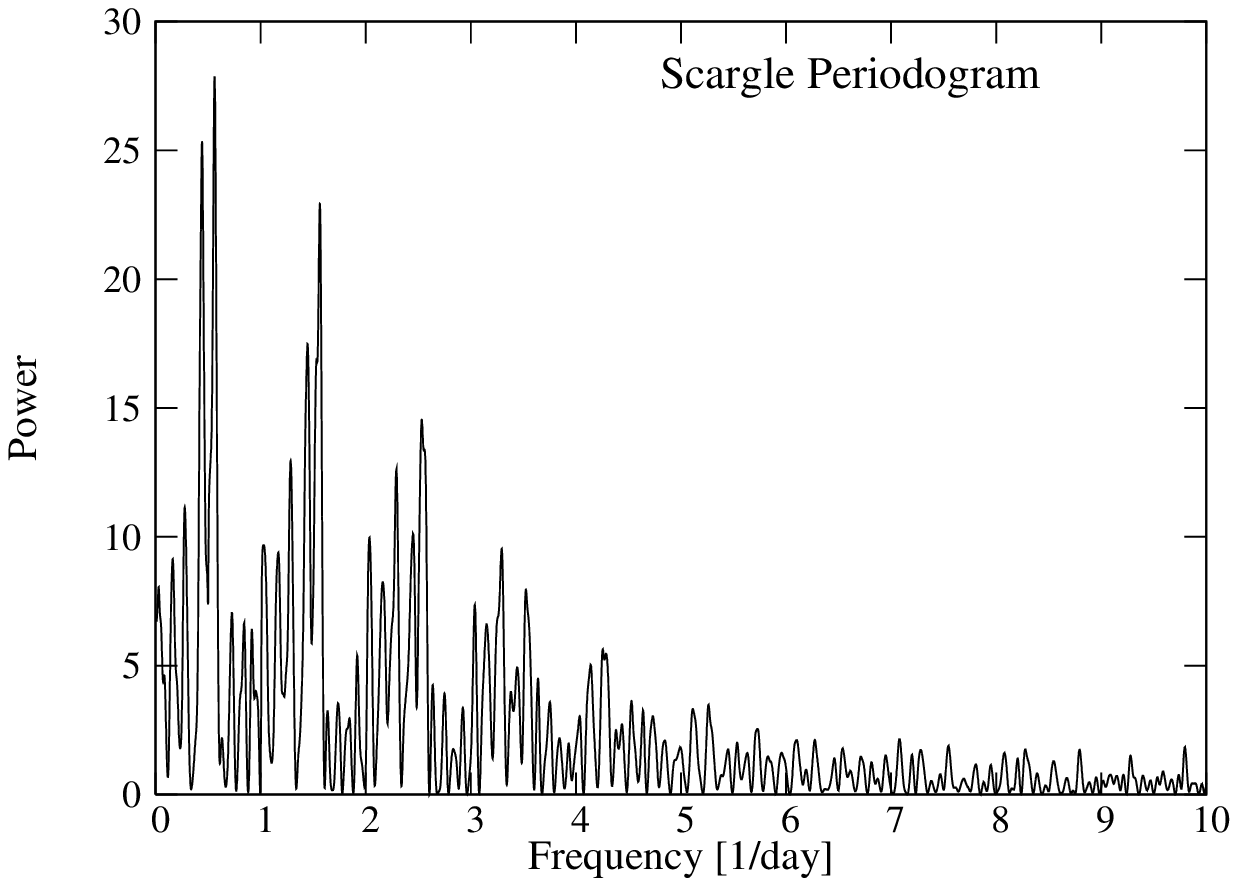}
\includegraphics[width=8cm]{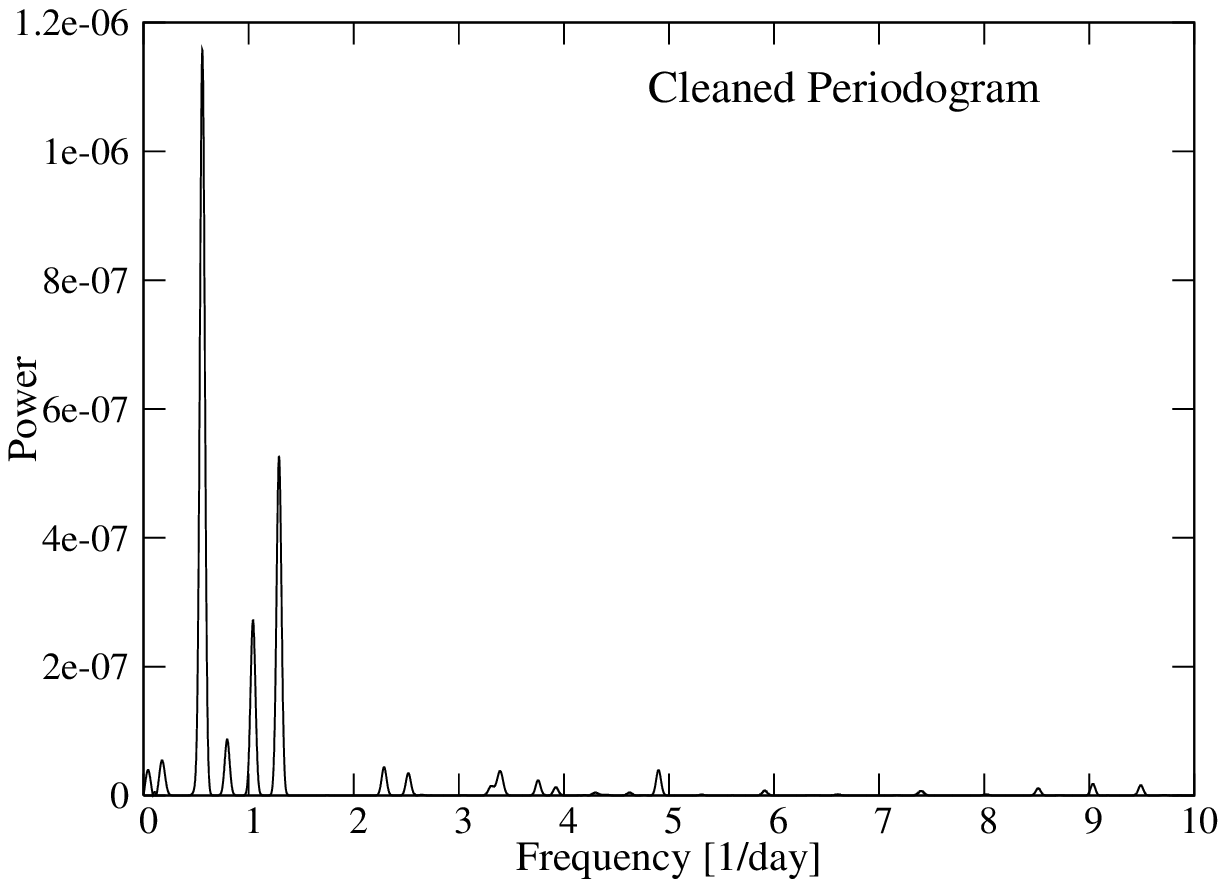}
\includegraphics[width=8cm]{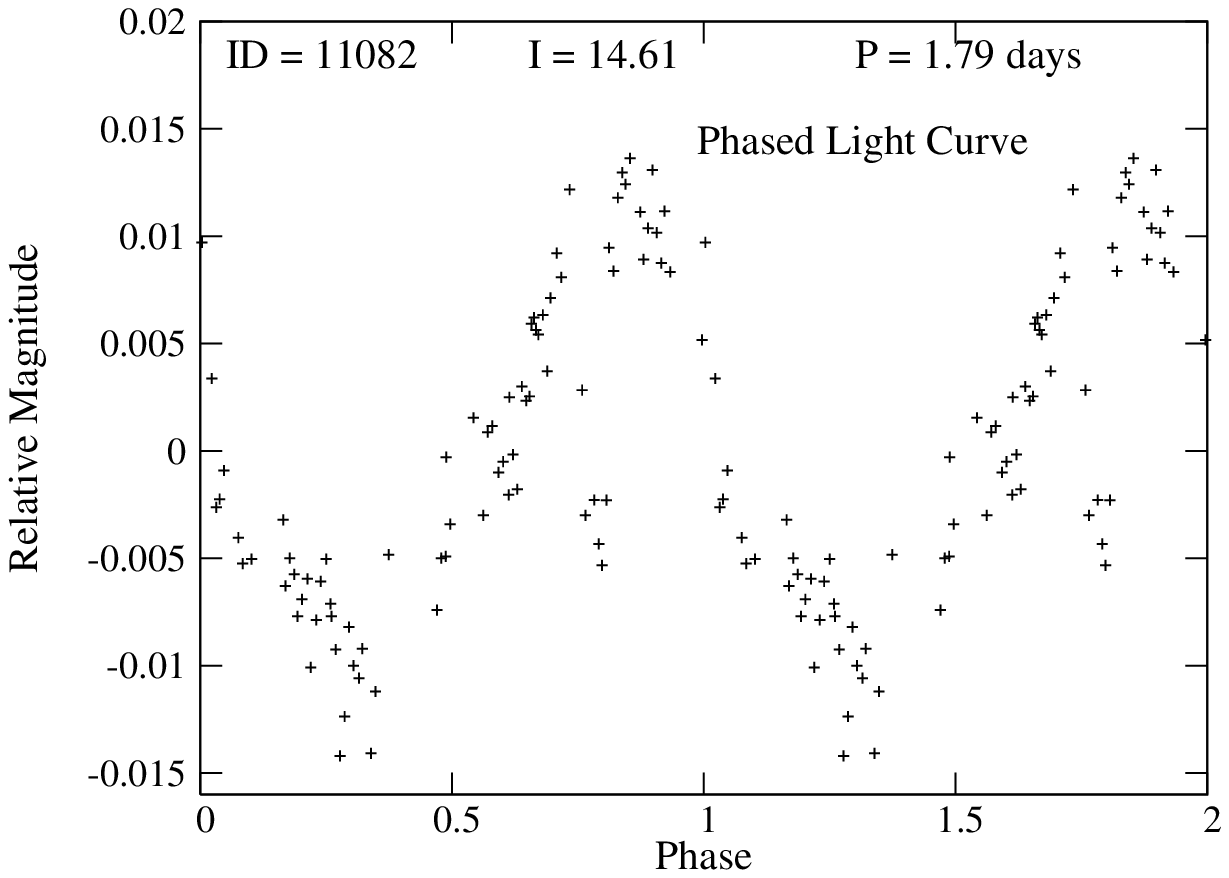}
\caption{Example of a Scargle periodogram, a CLEAN periodogram and the phased light-curve of a periodic variable with P\,=\,1.79\,days found by both techniques.
        }
\label{FigScargle}
\end{figure}
\section{Time series analysis}
\subsection{Methods}
In order to obtain as large as possible a sample of young objects with known periods, all stars in the field were checked for periodic light modulations using a combination of two periodogram analysis techniques: the Scargle periodogram (Scargle 1982) and the CLEAN algorithm (Roberts 1987). The Scargle periodogram analysis is particularly suited to a time series in which the sampling is not uniformly distributed.  In order to decide whether there is a significant signal from a certain period in the power spectrum, the power at that period calculated by the Scargle algorithm has to be related to a false alarm probability (FAP), which is the probability that a peak with a power \textit{z} would appear in a random data set. We derived FAP$_\mathrm{Scargle}$ for all monitored objects and only those with FAP$_\mathrm{Scargle}$ below 1\% were considered in later steps of the analysis. The FAP$_\mathrm{Scargle}$ is given by
\begin{equation}
     FAP_\mathrm{Scargle}=1-\mathrm{([1 - exp(-z)]{^{N_i}}} ,
 \end{equation}
in which \textit{N$\mathrm{_{i}}$} is the number of independent frequencies and \textit{z} the height of the peak. \textit{N$\mathrm{_{i}}$} is estimated as
\begin{equation}
\textit{N$\mathrm{_{i}}$}= -6.3 + 1.2\textit{N} + 0.00098\textit{N$\mathrm{^{2}}$} 
\end{equation}
where \textit{N} is the number of data points (Horne \& Baliunas \cite{Horne}). In the case of clumped data, other authors (e.g. Scholz \& Eisl\"offel, 2004) preferred to used \textit{N$\mathrm{_{i}}$}=\textit{N}/2 for their first estimation of FAP${_\mathrm{Scargle}}$ although these two different approaches do not result in a significant change of FAP$_\mathrm{Scargle}$.\\
Because our data are not evenly spaced, but clumped, the values for FAP$_\mathrm{Scargle}$ are considered as a first estimate in the analysis. Additional FAP determinations, also on the basis of Monte Carlo simulations, were performed. As discussed more in detail below, the FAP$_\mathrm{sim}$ derived from the simulations are similar to those derived from Eq.\,7.\\
Due to the windowing of the data, the Scargle periodograms are contaminated by false peaks or aliases which could result in spurious period detections. To overcome this problem, the CLEAN algorithm by Roberts et al. (1987) was used to deconvolve the dirty spectrum and the window function, resulting in a clean periodogram in which false peaks due to sampling no longer appear, as shown in Fig.\,5. The combination of these two techniques provides a reliable period detection as outlined in several rotational period and variability studies (e.g. Bailer-Jones \& Mundt, 1999; 2001; L2004; L2005; Scholz \& Eisl\"offel, 2004). \\
Independent of the Scargle periodogram FAP calculation, we performed an additional test to control the significance of the periods found. We followed the idea proposed by Scholz \& Eisl\"offel (2004) and used the statistical F-test and a derived FAP from it. Its main idea is the following: if the detected period is not significant, the variance of the light-curve with a periodic light modulation and without it should not be significantly different. To perform this test, each detected period was fitted with a sine wave and then subtracted from the original light-curve. The variance of both the original and the subtracted light-curves were compared as follows
\begin{equation}
     \textit{F}=\frac{\sigma^{2}}{\sigma\mathrm{^{2}_{sub}}} ,
 \end{equation}
resulting in a measurement of the probability that the two variances are equivalent. Because FAP$_\mathrm{F-test}$ represents the probability that the period found is caused by variations in the photometric noise, it is completely independent of the previous periodogram analysis. The values of the FAP$_\mathrm{F-test}$ are larger than the FAP$_\mathrm{Scargle}$ values for the lower S/N cases and smaller for the high S/N ones. \\
We only include in our final list of periodic variables those objects which fulfil the following criteria:
\begin{itemize}
\item A significant peak was found by the Scargle Periodogram analysis with FAP$_\mathrm{Scargle}$\,$<$\,1\%.
\item The CLEAN algorithm confirms the peak found by the Scargle Periodogram analysis.
\item The FAP$_\mathrm{F-test}$\,$<$\,5\%.
\item The light-curves and the phased light-curves show the found period clearly (final visual inspection).
\end{itemize}
\begin{figure*}
\centering
\subfigure{\includegraphics [width=8.5cm]{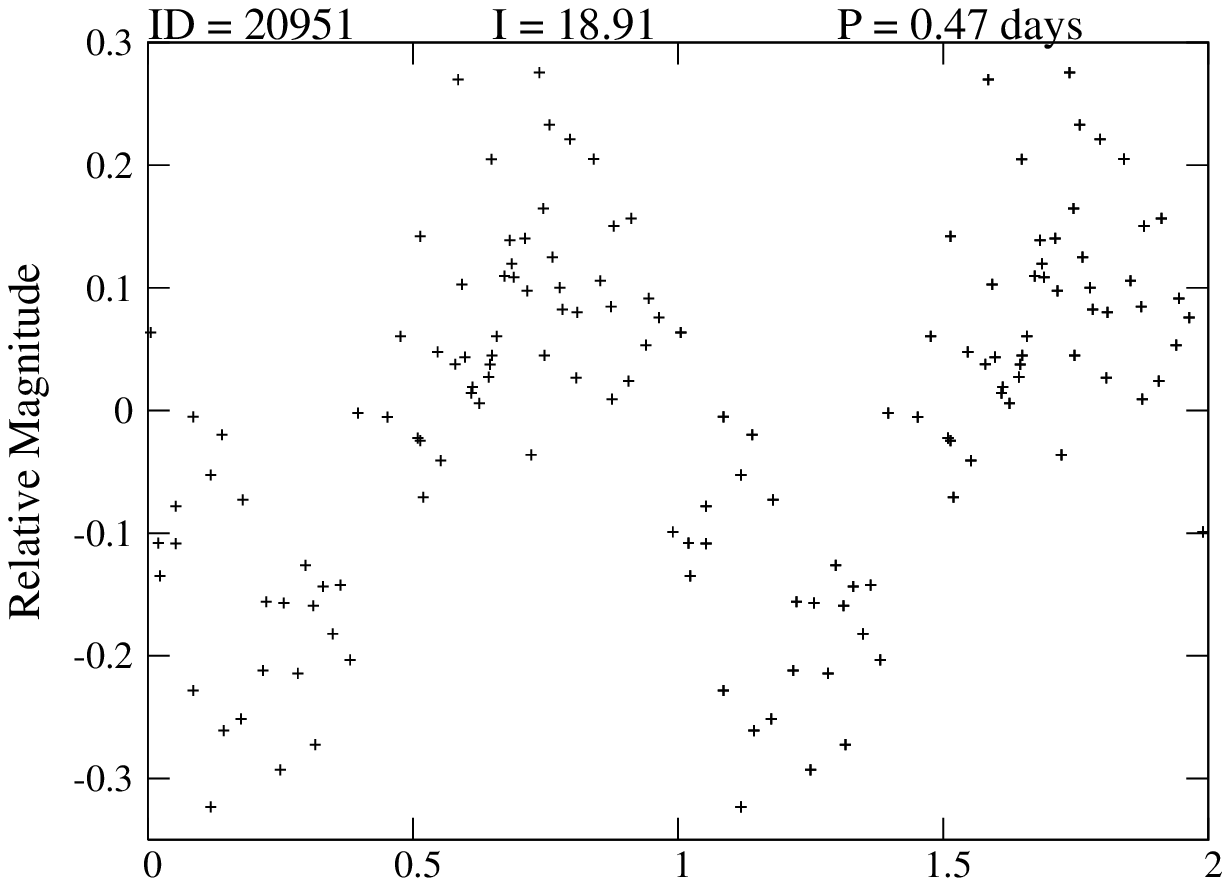}}
\subfigure{\includegraphics [width=8.5cm]{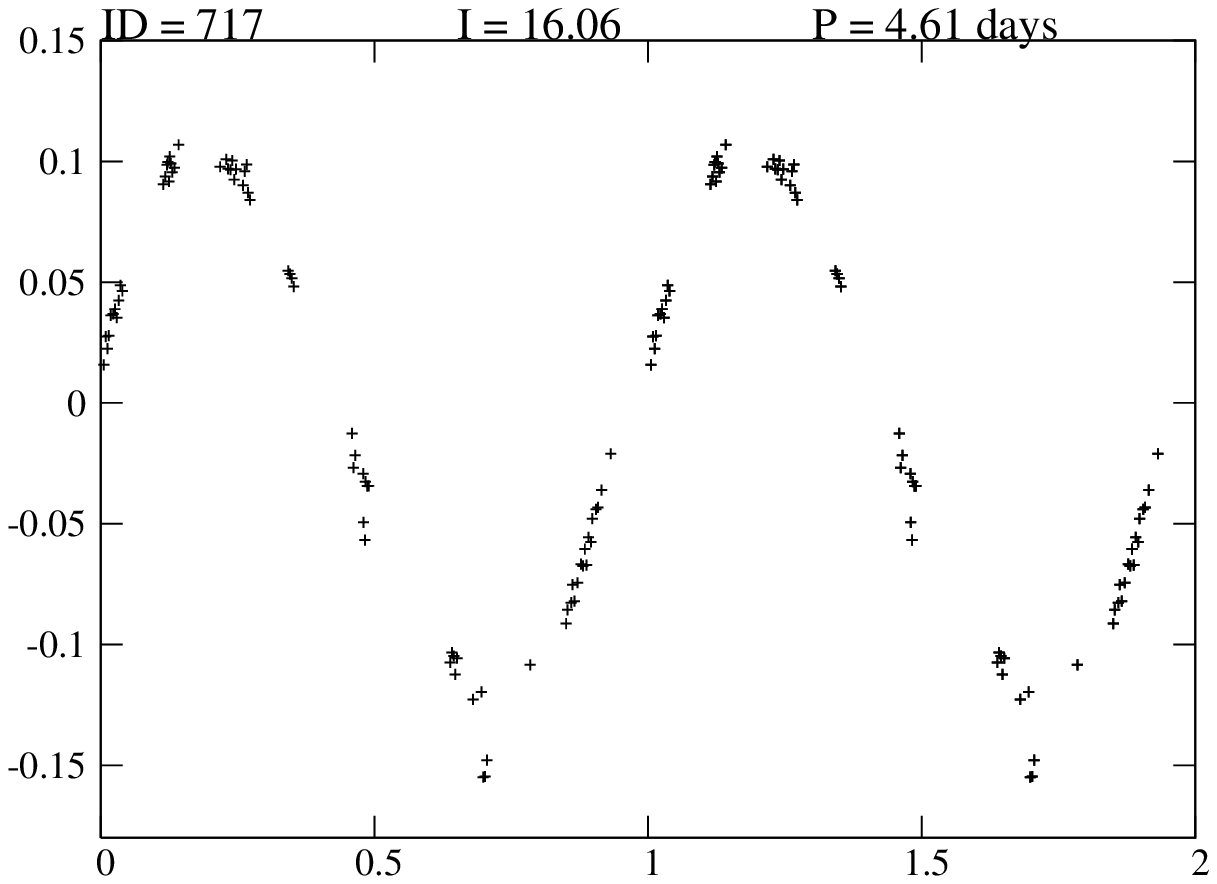}}\\
\subfigure{\includegraphics [width=8.5cm]{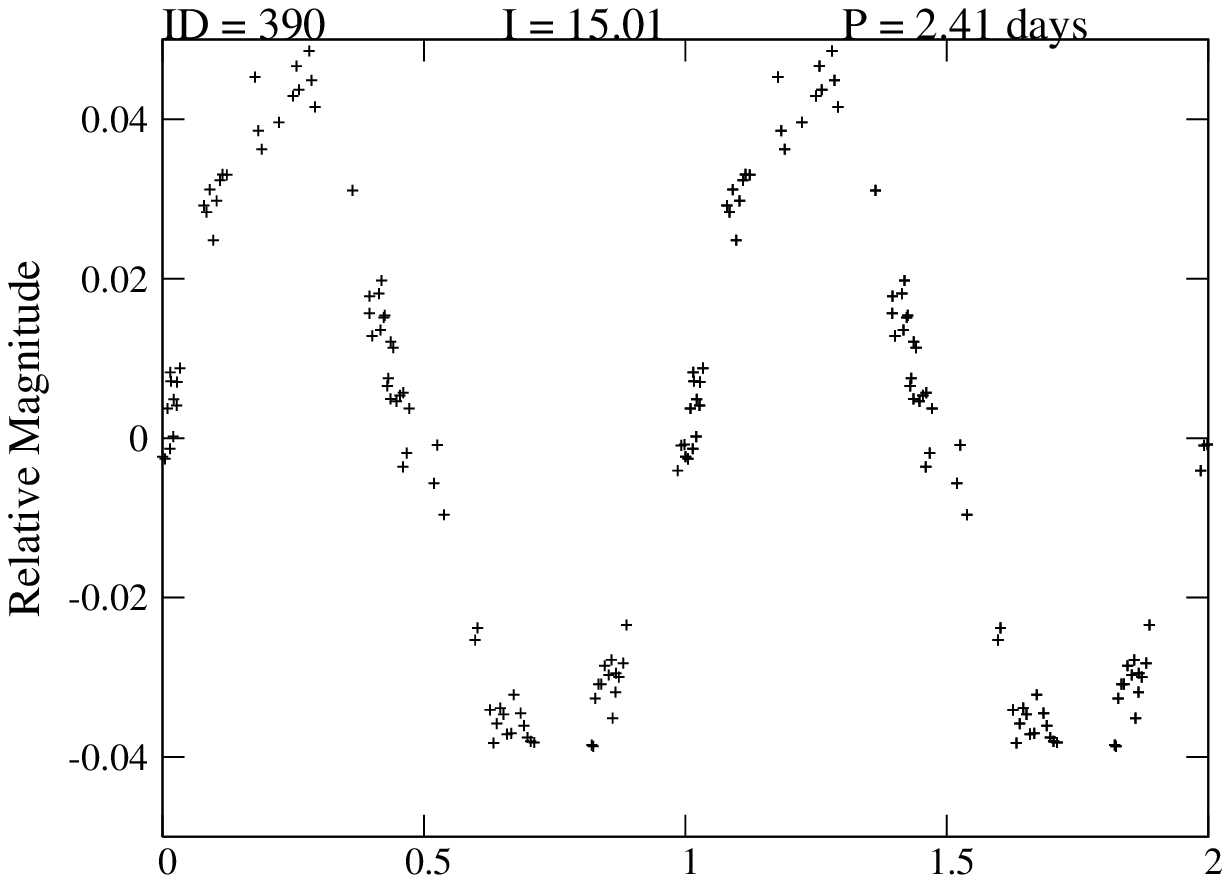}}
\subfigure{\includegraphics [width=8.5cm]{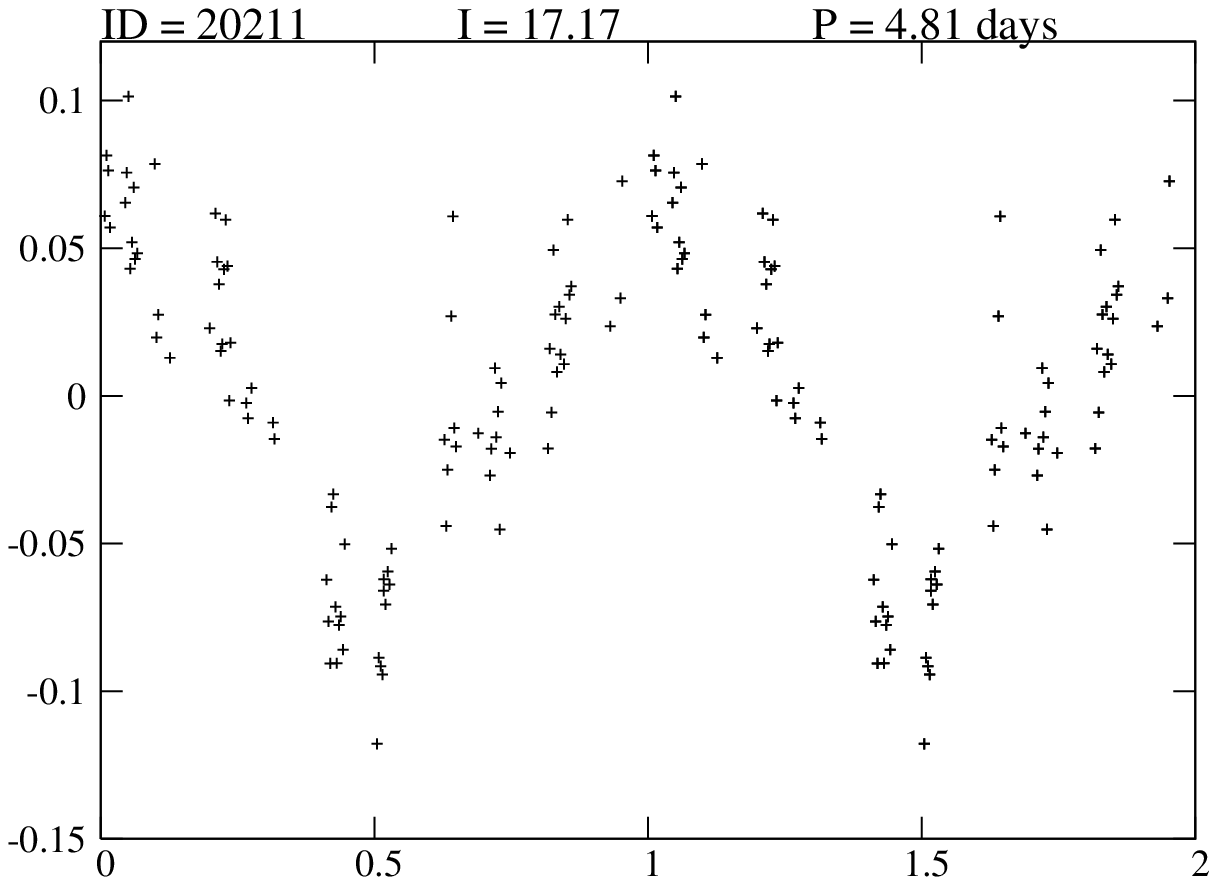}}\\
\subfigure{\includegraphics [width=8.5cm]{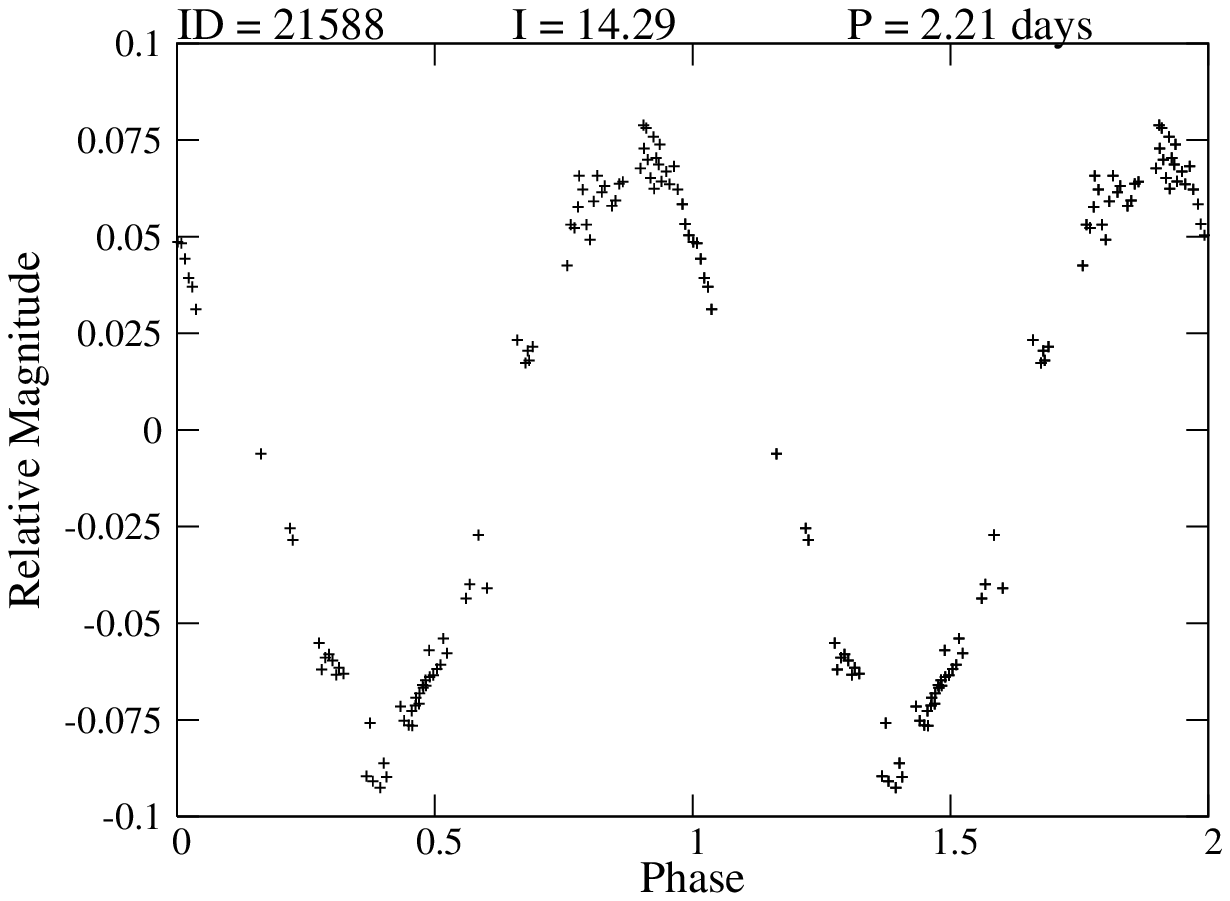}}
\subfigure{\includegraphics [width=8.5cm]{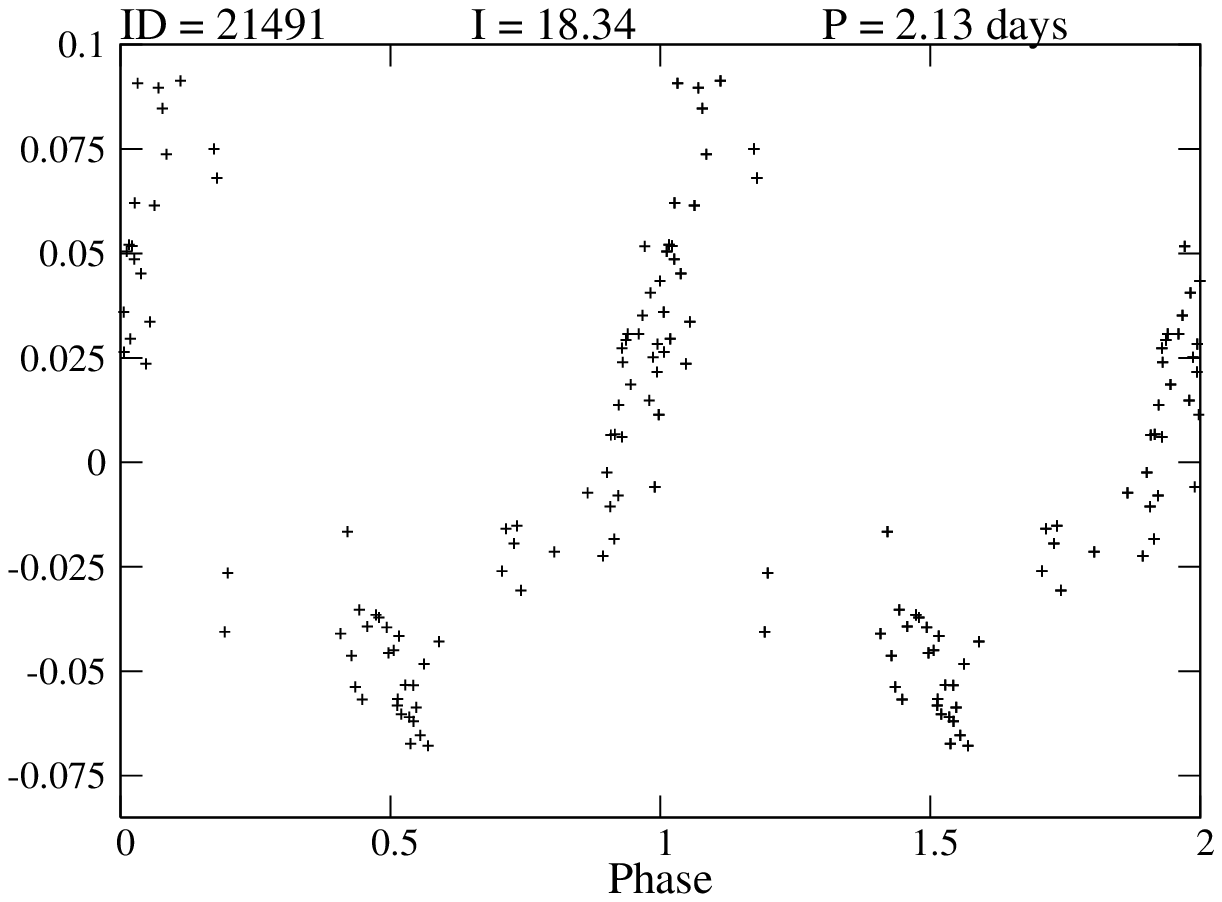}}
\caption{Phased light curves of six periodic variables, which show mostly a quasi-sinusoidal low level light modulation. The ID number of the stars as well as the measured I magnitude and the detected period (in days) are shown for each of the objects.
             }
        \label{FigPLCs1}
  \end{figure*}
\begin{figure*}
\centering
\subfigure{\includegraphics [width=8.5cm]{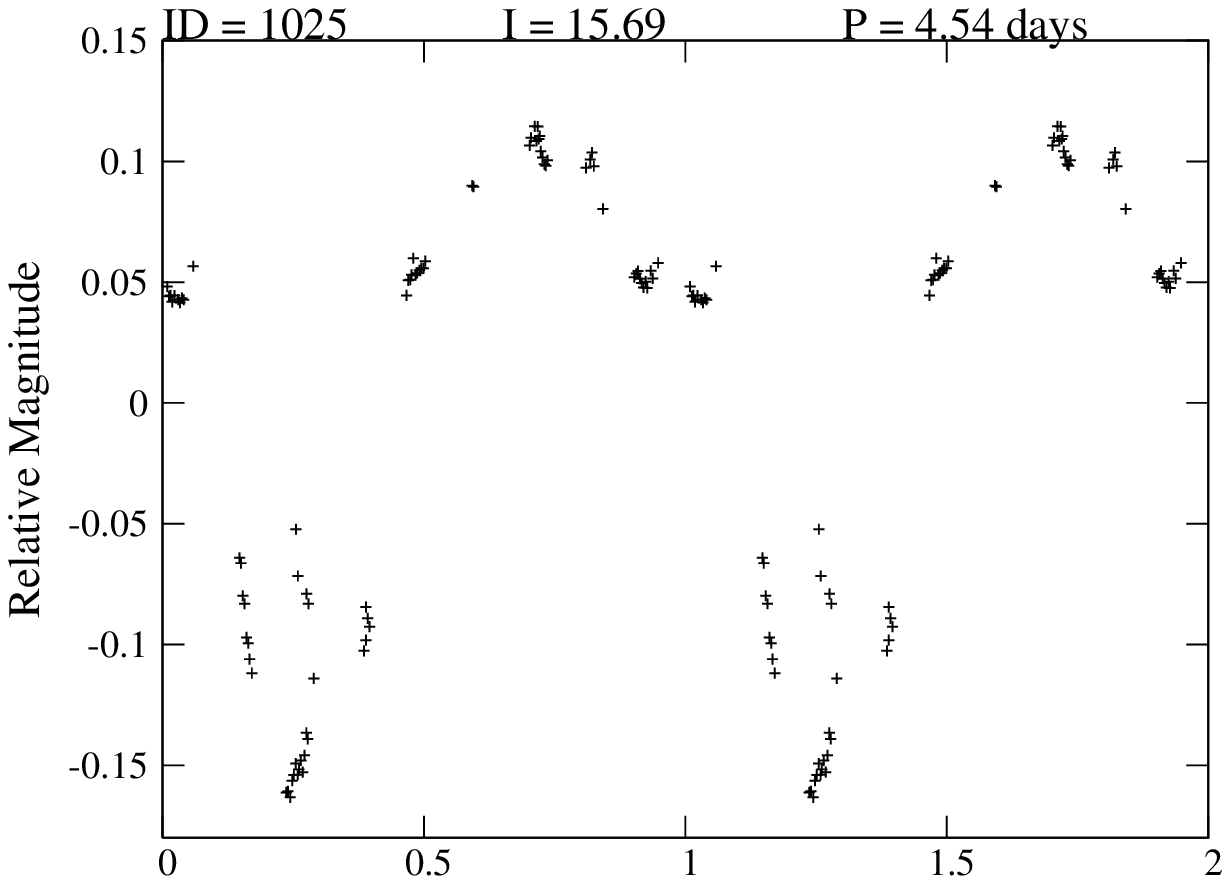}}
\subfigure{\includegraphics [width=8.5cm]{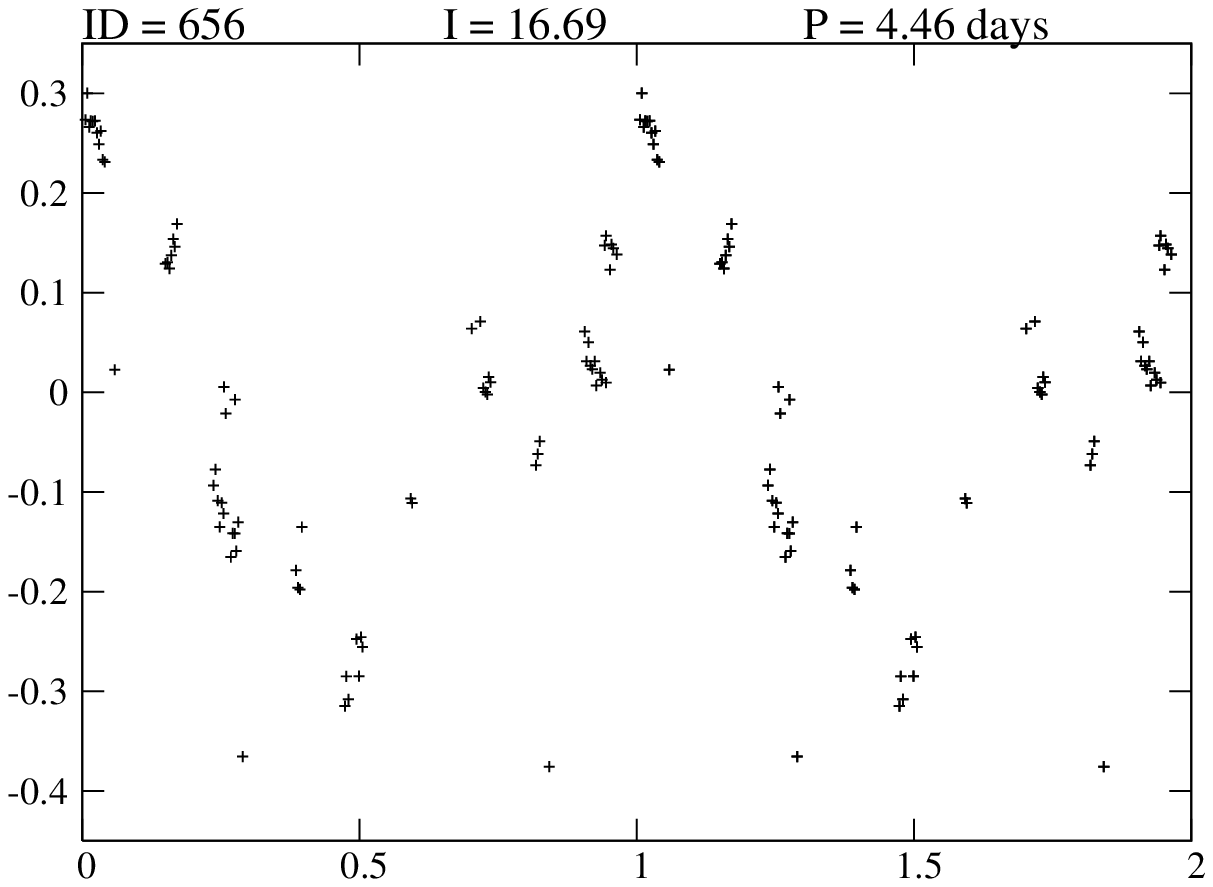}}\\
\subfigure{\includegraphics [width=8.5cm]{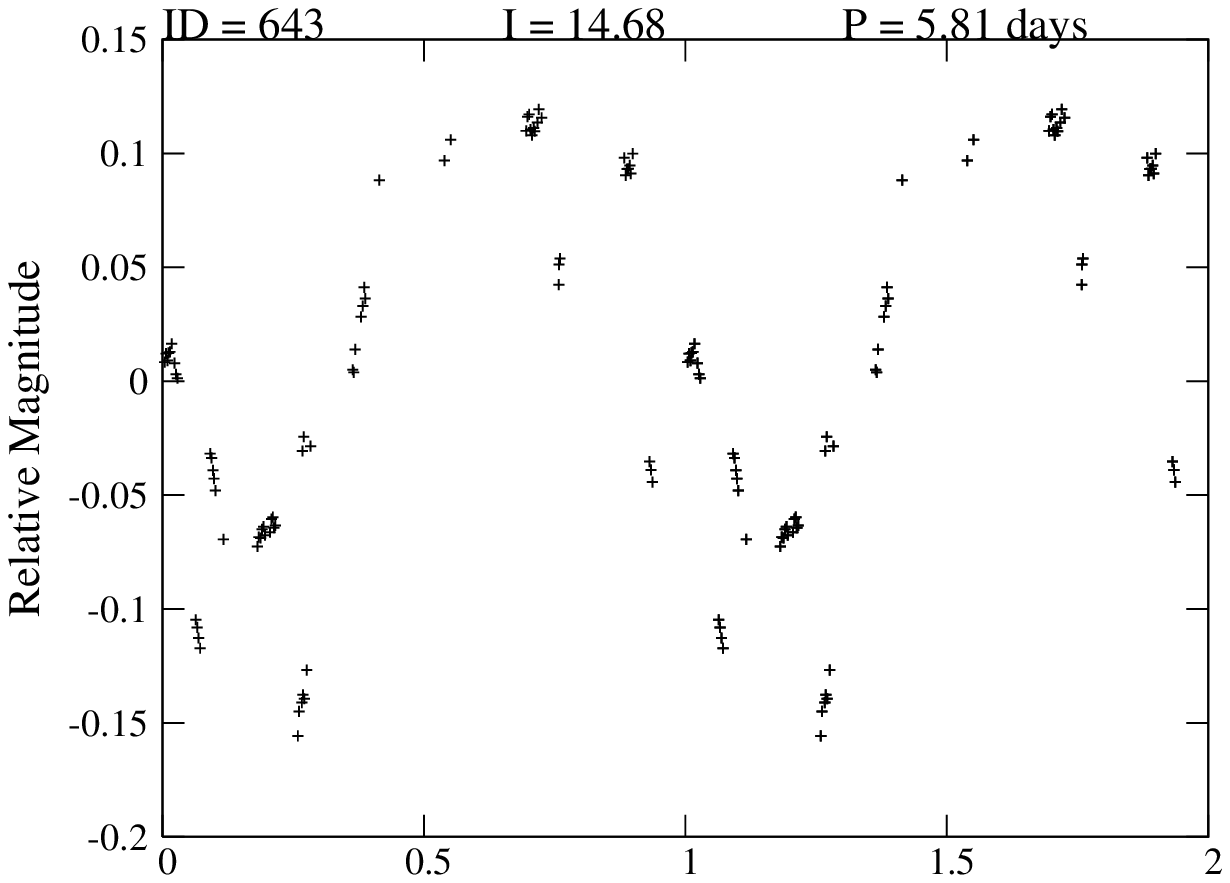}}
\subfigure{\includegraphics [width=8.5cm]{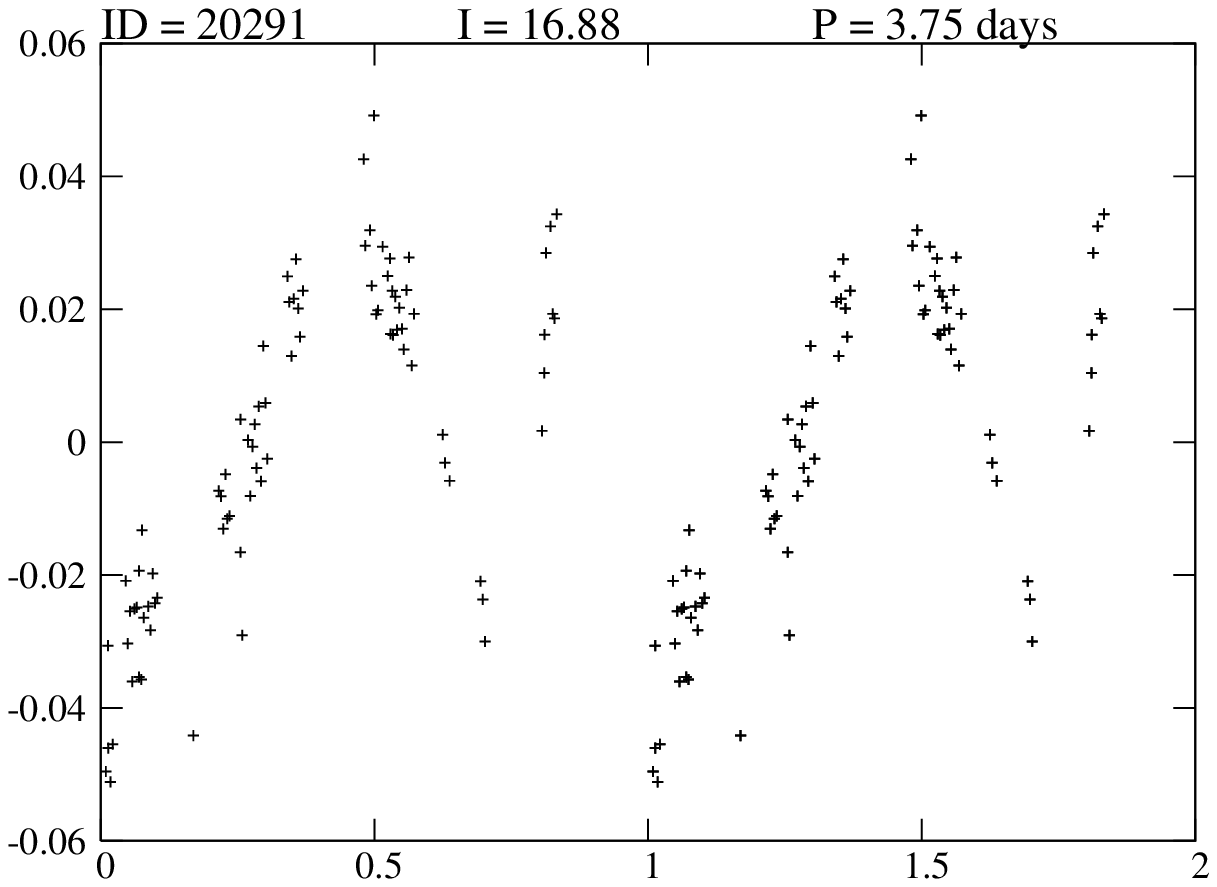}}\\
\subfigure{\includegraphics [width=8.5cm]{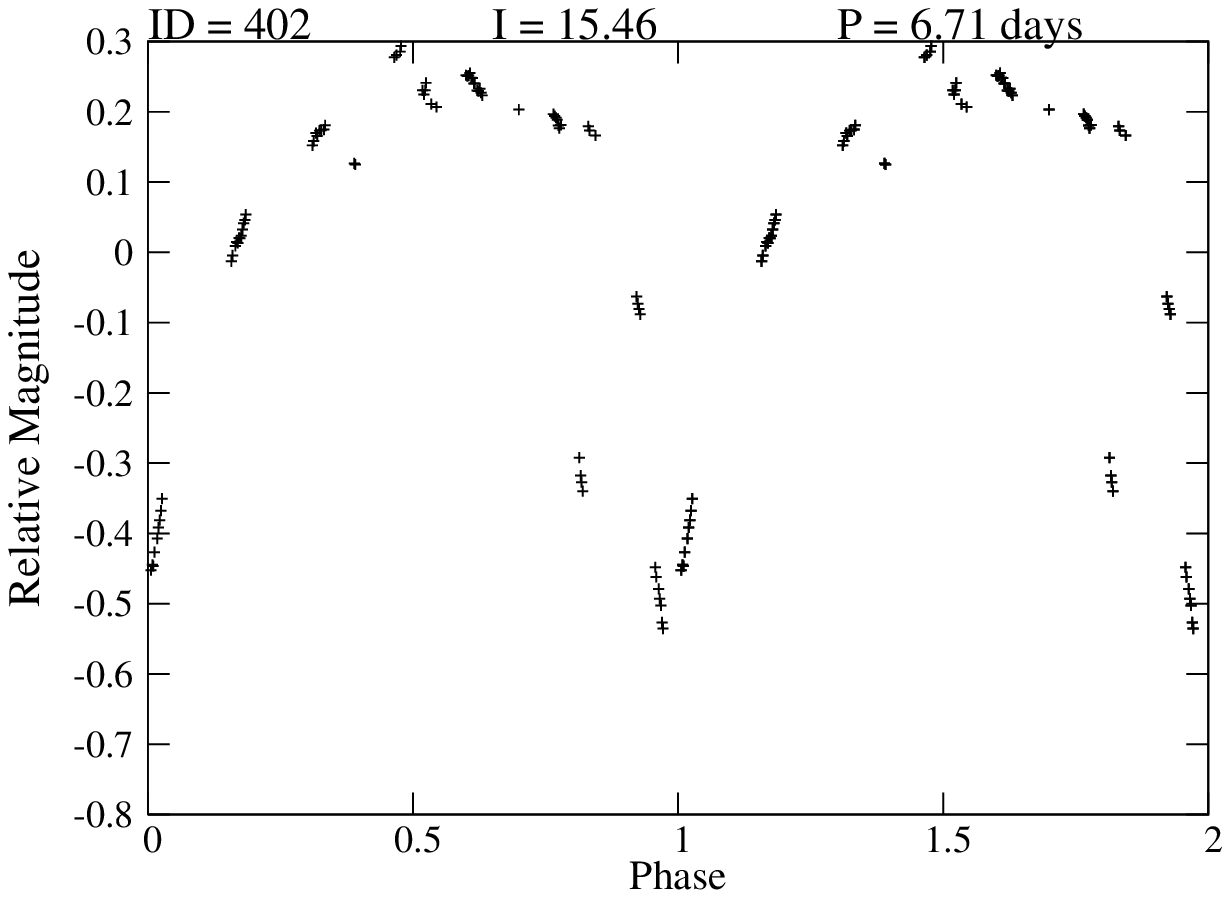}}
\subfigure{\includegraphics [width=8.5cm]{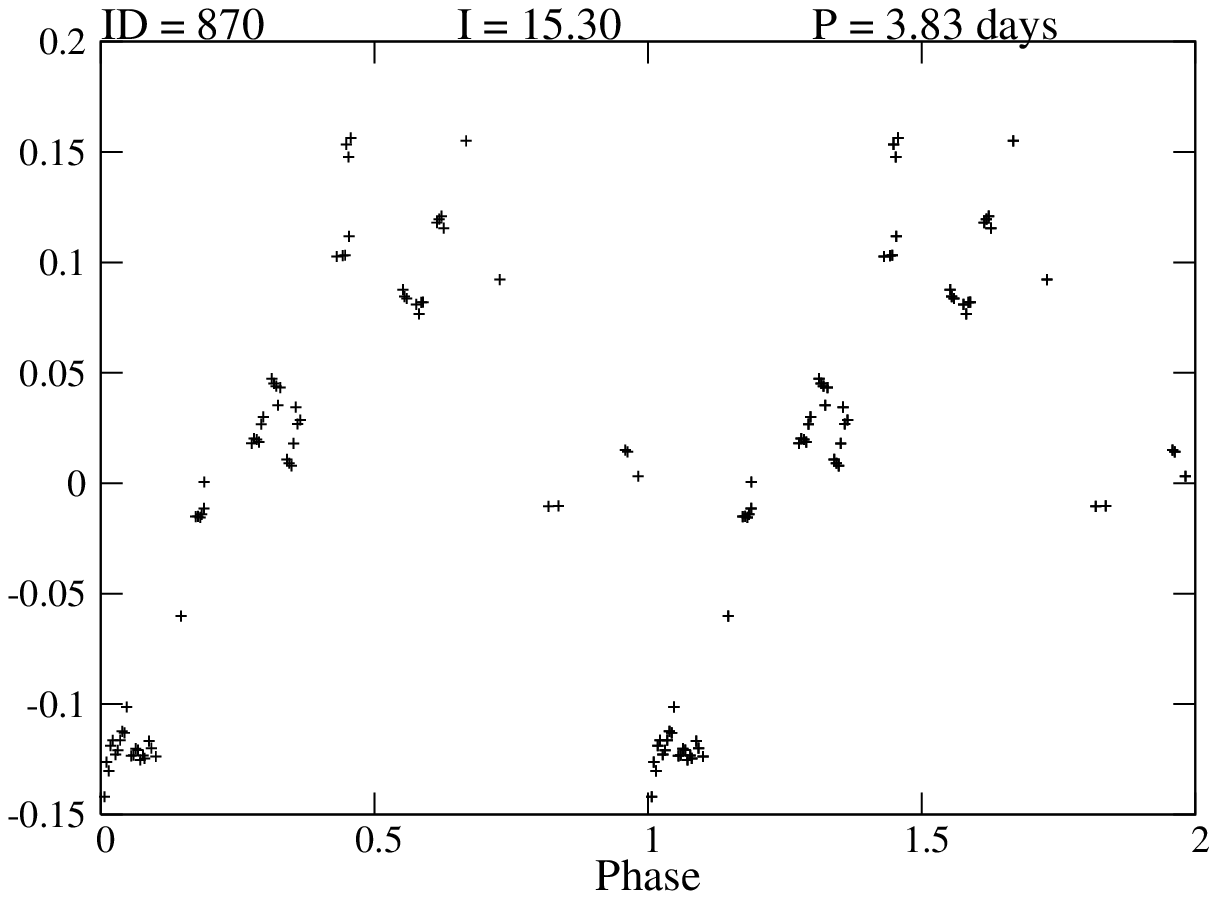}}
\caption{Phased light curves of six periodic variables showing irregular variations superimposed on their periodic light modulations, which is more typical for CTTs. As in Figure 6, ID, measured I and the period (in days) are shown.
             }
       \label{FigPLCs2}
 \end{figure*}
\begin{table*}
\renewcommand{\arraystretch}{1.2}  
\begin{minipage}[]{2\columnwidth}
    \centering
\caption{Periodic variable catalogue. The complete catalogue is available electronically at the CDS.}
\label{PVTtable}
\addtolength{\tabcolsep}{2pt}
\begin{tabular}{ccccccccc} 
\hline\hline             
ID & I[mag] & FAP$_\mathrm{scargle}$\footnote{False alarm probability computed from the Scargle time series analysis.} & $FAP_\mathrm{Ftest}$\footnote{False alarm probability computed from the Ftest} & P\footnote{Rotational period measured in this study.}[days] & P$_\mathrm{err}$\footnote{Error of the derived rotational period.}[days] & P$_\mathrm{H2002}$\footnote{Rotational period found by H2002.}[days] & P$_\mathrm{adopt}$\footnote{Final rotational period adopted.}[days] & PTP\footnote{Peak-to-peak amplitude of the modulation.}[mag]\\
\hline
1&	15.535&	1.13E-05&	2.87E-02&	0.710&	0.014&	... &  0.710 &0.019\\
11&	15.139&	6.34E-09&	1.08E-07&	1.740&	0.035&	1.730&  1.740 &0.030\\
87&	16.194&	6.34E-11&	1.65E-12&	3.100&	0.062&	3.110&  3.100 &0.020\\
1002&	15.019&	3.54E-09&	3.07E-09&	3.980&	0.080&	3.940 & 3.980 &0.572\\
1032&	15.993&	1.50E-09&	6.90E-09&	0.460&	0.009&	...  & 0.460 &0.061\\
3001&	15.533&	1.84E-10&	$<$1.00E-15&	3.240&	0.065&	3.310 & 3.240 &0.034\\
3002&	15.745&	2.22E-11&	$<$1.00E-15&	1.410&	0.028&	1.410 & 1.410 &0.043\\
10054&	20.031&	1.86E-03&	9.73E-02&	0.830&	0.017&	...  & 0.830 &0.258\\
10147&	19.428&	2.14E-06&	9.43E-03&	2.680&	0.054&	...  & 2.680 &0.118\\
10226&	17.643&	1.97E-12&	$<$1.00E-15&	2.390&	0.048&	...  & 2.390 &0.293\\
10310&	18.209&	1.13E-12&	$<$1.00E-15&	2.650&	0.053&	...  & 2.650 &0.072\\
10313&	17.766&	1.36E-09&	3.99E-09&	11.900&	0.238&	...  & 11.900 &0.062\\
10327&	19.120&	1.18E-08&	9.14E-09&	2.580&	0.052&	2.590 & 2.580 &1.240\\
10336&	16.958&	$<$1.00E-15&	$<$1.00E-15&	8.400&	0.168&	...  & 8.400 &0.256\\
10349&	18.237&	1.85E-09&	$<$1.00E-15&	9.900&	0.198&	9.560 & 9.900 &0.297\\
20033&	15.456&	1.25E-11&	1.67E-14&	1.260&	0.025&	...  & 1.260 &    0.090\\
20035&	16.823&	1.90E-08&	2.54E-05&	7.300&	0.146&	...  & 7.300 &    0.180\\
20124&	17.871&	3.60E-12&	$<$1.00E-15&	4.420&	0.088&	...  & 4.420 &	  0.807\\
20167&	17.682&	2.04E-11&	5.55E-15&	1.130&	0.023&	...  & 1.130 &	  0.377\\
20297&	19.436&	6.86E-05&	3.27E-02&	3.920&	0.078&	...  &  3.920&    0.433\\
20317&	17.882&	8.72E-11&	1.50E-13&	6.990&	0.140&	...  &  6.990&    0.121\\
20347&	19.013&	1.38E-05&	5.92E-02&	1.200&	0.024&	...  &  1.200&    0.134\\
20518&	18.456&	2.26E-12&	$<$1.00E-15&	4.110&	0.082&	...  &  4.110&    0.915\\
20532&	20.254&	2.16E-04&	4.06E-01&	0.460&	0.009&	...  &  0.460&    0.315\\
21614&	16.169&	9.00E-08&	5.20E-05&	0.330&	0.007&	...  &  0.330 &   0.048\\
\hline
\end{tabular}
\renewcommand{\footnoterule}{} 
\end{minipage}
\end{table*}
In total 487 stars fulfil these criteria and are therefore adopted as periodic variables. They are listed in Table 2. A very small fraction of the adopted periodic variables have FAP values that are close to the limits given above (i.e. only 1\% of them have 0.1\%\,$<$\,FAP$_\mathrm{Scargle}$\,$<$\,1\% and only 7\% of the periodic variables have 5\%\,$<$\,FAP$_\mathrm{Ftest}$\,$<$\,1\%).\\
In addition, a more reliable FAP calculation was done on the basis of Monte Carlo simulations similar to the approach by K\"{u}rster et al. \cite{Kurster}. We obtained 10\,000 randomised data sets in which the observing times were kept as in the original data and the relative magnitudes were randomly redistributed. Scargle periodograms were derived for each data set and new FAP$_\mathrm{sim}$ were calculated, which gives the fraction of data sets for which the power of the highest peak exceeds the power of the detected period. As it is a time-expensive simulation, we performed it for a few objects, with slightly different sampling, magnitudes and variability level to get a reliable estimate of the power corresponding to an FAP$_\mathrm{sim}$ of 1\% and 0.1\%. According to the simulations, an FAP$_\mathrm{sim}$ of 1\% and 0.1\% corresponds to peak powers of about 10.5 and 12.6, respectively. For all periodic variables in our sample, the power of the peak corresponding to the found period is higher than 12 and in about 80\% of them the powers exceed 20, this means that all the final periodic variables have FAP$_\mathrm{sim}$\,$<$\,1\% and 80\% of them even have FAP$_\mathrm{sim}$\,$<$\,0.0001\%. We would like to emphasise that not only in our Monte Carlo simulations do we find that the FAP$_\mathrm{sim}$ are similar (within a factor of about two) to the FAP$_\mathrm{Scargle}$ derived from Eq.\,7. The same was found by K\"{u}rster et al. (1997), Scholz \& Eisl\"{o}ffel (2004) and Lamm (2003). These results provide additional support for our period search procedure.\\
For comparison purposes we performed a Monte Carlo simulation similar to the one proposed by H2002 for the same few objects used in the previous simulation. In this case, we kept the decimal part of the observing days and the relative magnitudes while we randomly mixed the integer part of the observing dates. 10\,000 randomised data sets delivered an FAP of 1\% and 0.1\% for powers of 8.8 and 10.7, respectively.  If these simulations provide realistic power and FAP values, which seems unlikely due to their arbitrary time sampling, then our above derived FAP$_\mathrm{sim}$ would be even more conservative. That such a type of simulation presumably does not provide realistic FAP estimates was already stated by Scholz \& Eisl\"{o}ffel (2004).\\
Among the found periodic variables, we can distinguish between two different kinds of variables. The first and much larger group includes objects with low-level variability (i.e. typical peak-to-peak amplitudes below 0.25\,mag) which are relatively well-approximated by a sine wave. These variations are most probably caused by magnetic cool spots co-rotating with the object. As an example, Fig.\,6 shows six phased light curves of ``well-behaved'' periodic variables. The other group presents a superposition of short-term irregular variations (``noise'') with a periodic light modulation as shown in Fig.\,7. The irregular variations are often observed in classical TTauri stars (CTTs). These kind of objects are not very numerous in our sample of periodic variables. Only about 10\% of the periodic variables outside the so-called cluster radius (see Sect. 7) belong to this group, while near the cluster center this number increases to about 15\%. This increase probably results from a larger number of CTTs in that region.
We also classified 75 objects as possible periodic variables, as well as 7 objects as possible eclipsing systems. In the first case, the objects often show unusual phased light-curves and the false alarm probabilities give values close to the limits adopted above. Both the possible periodic and eclipsing objects are listed in Table 1, marked with 'PPV' and 'PES' in the Variability Flag column, respectively. They are not taken into account in any further analysis.
\subsection{Accuracy of the period determination}
The uncertainties in the measured periods are determined mainly by the frequency resolution in the power spectrum, and the sampling error, i.e. the typical spacing between data points, which is most relevant for very short periods. \\
We investigated the ``sensitivity'' range of our period determination, i.e. the smallest and highest frequencies we can in principle detect with acceptable accuracy. The Nyquist frequency $\nu_{max}=\frac{1}{2\Delta}$ (with $\Delta$ the distance between two data points) defines the upper frequency limit in a uniform sampling. The lower frequency limit is given by the total time interval of the sample. For non-regularly spaced data, the limits given above are only rough approximations. For a more accurate determination of the period range over which we can determine rotational periods with an accuracy below $\sim$\,2 \%, we carried out a simulation following the procedure described in Scholz \& Eisl\"offel (2004) and summarised in the following. We added sine waves corresponding to periods between 0.1 days and 25 days to the light-curve of one of the non-variable flux reference stars. We assumed a S/N of 5 (i.e. ratio between the amplitude imposed and the noise of the light-curve), corresponding to an imposed signal with an amplitude of 0.02 magnitudes. We used the Scargle periodogram in order to obtain the frequency of the highest peak in each case and we compared the differences between the imposed and the measured period. This procedure delivers information on the accuracy of the found periods. The results of the simulation are shown in Fig.\,8, in which also a 2\% period error is shown as a dotted line. The dependence of the accuracy of the recovered periods as a function of period is a consequence of the irregular time sampling. We performed these simulations for several values of the S/N (2.5, 5 and 10), but we did not obtain any significantly different results between various simulations. Therefore, we conclude that the method is not highly dependent on the amplitude of the modulation, at least for S/N values higher than 2, which is in agreement with Scholz \& Eisl\"offel (2004). Taking into account very high S/N (more than 10) will decrease only slightly the height of the higher peaks in Fig.\,8, resulting in a slightly better accuracy in these regions. The simulation also shows that it is possible to detect periods even below the Nyquist limit, which in our case is $\sim\,0.45$\,days, if we consider the median time differences of the data points in the light-curve of the reference star used in the simulation. This is consistent with Eyer \& Bartholdi (1999) who proved that in time series with irregular spacing the Nyquist limit is only an upper limit for the P$_\mathrm{min}$. The same was found by Scholz \& Eisl\"{o}ffel (2004).\\
\begin{figure}
 \centering
 \subfigure{\includegraphics[width=9cm]{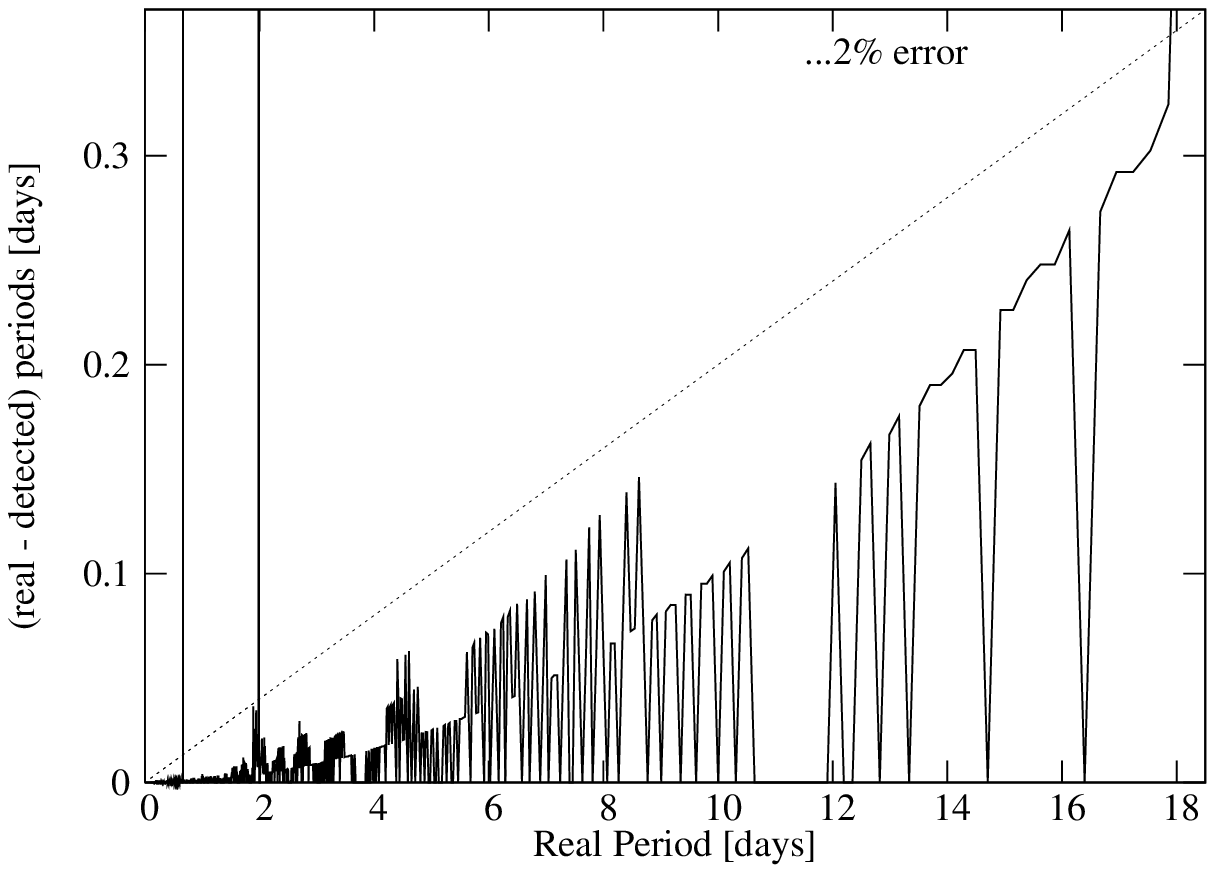}}\\
 \subfigure{\includegraphics[width=9cm]{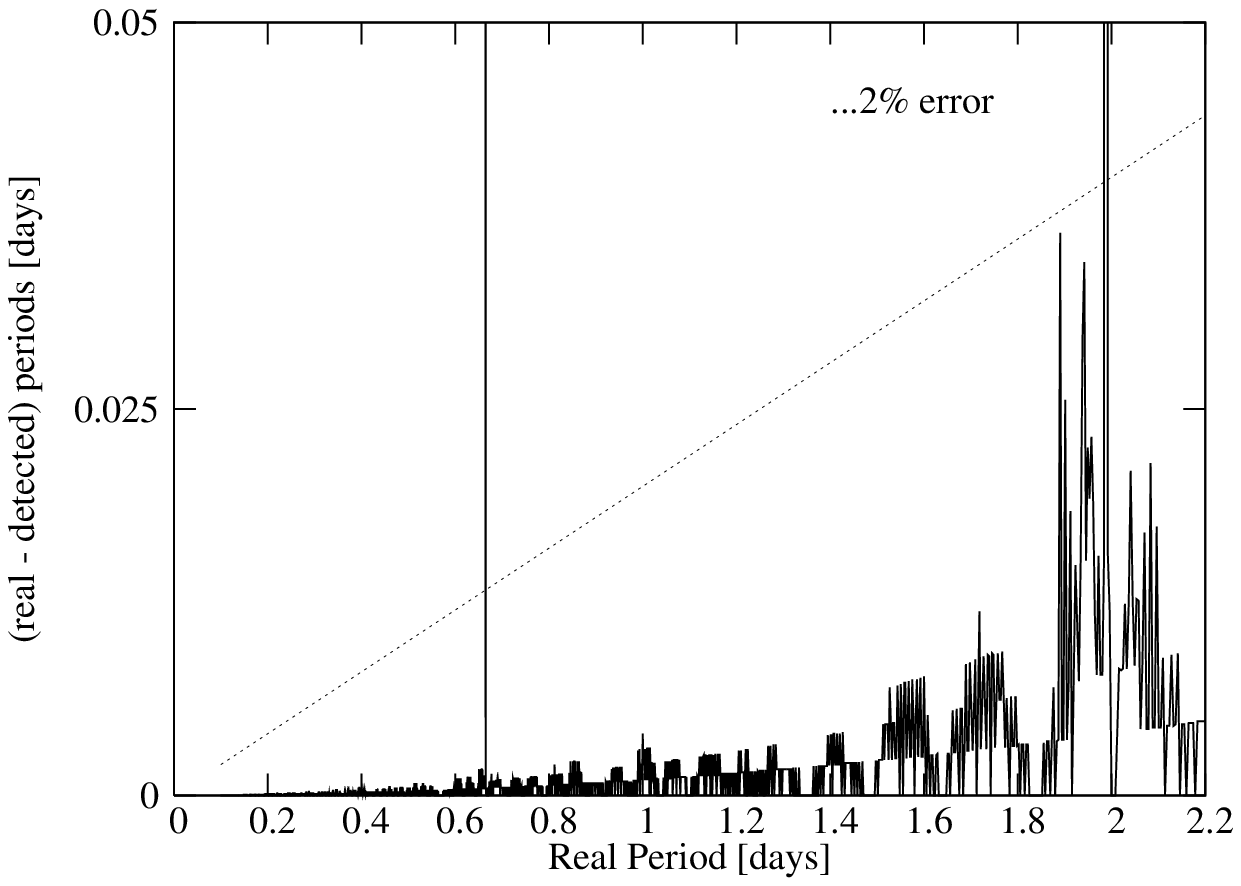}}
 \caption{ Results of a simulation showing the difference between detected and real period as a function of real period. This simulation reflects the accuracy of the method used for searching for rotational periods (see text for details). The dotted line corresponds to a period error of 2\%. The lower panel shows a zoom into the short period regime.
         }
  \label{FigCompleteness}
  \end{figure}
We finally considered only periods below 17 days, since longer periods may be affected by errors larger than 2\% in the period determination. Only 12 of our periodic variable candidates had to be removed from our sample because of periods longer than 17\,days. Fig.\,8 also shows a zoom into the shorter periods. The simulation shows several peaks around 2\,days as well as a high peak at 0.67 days which is directly related to the sampling of our data. Note that this test is only based on the Scargle periodogram technique, while as outlined in the previous Sect., we only take into account the Scargle peak which remains after running the CLEAN algorithm which corrects for the non-uniformity in the time series. This means that we should be able to exclude spurious detections by means of our whole period analysis procedure. We considered the 2\% error curve in Fig.\,8 and used these values as the estimated errors in the period determination. Note that these are upper limits and therefore the real errors are often much smaller.
\section{Irregular variables}
Many of the objects in our sample for which no significant period was detected nonetheless show relatively large brightness variations in their light curves. The peak-to-peak amplitudes for these objects are usually much higher than for the periodic ones, reaching up to 3 magnitudes. We performed the $\chi^{2}$ variability test on all the 2908 monitored stars in which the time series data contain at least 30 data points. The $\chi^2$ test gives the probability that the deviations in the light curves are due to the photometric errors and not due to an intrinsic variability of the object (e.g. Bailer-Jones \& Mundt, 1999). Therefore we compute
\begin{equation}
 \chi^2= \displaystyle\sum^{N_B}_{j=1} (\frac{m_{rel}(j)}{\delta m_{rel}(j)})^2
\end{equation}
with m$_{rel}$ the (mean subtracted) relative magnitude and $\delta m_{rel}$ the error of the \textit{jth} star. N$_B$ is again the number of data points in the light-curve.
The probability that the star is variable is given by Press et al. (1992)
\begin{equation}
p=1-Q(\chi^2|N_{B})
\end{equation}
where Q($\chi^2|$N$_{B}$) is the probability that a non-variable star with \textit{N$_B$} time series measurements has a $\chi^{2}$  value higher than the one given by Eq. 10. \textit{Q($\chi^2|$N$_{B}$)} is described mathematically in the following Eq.,
\begin{equation}
 Q(\chi^2|N_{B})=\frac{\Gamma(\frac{N_{B}-1}{2},\frac{\chi^2}{2})}{\Gamma(\frac{N_{B}-1}{2})}
\end{equation}
where we have used \textit{N$_B$-1} instead of \textit{N$_B$} since our time series has been mean subtracted and therefore the degrees of freedom are reduced by 1.\\
As stated above, we have included all the monitored stars in this analysis and since this test is also sensitive to periodic variations we have to exclude from the resulting 1186 objects that are found to be variables by the $\chi^{2}$ variability test the ones that are known to be periodic variables. 76\% of the periodic variables (this number includes also the possible periodic variables) are variables by means of the $\chi^{2}$ test. The reason why not all of the periodic variables are variables according to this test is that the periodogram analysis is sensitive to much lower amplitude variations than the $\chi^{2}$ test and therefore the low amplitude periodic variables are only detected by the time series analysis described in Sect. 5 .\\
As a result, after removing the periodic and possible periodic variables from the sample, we found that 808 objects present irregular variations with a probability higher than 99.9\%. 
\begin{figure}
\centering
\rotatebox{270}{\resizebox{!}{8cm}{%
\includegraphics[width=9cm]{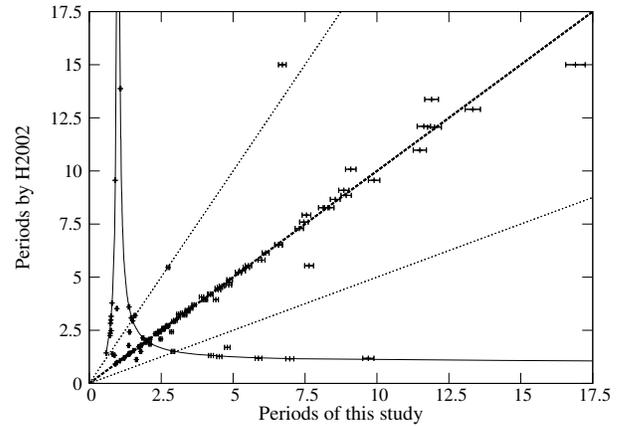}}}
\caption{ Comparison of the periods measured for the 110 stars common to this and the H2002 study. Stars with the same periods found in both studies are located on the dashed line. They represent 74\% of the total. The beat periods and the harmonics curves are shown as solid and dotted lines, respectively. The error bars correspond to a 2\% error in the period determination, which is an upper limit to the measured errors (see Fig.\,8) 
         }
 \label{FigCompH2002}
 \end{figure}
\section{Properties of the periodic variables}
\subsection{Period distribution, overview}
In the following, we investigate in detail the period distribution of the already known and newly found periodic variables in the ONC and also their peak-to-peak amplitude distribution. In addition, we compare our results with the ones by L2004 in NGC 2264. In particular we investigate the dependence of these two distributions on mass and position of the objects in the field. By including the H2002 results, we have a more complete sample of periodic variables and in particular a much broader range of masses, ranging from about 1.5\,$M_\odot$ (H2002) down into the BD regime with the faintest objects having an estimated mass of only $\sim$\,16\,$M_{jup}$. As outlined in more detail below, 124 of our periodic variables and 139 in the combined data-set (i.e. this study + H2002) are potential BDs, which is by far the largest rotational period sample of BDs and BD candidates.\\
\begin{figure*}
\centering
\subfigure[Magnitude distribution of the periodic variables.]{\includegraphics[width=8.5cm,height=11cm]{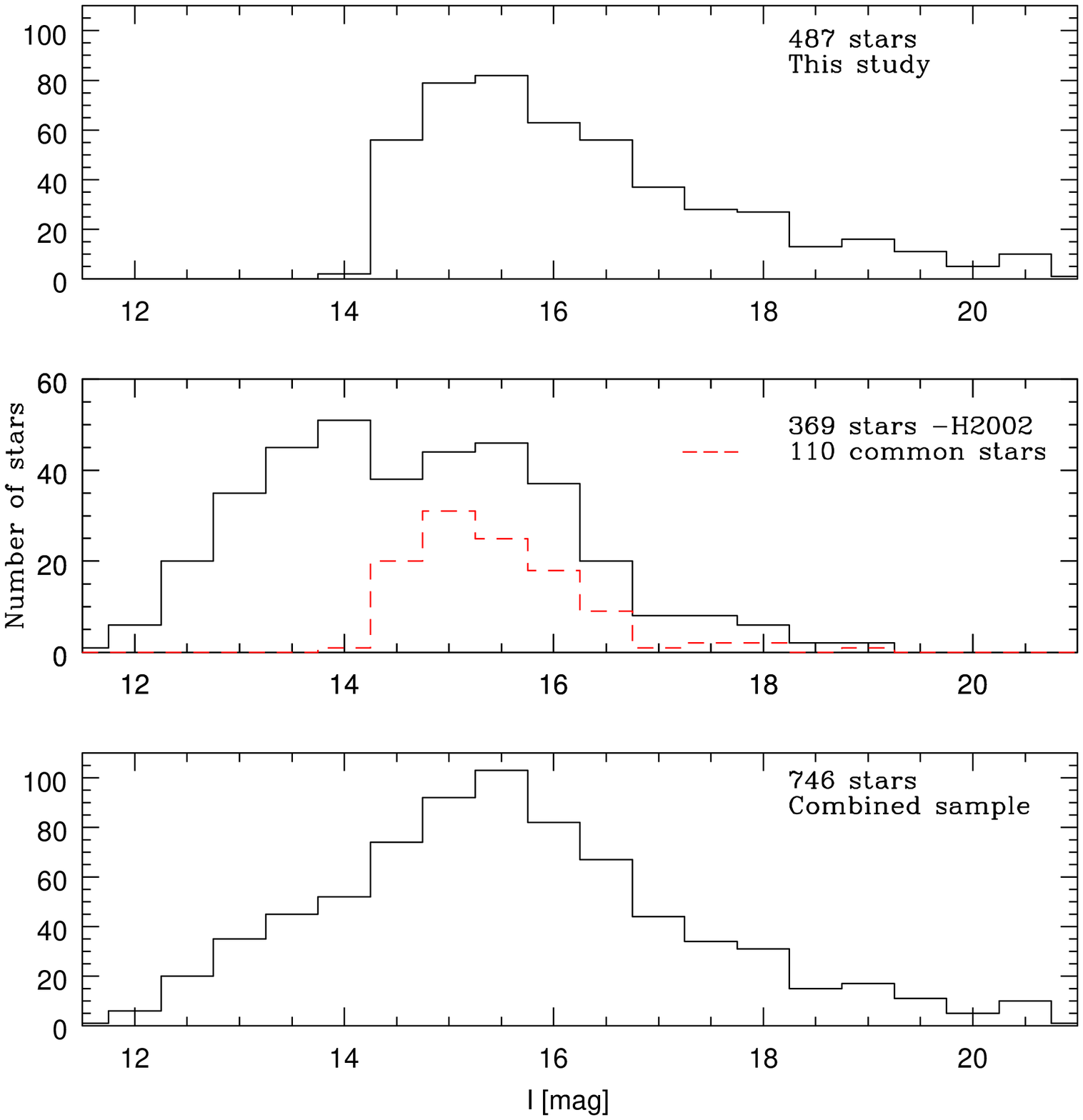}}
\subfigure[Period distribution of the periodic variables.]{\includegraphics[width=8.5cm,height=11cm]{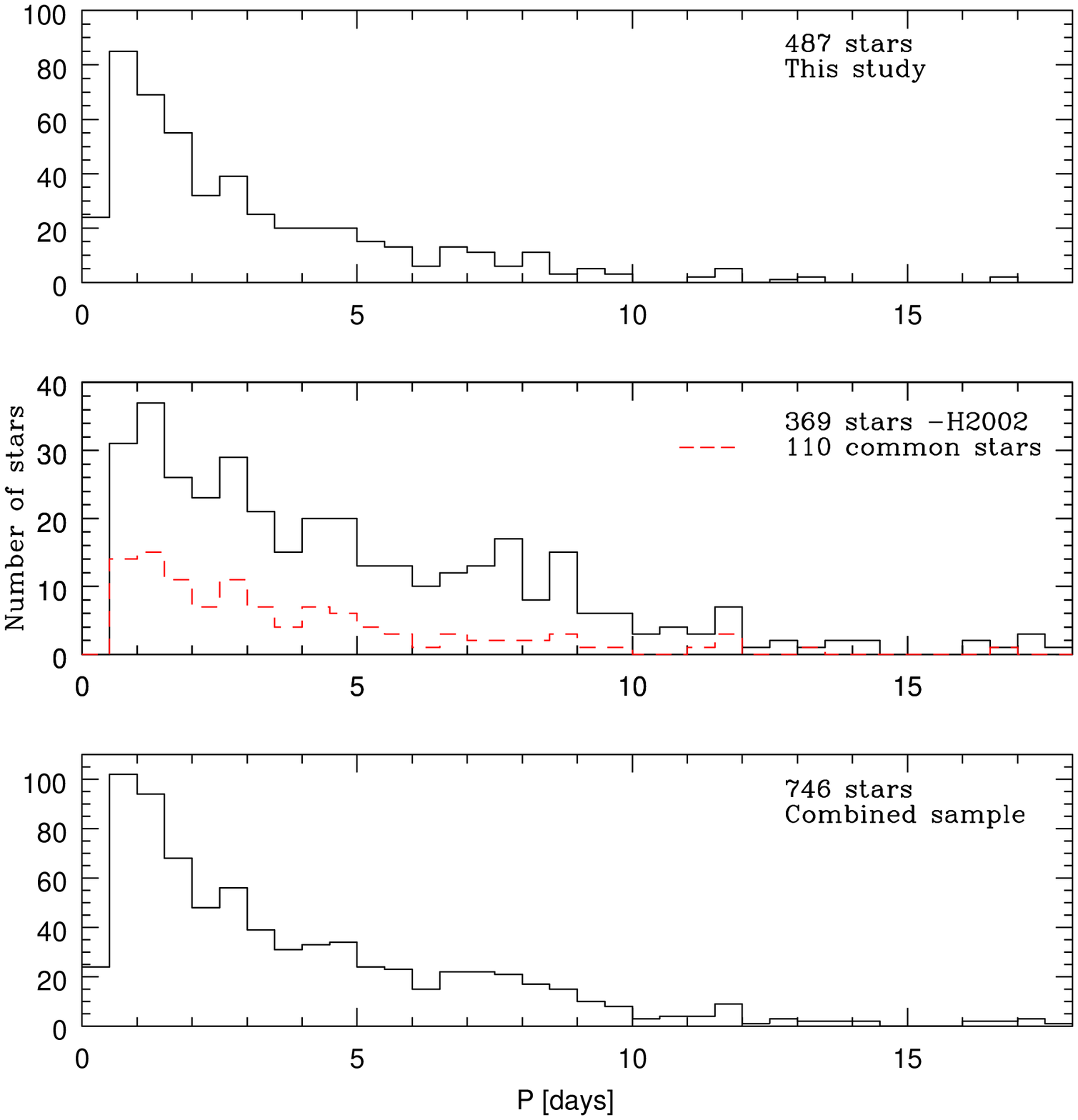}}
\caption{\textit{a}) Magnitude and period distribution of the periodic variables. The top panel shows the magnitude distribution of the 487 periodic variable stars detected in this study. In the middle panel, the variables found by H2002 are displayed (solid line), where the dashed line indicates the 110 objects which are common for both studies. The bottom panel shows the distribution for the combined sample of 746 periodic variables. \textit{b}) Same as \textit{a}) but for the period distribution of the periodic variables.}
\label{FigIPdistri}
\end{figure*}
We compared our results with the rotational data published by H2002. The result of this comparison is shown in Fig.\,9. We found that 110 of the periodic variables reported by H2002 are also found to be periodic in this study. From the 369 periodic variables found in H2002, 143 are brighter than 14.1 in I (which is the brightest PV in our study due to
our CCD saturation limit) and 14 are located in the gaps of our survey
(since the positioning is not exactly the same). This results in 212
periodic variables that in principle can be recovered, although we
recovered 110 of them, which corresponds to 52\%. The remaining 48\% consist of 34\% irregular variables, 5\% possible periodic variables and 7\% non-variables, which means that a small fraction of the non-recovered periodic variables are found to be non-variables in our study. Moreover, we found that 74\% of the recovered periodic variables have periods which agree within the errors in both epochs. The high level of agreement implies that many of these objects have spot groups stable enough to be detectable by our method over many years. This is quite different to Scholz et al. (2008), who found that in the much older (40 Myr) cluster IC4665 only 10\% of the rotational periods could be recovered after 1 year. This may suggest that the fraction of objects having detectable spot groups over many years strongly decreases with age.\\
Most of the periods that show no agreement can be understood as aliasing or harmonics. Aliasing or ``beat'' phenomena result from the one day sampling interval in data taken at the same longitude (H2002). This means that a periodic variation of period P will result in a beat period B:
\begin{equation}
\frac{1}{B}=\frac{1}{P_\mathrm{sampling}} - \frac{1}{P} 
\hspace*{3cm}
( P > P_\mathrm{sampling})\\
\end{equation}
\begin{equation}
\frac{1}{B}=\frac{1}{P} - \frac{1}{P_\mathrm{sampling}} 
\hspace*{3cm}
( P < P_\mathrm{sampling})
\end{equation}
with $P_{sampling}$ equals 1\,day in our case.\\
Only five possible beat periods were identified in our sample as shown in Fig.\,9, while in the H2002 sample there are 12 possible long beat periods. This result is as expected since we took into account the data sampling (see Sect.\,5) and therefore the beat effect should be much smaller on our period determinations.\\
In Fig.\,9, about 3\% of the 110 common periodic variables lie on two straight lines of slope 1/2 or 2. The effect in which a measured period is found to be a double or half of the true period is called harmonics. The beat periods and the harmonics curves are shown in Fig.\,9 as solid and dotted lines, respectively. \\
Because of our more sophisticated time series analysis (e.g. combination of Scargle and CLEAN algorithms and several FAP calculations), we regard our period determination as more reliable and we used our measured periods for further analysis. Only for the five objects in which the measured periods seem to be (long) beat periods of the true (short) ones do we adopt the periods determined by H2002.\\
The magnitude as well as the period distribution of the periodic variables detected in this study and in the one by H2002 are shown in Fig.\,10. In the top panels of Fig.\,10\textit{a} and 10\textit{b}, the distribution of the 487 periodic variables from our study are shown. The middle panels show the distributions of all the periodic variables found by H2002 and we discriminate between the objects found to be periodic in both studies (dashed red lines) and the stars found to be periodic only in H2002 (solid line). Fig. 10 clearly shows that the overlapping magnitudes mostly range from 14 to 17 magnitudes. The combined samples are shown in the two bottom panels of Fig.\,10. Note that we do not consider extinction-corrected magnitudes for this analysis.
\begin{table}[t]
\begin{minipage}[c]{\columnwidth}
\renewcommand{\arraystretch}{1.2}  
\caption{Number of variable objects found in the ONC ($\sim$ 1\,Myr) by this study and in NGC 2264 ($\sim$\,2\,Myr) by L2004.}
\label{ONC-NGC2264table}
\begin{tabular}{c|ccc|c} 
\hline\hline             
VarFlag\footnote{As in Table 1, PV are periodic variables and IV irregular variables. All Stars include the total number of monitored stars independently of whether they are variables or not. }&\multicolumn{3}{c}{ONC ($\sim$\,1\,Myr)} & NGC 2264 ($\sim$\,2\,Myr)\\
\hline
& R$<$R$_\mathrm{cluster}$ & R$>$R$_\mathrm{cluster}$ & Total & Total\\
\hline
PV & 196 & 291 & 487 & 405\\
IV & 335 & 473 & 808 & 184\\
All Stars& 847 & 2061 & 2908 & 10554\\
\hline
\end{tabular}
\renewcommand{\footnoterule}{} 
\end{minipage}
\end{table}
\begin{figure*}
 \centering
 \includegraphics[width=17.5cm, height=15cm]{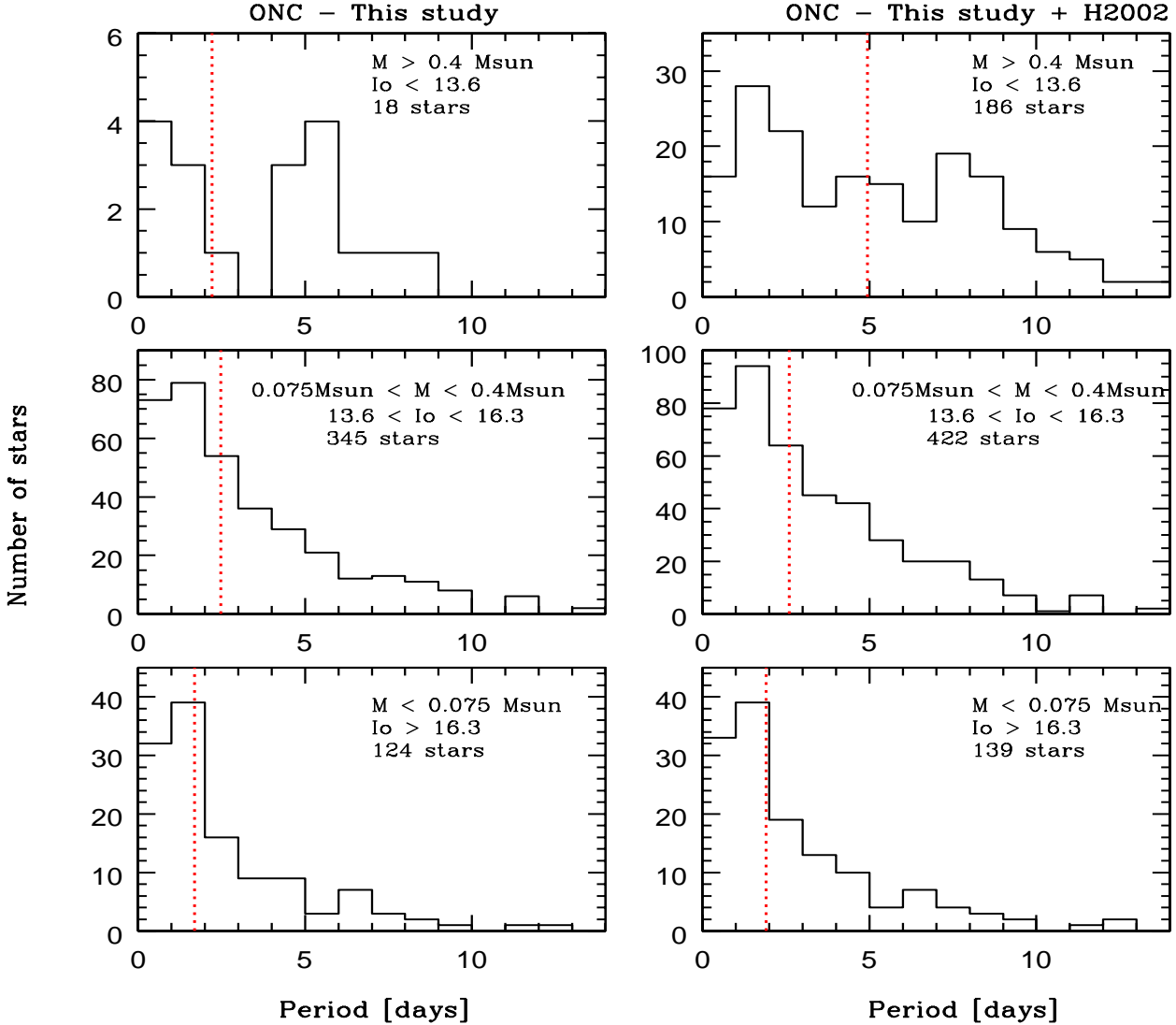}
 \caption{The left panels show the period distribution for the periodic variables in our study, while the right panels show the period distribution of the periodic variables in the combined data set (This study + H2202) for the whole region covered by the CCD image displayed in Fig.\,1. The mass regime for each panel is indicated and is based on models by Baraffe et al. (1998) and Chabrier et al. (2000), assuming a constant \textit{A$\mathrm{_{v}}$}\,=\,1.4 mag for all stars (see text for details). In the bright regime (top left panel) we have few stars and a meaningful analysis can only be done in combination with the H2002 data (top right panel). The medians of the periods for each panel are shown as dashed lines.}
 \label{Pwholefield}
 \end{figure*}
\subsection{Period distribution and its dependence on mass}
In order to estimate masses of the periodic variables in the ONC from evolutionary models (Baraffe et al., 1998 and Chabrier et al., 2000), knowledge of the individual extinction values is necessary. This requires knowledge of the spectral type and at least one colour, which is difficult to measure for some wavelength bands (e.g. R-band) due to the strong nebular background. Since for most objects we do not know a colour and their spectral type, individual extinction values cannot be determined. For simplicity we therefore assumed a uniform extinction for all stars in the field. Several authors ( Jones \& Walker 1988, Bautista 1995, Hillenbrand 1997) have estimated the average visual extinction \textit{A$\mathrm{_{v}}$} within 1\,pc of the Trapezium stars and derived values between 1.3 and 1.5 magnitudes. We adopt a constant value of \textit{A$\mathrm{_{v}}$}\,=\,1.4 mag, which corresponds to an \textit{A$\mathrm{_{I}}$} of 0.8\,mag. We are aware that with this simplified assumption we may underestimate the intrinsic brightness (and therefore the mass) of some objects, particularly in the inner part of the field (because their true \textit{A$\mathrm{_{v}}$} can be much larger than 1.4\,mag) as well as overestimate the intrinsic brightness of objects, particularly at the eastern edge of the field because their \textit{A$\mathrm{_{v}}$} is certainly much lower (e.g. Castets et al. \cite{Castets}). We are also aware that local extinction (e.g. due to circumstellar disk) can vary strongly from object to object and in addition intrinsic brightness variations can mimic different masses than implied by the flux in the I-band. We will denote the extinction-corrected magnitudes as I$_\mathrm{0}$ from now on.\\
In order to investigate the dependence of the period distribution on mass, we selected the following three brightness bins:
\begin{itemize}
\item A bright bin containing stars with I$_\mathrm{0}$\,$\leq$\,13.7\,mag, corresponding to \textit{M}\,$\geq$\,0.4\,\textit{M$\mathrm{_{\odot}}$}.
\item An intermediate bin with 13.7\,$<$\,I$_\mathrm{0}$\,$\leq$\,16.3, corresponding to 0.4\,\textit{M$\mathrm{_{\odot}}$}\,$<$\,\textit{M}\,$\leq$\,0.075\,\textit{M$\mathrm{_{\odot}}$}.
\item A faint bin in which all objects have I$_\mathrm{0}$\,$>$\,16.3, corresponding to \textit{M}\,$<$\,0.075\,\textit{M$\mathrm{_{\odot}}$}.
\end{itemize}
This selection is not arbitrary. The limit for the brighter bin corresponds to the same limit used by H2002 to separate the high and low mass regime, in which the high mass regime showed a bimodal period distribution and the low mass did not. H2002 used Hillenbrand (1997) mass estimates based on D'Antonna and Mazzitelli (1994) models and the dividing line between the two populations corresponds to 0.25\,\textit{M$\mathrm{_{\odot}}$} with this model. However, we used models by Baraffe et al. (1998) and Chabrier et al. (2000) to estimate masses from the measured magnitudes after assuming \textit{A$\mathrm{_{v}}$}\,=\,1.4 mag. The corresponding mass limit for the brightest bin when using Baraffe et al. (1998) models is 0.4\,\textit{M$\mathrm{_{\odot}}$}. The middle mass bin corresponds to a mass range between 0.4\,\textit{M$\mathrm{_{\odot}}$} and 0.075\,\textit{M$\mathrm{_{\odot}}$}, while the last bin is for substellar objects, i.e. \textit{M}\,$<$\,0.075\,\textit{M$\mathrm{_{\odot}}$}. This last mass bin contains 124 potential BDs (only our study) which is at least 10 times larger than the existing number known in any other cluster, e.g. $\sigma$ Orionis (Scholz \& Eisl\"offel, 2004). Although we have only a few objects in the first bin, this bin is very useful when analysing the combined data of this study and of H2002. \\
The resulting period distributions for the periodic variables measured in this study and in the combined data (This study + H2002) are shown in Fig.\,11, in the left and right columns, respectively. The median values for each set of data, plotted as vertical dashed lines, show the already known tendency to have faster rotators towards lower masses, which is expected from previous studies (e.g. H2001; H2002; L2005), but has not yet been investigated for a large sample of objects in the BD mass regime. In order to give statistical significance to the mentioned tendency we perform a K-S test for our and the combined data set. Due to the small number of objects in the first mass bin in our sample, we only performed the K-S test for the intermediate and low mass bins, resulting in a 96\% (2\,$\sigma$ level) probability that the two distributions are different. For the combined data, the K-S test confirmed the different distributions at a very high confidence level, $>$\,99.9\% for the high-intermediate mass-bins and at a 2\,$\sigma$ level (95\%) for the intermediate-low mass bins. These results indicates that the already known trend of having fast rotators towards lower masses extends also into the substellar regime.
\begin{figure*}
\centering
 \subfigure{\rotatebox{270}{\resizebox{!}{14cm}{\includegraphics[width=9.8cm,height=13.3cm]{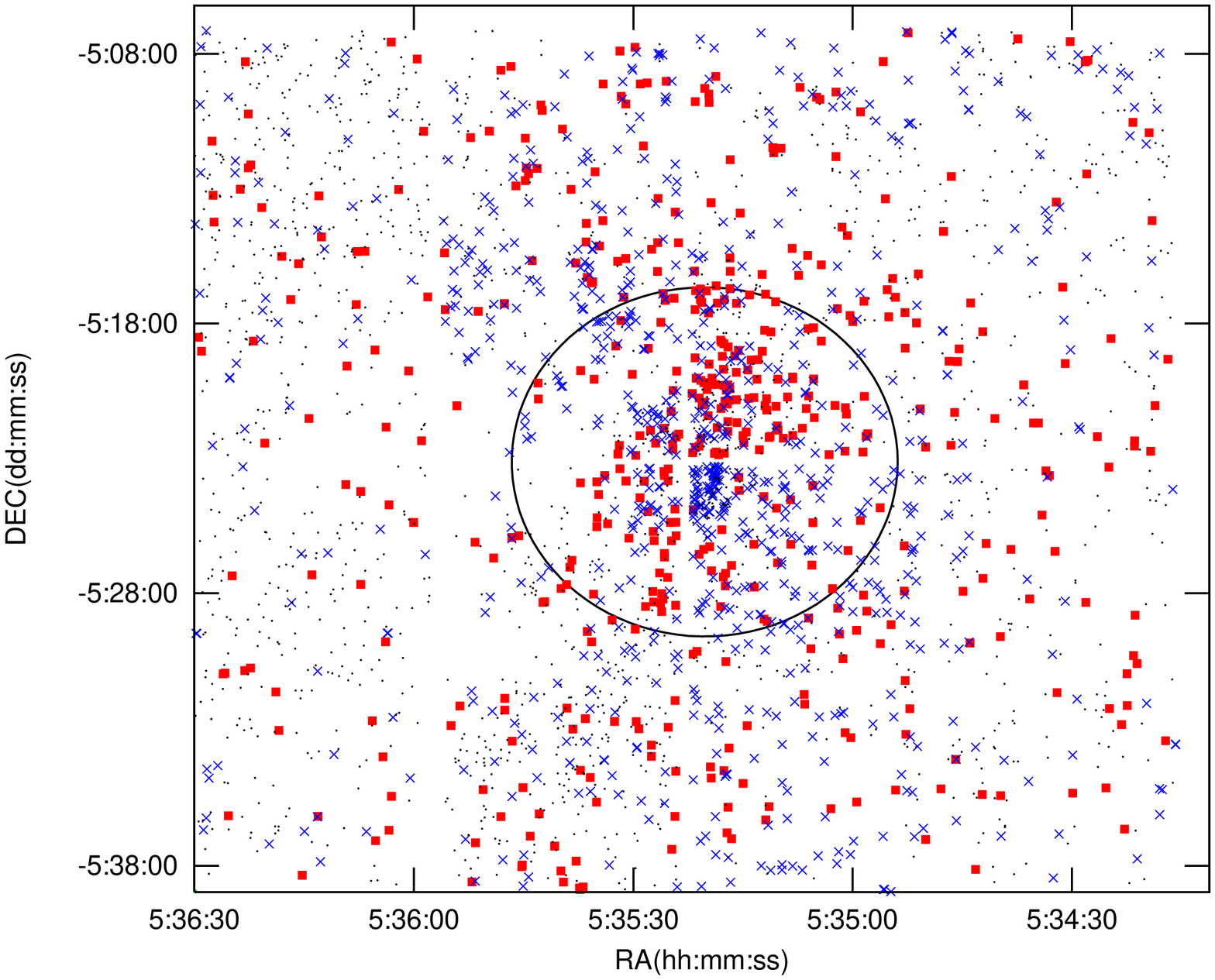}}}}\\
\vspace{1cm}
\subfigure{\includegraphics[width=5.9cm,height=5.5cm]{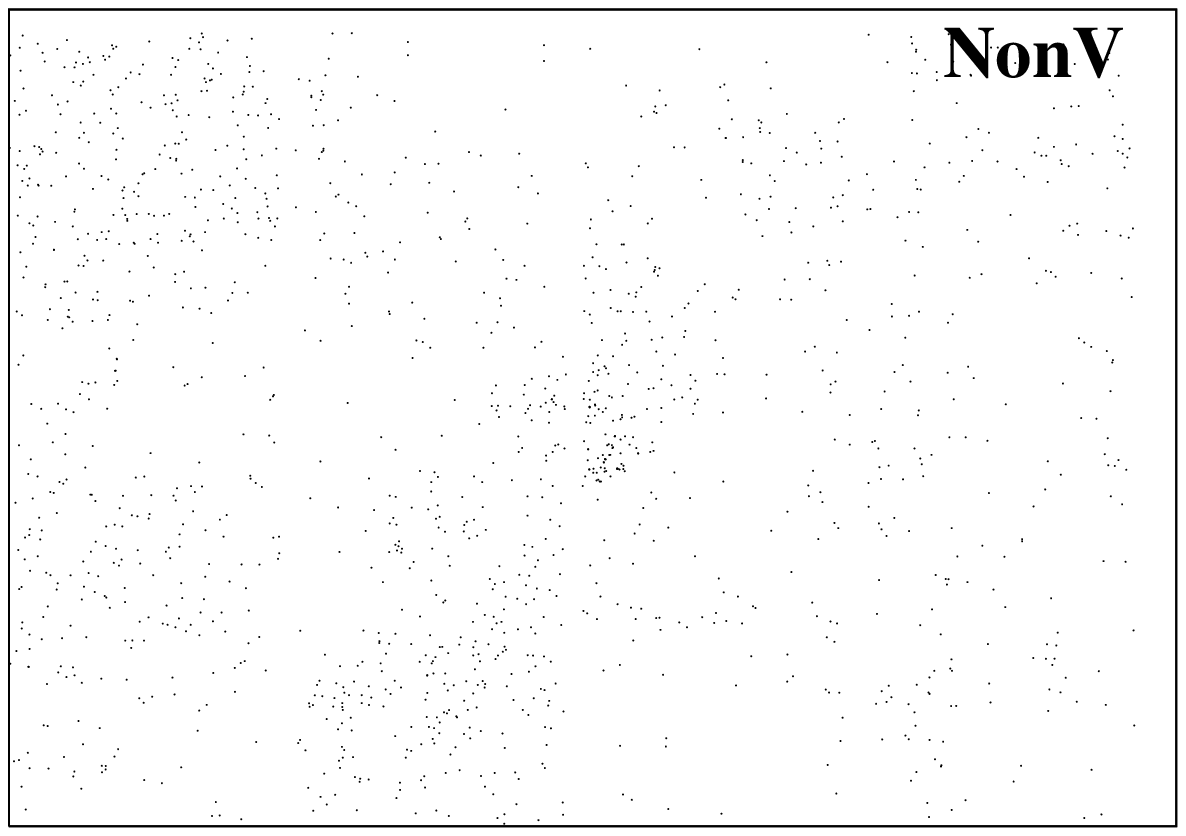}}
\subfigure{\includegraphics[width=5.9cm,height=5.5cm]{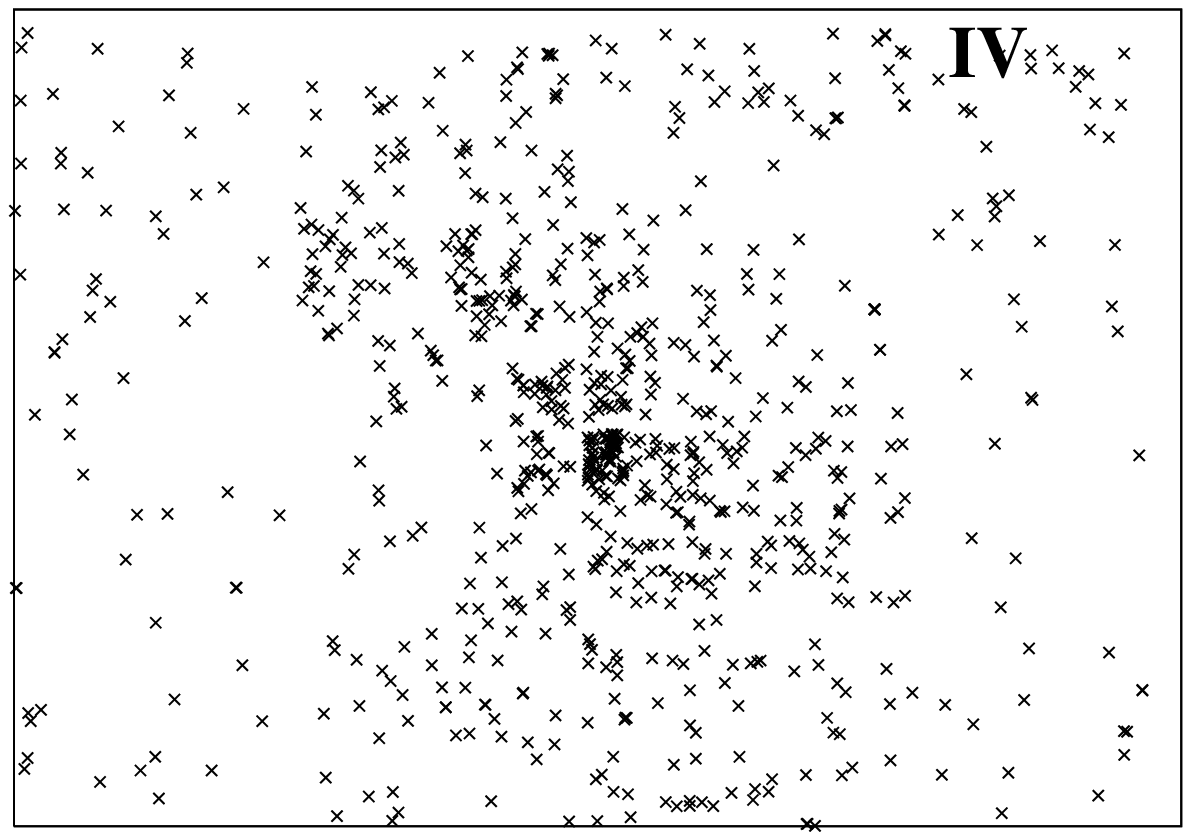}}
\subfigure{\includegraphics[width=5.9cm,height=5.5cm]{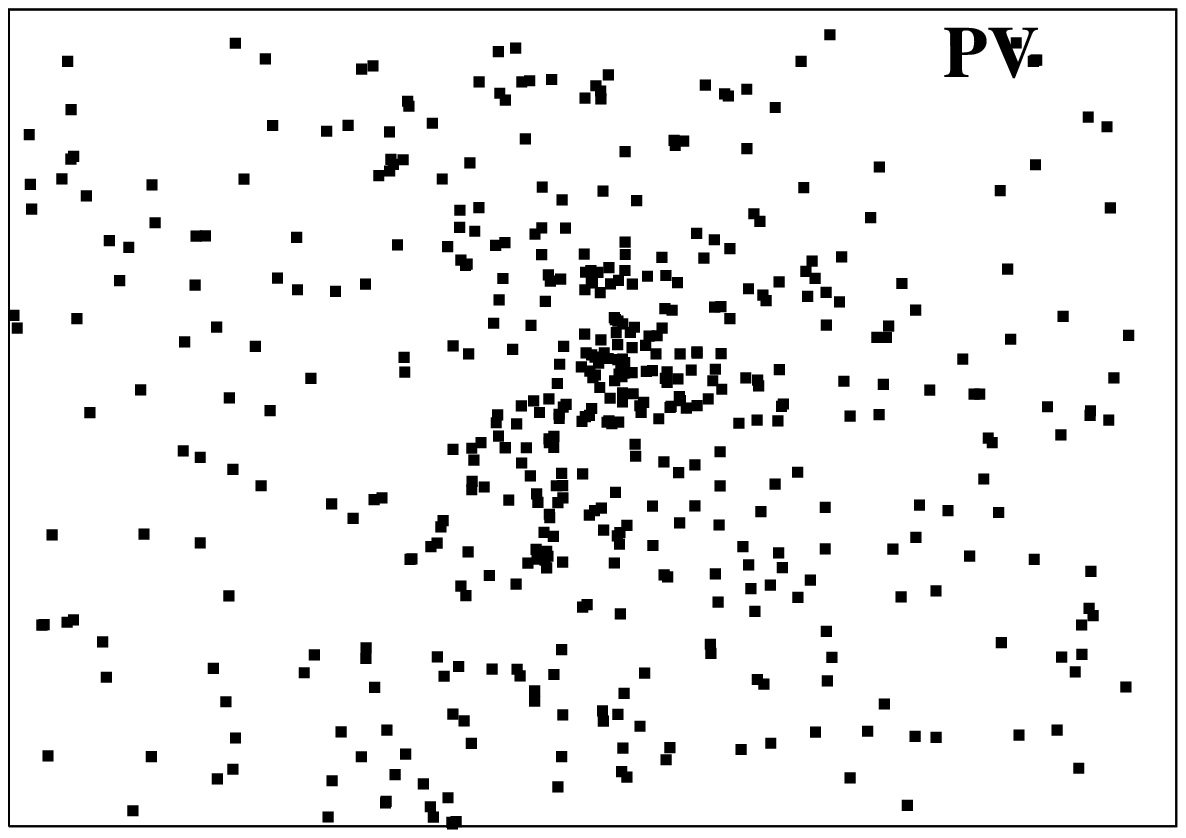}}
\caption{Spatial distribution of all variable and non-variable objects in our photometric catalogue, i.e. 487 periodic variables (squares), 808 irregular variables (crosses) and 1531 non-variables (dots). The circle in the center of the field in the top panel represents the cluster radius, and is used as a dividing criterion in the analysis of the properties of the periodic variables (see text). It is centered on $\Theta^1$ Ori and has a radius of 6\farcm7 ($\sim$\,1\,pc). The top panel shows the spatial distribution of all objects, while the three lower panels show the corresponding distributions of the non-variable objects, the irregular objects and the periodic variables, from left to right. These lower panels illustrate the rather different spatial distributions between variables and non-variable objects. 
              }
         \label{FigSpatdistri}
   \end{figure*}
\subsection{Spatial distribution of variable objects}
We analysed the spatial distribution of the variable and non-variable objects in our field. As a meaningful, although somewhat arbitrary, spatial separation within the studied field, we chose the half-mass radius of the cluster (also called cluster radius, Hillenbrand, 1997) of 6\farcm7 from $\Theta^1$ Ori, which corresponds to about 1 pc distance from that star. The spatial distribution of the catalogued objects is shown in Fig. 12. The top panel shows the position of all stars observed by us. Periodic variables (487) are represented by squares, irregular variables (808) by crosses and non-variable objects (1531) by dots. The lower panels show the spatial distribution of non-variable, irregular, and periodic variable objects from left to right. It is evident that the distribution of the non-variable objects is much smoother than the distribution of either the irregular or periodic variables. It is interesting that the irregular variables and the periodic ones are not located in exactly the same regions of the field, although an obvious clustering inside the cluster radius is evident in both cases. Most of the irregular variables are clumped in a small region south of $\Theta^1$ Ori, while the periodic variables are mainly located north-west and south-east of that star. A K-S test resulted in a 99\% probability that the distributions of the declination for irregular and periodic variables are different from one another and 43\% for the right ascension values.\\
We found 198 objects with periodic brightness modulations inside the cluster radius, and 291 outside of it. The number of irregular variables is 335 inside R$_\mathrm{cluster}$ and 473 outside. Relative to all stars observed in each spatial region, we found a higher fraction of irregular variables inside R$_\mathrm{cluster}$ than outside. Since this finding may be a consequence of the decreasing cluster membership probability toward the outer region, we took into account typical membership probabilities (Jones \& Walker \cite{Jones}) of 97\% in the inner region and 70\%\footnote{From R\,=\,R$_\mathrm{cluster}$ towards the limits of our WFI field the membership probability varies from $\sim$\,90\% to $\sim$\,60\%.} outside R$_\mathrm{cluster}$. We found that $\sim$\,40\% of the objects in the inner region are irregular variables while a smaller fraction of $\sim$\,28\% irregular variables is found outside R$_\mathrm{cluster}$. Moreover, a higher fraction of periodic variables that show a superposition of short-term irregular variations is found inside R$_\mathrm{cluster}$ than outside. Although these analyses may suggest a younger age for stars inside Rc relative to the objects located outside it, the observed ratio between PVs and IVs only marginally supports this idea. Therefore no unambiguous conclusion can be derived from the number or ratio of IVs and/or PVs on a possible age spread in the ONC.\\
In addition, the fraction of irregular and periodic variables measured can be slightly biased towards the irregular variables.  As stated by other authors (e.g. L2004), the photometric method for measuring periodic brightness modulations is more efficient among the WTTs than for CTTs, since irregular variations can add a strong noise to the periodic modulation, preventing the detection of the periodic signal. Due to this noise we are definitely missing periodic variables among the young highly active CTTs.\\
\subsection{Period distribution and its dependence on position and mass}
We investigate if there is a dependence of the period distribution on position in the field. Table 4 shows the number of periodic variables in the various mass intervals for the ONC and NGC 2264. In the case of the ONC, both this study and the combined data (this study + H2002) are considered, taking into account the two regions (i.e. inside and outside R$_\mathrm{cluster}$).\\
\begin{figure*}
    \centering
    \includegraphics[width=17.5cm, height=15cm]{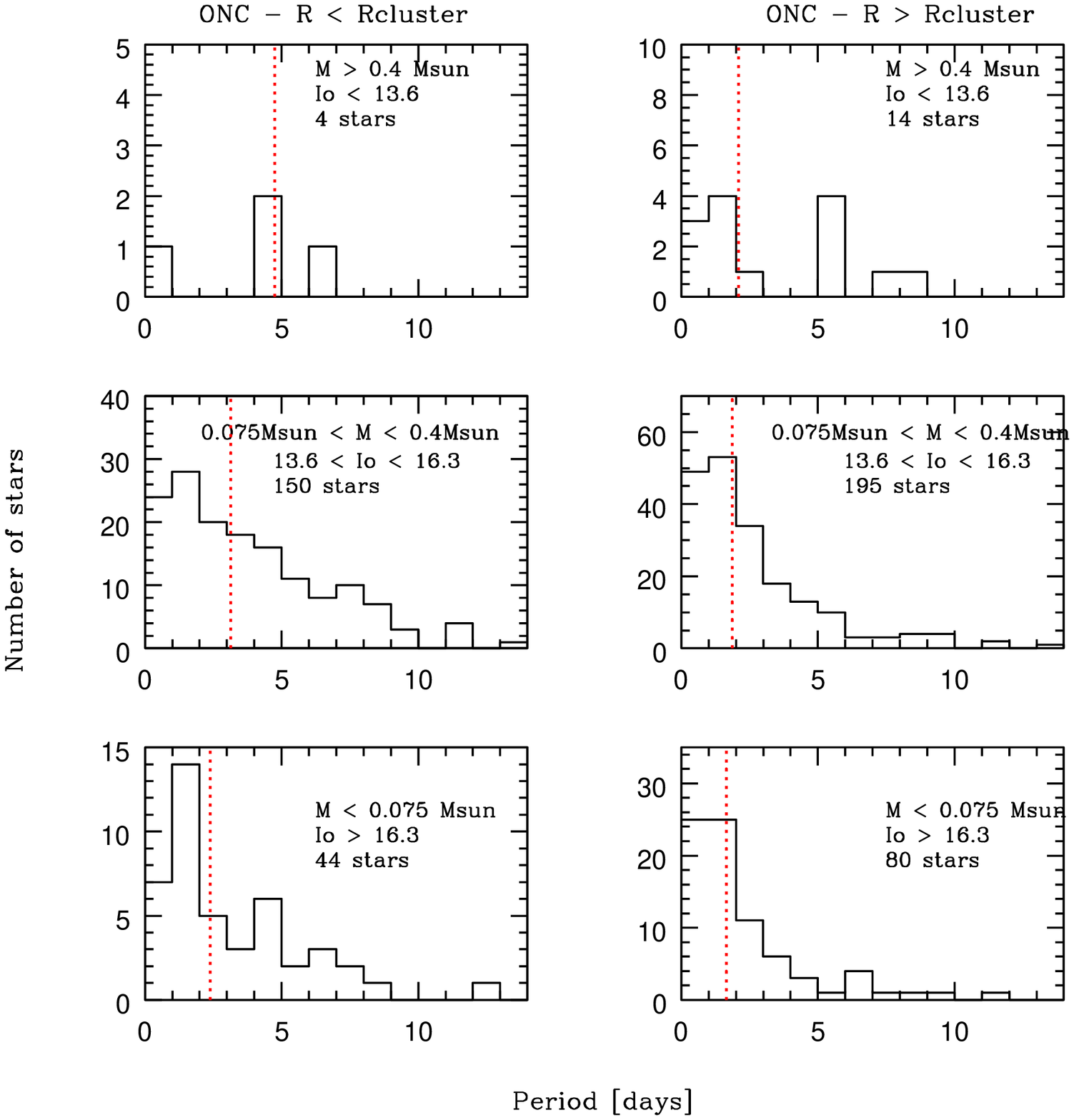}
  \caption{ The left panels show the period distribution for the stars located inside the cluster radius (6\farcm7 i.e. $\sim$\,1\,pc), while the right panels show the period distribution of the stars outside this radius for stars from our study. The mass regime for each panel is indicated and is based on models by Baraffe et al. (1998) and Chabrier et al. (2000), assuming a constant \textit{A$\mathrm{_{v}}$}\,=\,1.4\,mag for all stars (see text for details). In the bright regime (top panel), we have few stars and a meaningful analysis can only be done in combination with the H2002 data (see Fig.\,13). The medians of the periods for each panel are shown as dashed lines.}
         \label{Pspatial1}
   \end{figure*}
\begin{figure*}
   \centering
   \subfigure{\includegraphics[trim = 0mm 0mm 5mm 5mm, clip,width=12cm, height=15cm]{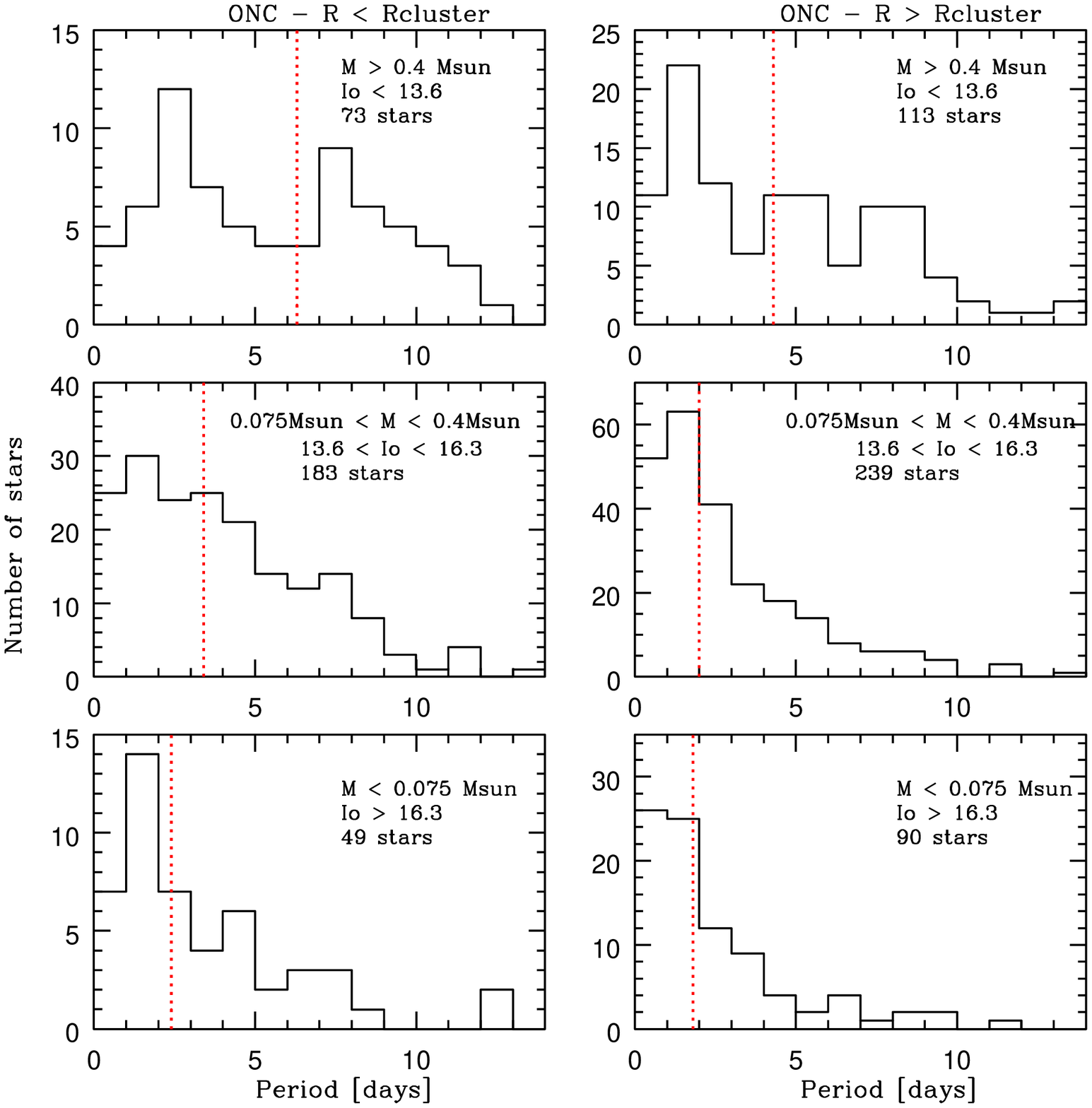}}
   \subfigure{\includegraphics[trim = 7mm 0mm 100mm 5mm, clip,width=5.5cm, height=15cm]{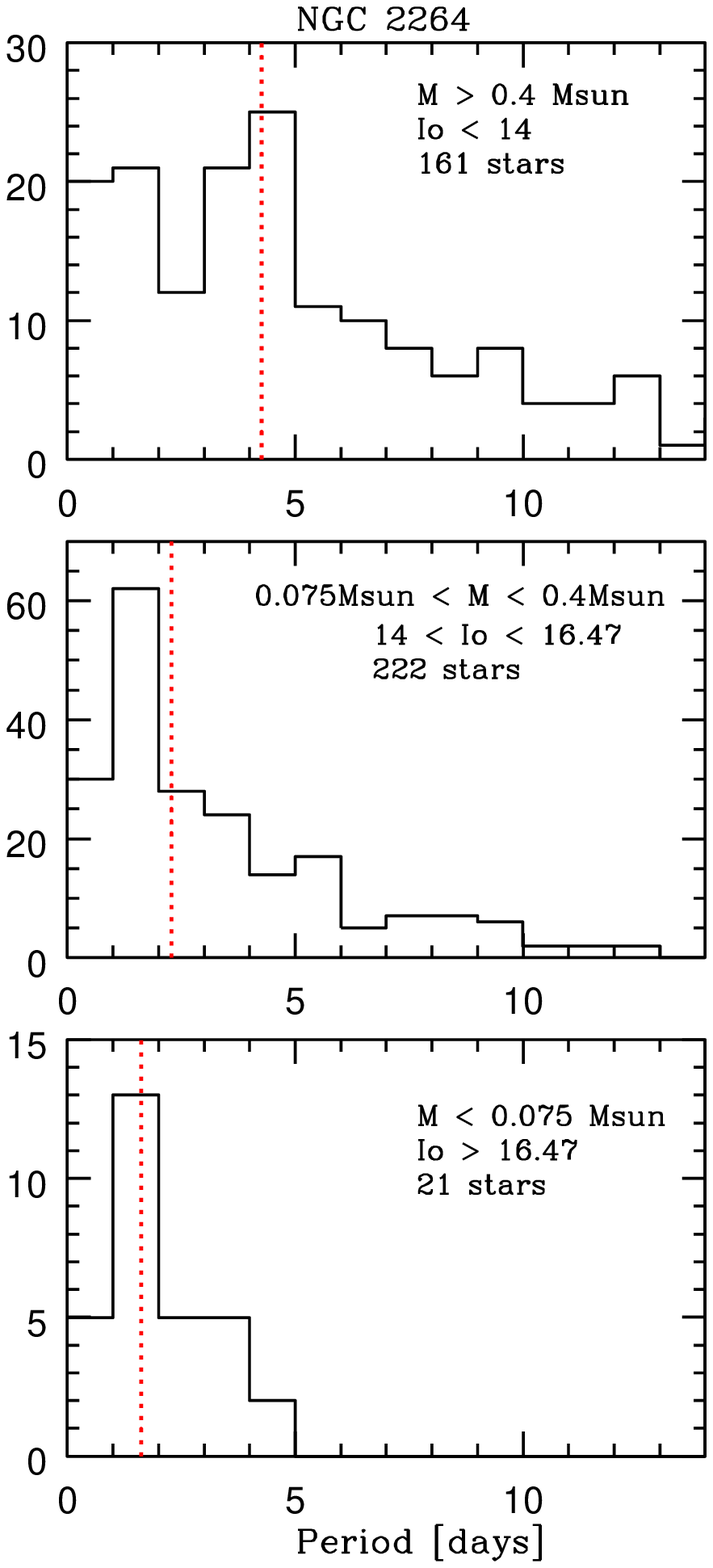}}
  \caption {The same as Fig.\,11, but for the combined ONC data of H2002 and this study in both left and middle columns, and for the periodic stars in NGC 2264 (L2004) in the right column. The top left and middle panels consist mostly of H2002 data. The four lower panels in the ONC show the same trend as in Fig.\,11, i.e. there is a clear difference in the distributions in and outside R$_\mathrm{cluster}$ as well as a clear trend of shorter periods towards lower masses. The period distribution for the stars in NGC 2264 (L2004) for which their measured I magnitudes were shifted to the corresponding ones at the distance of the ONC for a proper comparison are shown in the right column.  As in the case of the ONC, the mass estimations for each panel are based on models by Baraffe et al. (1998) and Chabrier et al. (2000) after assuming a constant textit{A$\mathrm{_{v}}$}\,=\,0.18\,mag (Perez et al., 1987). The same trend found in the ONC in which the less massive objects have on average shorter periods is seen here. The median periods for each panel are indicated as dashed lines.}
\label{Pspatialall}
   \end{figure*}
The resulting period distributions for the periodic variables measured in this study are shown in Fig.\,13, while the left and middle columns in Fig. 14 shows the combined data. In both figures we show the period distribution inside and outside R$_\mathrm{cluster}$. The median values for each set of data are plotted as vertical dashed lines. A first look at Figs.\,13 and 14 suggests different period distributions for stars located in the central and the outer region. For each mass bin, the median of the periods is always higher for objects inside R$_\mathrm{cluster}$ than for objects outside R$_\mathrm{cluster}$. The left and middle top panels in Fig.\,14 furthermore illustrate that the distributions are smoother in the outer region than in the inner one for the highest mass-bin. We performed a two-sided K-S test to corroborate these differences in the combined data shown in Fig.\,14. The result of the K-S test, which indicates the probability that the distributions come from the same population, is 0.07, 0.0001 and 0.13  for the top, middle, and lower panels, respectively. These numbers indicate that the differences in the top and bottom panels are not highly significant ($<$\,2$\sigma$ level), while middle panels are statistically significant at a very high confidence level, $>$\,99.9\%.\\
The same mass dependence found for the whole ONC field, in which towards lower masses the medians of the period distributions shift towards shorter periods, is seen in Figs.\,13 and 14 for the two spatial regions considered. In addition, the period distributions of the very low mass and substellar objects (second and third mass bin) are smoother than for the highest mass regime.\\
Our study suggests that the spatial position within a star forming region influences the period distribution as illustrated by Fig.\,14. For example, it is interesting that in the inner region a bimodal distribution for stars with \textit{M}\,$>$\,0.4\,\textit{M$\mathrm{_{\odot}}$} (two peaks around 2.5 and 8\,days) is clearly seen, while for the outer region no such bimodal distribution is evident (there are may be 2 peaks at 5 and 8\,days in addition to a 1.5\,day peak). In the middle left panel there might be a small bump around 8 days in addition to the 1.5\,day peak (but this could result from more highly extincted, more massive periodic variables). For the same mass bin, but in the outer region, the highest peak around 1.5\,days is the most prominent and the period distribution declines more rapidly towards longer periods. The same trend extends to the lowest mass bin (i.e. substellar bin). There, the highest peak is again around 1.5\,days for R\,$<$\,R$_\mathrm{cluster}$ and around 1 day for R\,$>$\,R$_\mathrm{cluster}$.\\
We are aware that our results are possibly affected by an observational bias in the mass estimations. As stated above, deriving masses assuming an average \textit{A$\mathrm{_{v}}$} leads to overestimated masses for stars in the outer regions and underestimated masses for stars in the more extincted inner region. This bias would result in a certain fraction of highly extincted higher mass objects, which are on average slower rotators, contaminating the lower mass bins in the inner region, while some of the low mass objects, which are on average fast rotators, would contaminate the higher mass bins in the outer region. To check whether this bias could affect our results signifficantly, we performed the following test: we assumed an avergae extiction \textit{A$\mathrm{_{v}}$\,=\,1} in the outer region, \textit{A$\mathrm{_{v}}$\,=\,2} in the inner one and we recomputed the masses of the stars. No signifficant differences were found between the new period distributions and the ones based on a constant \textit{A$\mathrm{_{v}}$\,=\,1.4}. The median rotational periods were slightly different only in the highest mass. We then conclude that although this bias may be present, it probably does not affect our results and conclusions.\\
In summary, we find that, on average, objects situated outside R$_\mathrm{cluster}$ tend to rotate faster than their inner counterparts. As observed in previous studies, lower mass objects tend to rotate faster than their massive conterparts and for the first time, this trend is shown to extend into the substellar mass regime.
\begin{table*}
\begin{minipage}[c]{2\columnwidth}
\centering
\renewcommand{\arraystretch}{1.4}  
\caption{Number of periodic variables in various mass intervals.}
\label{ONCH2002pv}
\addtolength{\tabcolsep}{0.7pt}
\begin{tabular}{c|ccc|ccc|c}
\hline \hline 
&\multicolumn{3}{|c|}{ONC-This Study} &\multicolumn{3}{c|}{ONC-Combined data (H2002 + This study)} & NGC 2264\\[1ex]
\hline
Mass Bins &R\,$<$\,R$_\mathrm{cluster}$ & R\,$>$\,R$_\mathrm{cluster}$ & Total & R\,$<$\,R$_\mathrm{cluster}$ & R\,$>$\,R$_\mathrm{cluster}$ & Total& \\[1ex]
\hline
I$_\mathrm{0}$\,$\leq$\,13.6 or \textit{M}\,$\geq$\,0.4\,\textit{M$\mathrm{_{\odot}}$} & 4 & 14 & 18 & 73 & 113 & 186 & 161\\
\hline
13.6\,$<$\,I$_\mathrm{0}$\,$\leq$\,16.3 or 0.4\,\textit{M$\mathrm{_{\odot}}$}\,$\leq$\,\textit{M}\,$<$\,0.075\,\textit{M$\mathrm{_{\odot}}$}& 150 & 195 & 345 & 183 & 239 & 422 & 222\\
\hline
I$_\mathrm{0}$\,$>$\,16.3 or \textit{M}\,$<$\,0.075\,\textit{M$\mathrm{_{\odot}}$}& 44 & 80 & 124 & 49 & 90 & 139 & 21\\
\hline
\end{tabular}
\renewcommand{\footnoterule}{} 
\end{minipage}
\end{table*}
\subsection{Comparison with NGC 2264 period distribution}
\begin{figure*}
   \centering
   \includegraphics[trim = 5mm 0mm 0mm 60mm, clip, width=17cm, height=12cm]{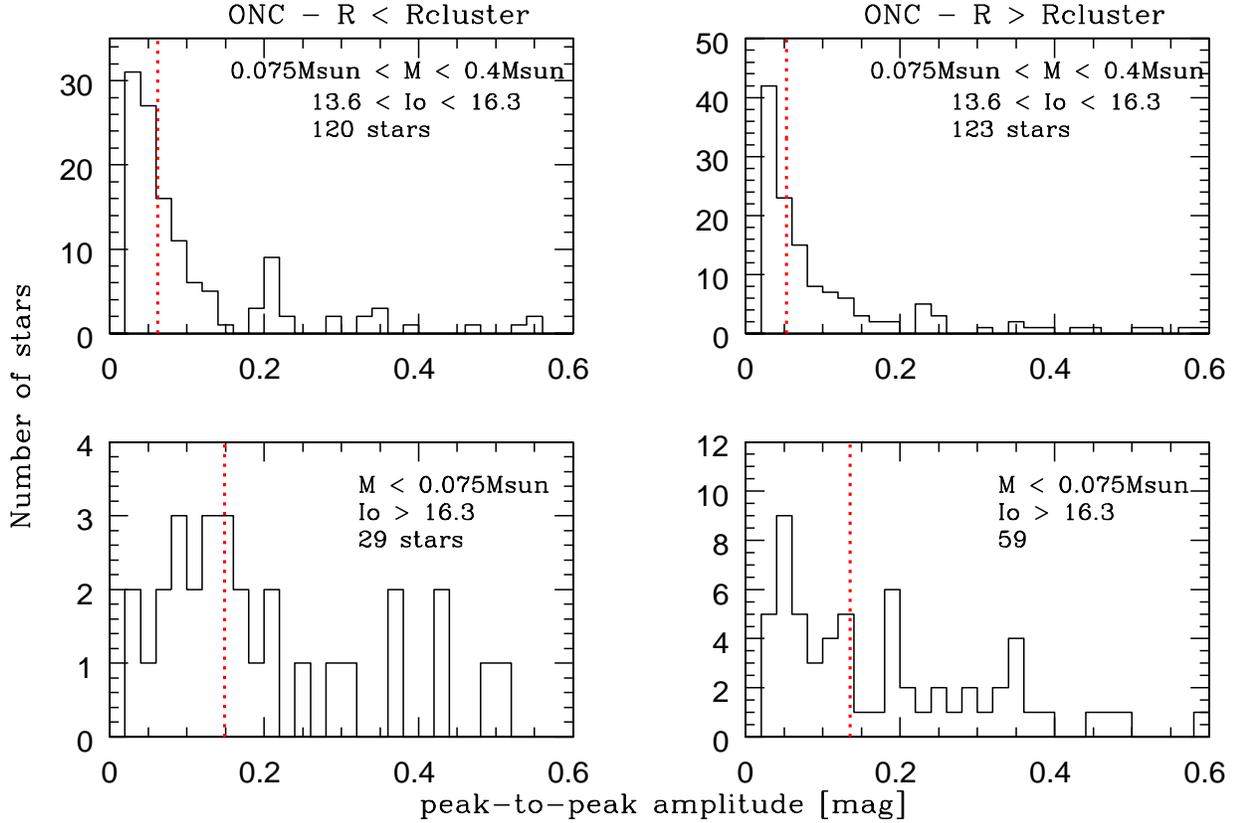}
\caption{ Peak-to-peak amplitude distribution of 331 periodic variables with \textit{M}\,$<$\,0.4\,$M_{\odot}$ and peak-to-peak amplitudes up to 0.5\,mag and above a 1\,$\sigma$ confidence limit. The left and right panels show the resulting amplitudes for stars inside and outside R$_\mathrm{cluster}$, respectively. The top and lower panels show the distributions for \textit{M$\mathrm{_{\odot}}$}\,$<$\,\textit{M}\,$leq$\,0.075\,\textit{M$\mathrm{_{\odot}}$} and 
 \textit{M}\,$<$\,0.075\,\textit{M$\mathrm{_{\odot}}$}, respectively. The medians of each distribution are shown as a vertical dotted line.}
         \label{Ampdistri_spatial}
   \end{figure*}
Although not highly significant, the differences found in the rotational period distributions for objects located inside and outside  R$_\mathrm{cluster}$ might be attributed to an age spread in the field. This means that objects inside R$_\mathrm{cluster}$ rotate on average slower because they are somewhat younger and therefore have had less time to spin up. H2000 reached a similar conclusion for higher mass objects (i.e. above 0.25\,M$\mathrm{_{\odot}}$) when analysing the correlations that exist between rotational period and projected radius (Hillenbrand, 1997), and also taking into account that the fraction of stars with disks is higher toward the center of the ONC (Hillenbrand, 1997; Hillenbrand \& Hartmann, 1998).\\
The best way to test our hypothesis is to look at different young regions, such as NGC2264, which probably has twice the age of the ONC. L2005 compared the ONC (H2002 data) and NGC2264, and found they could explain the different period distributions by different ages. They found that the median values of the period distributions in NGC 2264 are shifted towards shorter periods by a factor of $\sim$\,1.5 relative to the ONC and they found this amount of spin-up consistent with expectations based on PMS contracting models and conservation of angular momentum. \\
\begin{table}
\centering
\begin{minipage}[c]{\columnwidth}
\renewcommand{\arraystretch}{1.2}  
\caption{Median values of the rotational periods given in days for objects in the ONC (combined data) and NGC 2264.}
\label{ONC-NGC2264table2}
\addtolength{\tabcolsep}{2pt}
\begin{tabular}{c|ccc|c} 
\hline\hline             
Mass Bins\footnote{The same mass bins of Table\,4, described also in the text. A, B and C corresponds to the highest, intermediate and substelar mass interval respectively.} & \multicolumn{3}{c|}{ONC}  & NGC 2264 \\
\hline
& R$<$R$_\mathrm{cluster}$ & R$>$R$_\mathrm{cluster}$ & Total &\\
\hline
A & 6.2 & 4.3 & 4.9 & 4.2\\
B & 3.4 & 2.0 & 2.6 & 2.3\\
C & 2.4 & 1.8 & 1.9 & 1.6\\
\hline
\end{tabular}
\renewcommand{\footnoterule}{} 
\end{minipage}
\end{table}
Their conclusion was based on the ONC data from H2002 and therefore the very low mass regime as well as the substellar population were not studied. In addition, when re-evaluating the NGC 2264 data set, we used models by Baraffe et al. (1998) and Chabrier et al. (2000) while L2005 used models by D'Antona and Mazzitelli (\cite{Dantona07}) in the mass estimations, which results in a slightly larger number of VLM objects and BD candidates in NGC 2264 than reported by L2005. This new mass estimate allows a reasonable comparison between the ONC and NGC 2264 possible. The right column in Fig.\,14 shows the period distribution of the periodic variables in NGC 2264 from L2004 for the same three mass bins used in the analysis of the ONC (i.e. we considered an average \textit{A$\mathrm{_{v}}$}\,=\,0.18 in NGC 2264, P\'{e}rez et al., \cite{Perez}). The median values of each distribution are shown as red dashed lines. It is evident that the trend towards shorter periods among lower masses is true in NGC 2264 even for the 21 substellar objects (bottom panel). The highest mass objects (top panel) show the previously reported bimodal period distribution, while both the VLMs and BDs (i.e. middle and bottom panels) show a smoother, unimodal period distribution with the highest peak at about 1.5\,days in each case. The substellar objects in NGC 2264 were found to be entirely rapid rotators with periods below 5\,days. Table 5 lists the median values of the period distributions for NGC 2264 and the ONC. The comparison of the medians between NGC 2264 and the periodic variables in the whole ONC field delivers a marginal difference, suggesting that objects in the ONC rotate on average slightly slower (by a factor of about 1.2) than the objects in NGC 2264. But the comparison with the periodic variables located inside R$_\mathrm{cluster}$ results in a factor $\sim$1.5 faster rotators in NGC 2264 for all three mass bins. This is in agreement with the previous results by L2005 and is consistent with the age ratio of the clusters. Moreover, when we compare NGC 2264 with objects outside R$_\mathrm{cluster}$ in the ONC we find that the median values are almost identical for all three mass regimes. This result is a strong argument in favour of an age spread in the ONC, where the objects outside R$_\mathrm{cluster}$ are on average older than their counterparts inside it. In addition, this comparison allows us to give a rough estimate of the age difference, since the objects outside R$_\mathrm{cluster}$ in the ONC rotate on average at the same rate as the ones in NGC 2264. Our results aparently suggest that the outer region is about twice as old as the inner one\footnote{Under the assumption of a 2\,Myr old population in the outer region, we recalculated the correspnding mass intervals to test whether this would have change our results based on a 1\,Myr old region. We found that only 7 objects from the lower mass bin move to the intermedite one and 10 from the intermediate mass bin move to the higher mass bin. The new period distributions and their corresponding median values were found to be identical to those previously derived for a 1Myr old outer population, without affecting the whole mass-dependence analysis.}.
\subsection{Peak-to-peak amplitude distribution - high and low level variations}
In the following we will discuss how the peak-to-peak (ptp) amplitude distribution of the periodic variables depends on mass and on their location with respect to R$_\mathrm{cluster}$. For the determination of the ptp variation of the periodic variables we used the phased light-curve of each object and divided it into 10 equally spaced phase bins as described by Lamm (2003). In each bin, the median of the relative magnitudes was calculated. The largest and smallest median values for each object were used to compute the ptp amplitude as the difference between these two extreme median values.\\
\begin{figure*}
\centering
\subfigure{\rotatebox{270}{\includegraphics[height=9cm]{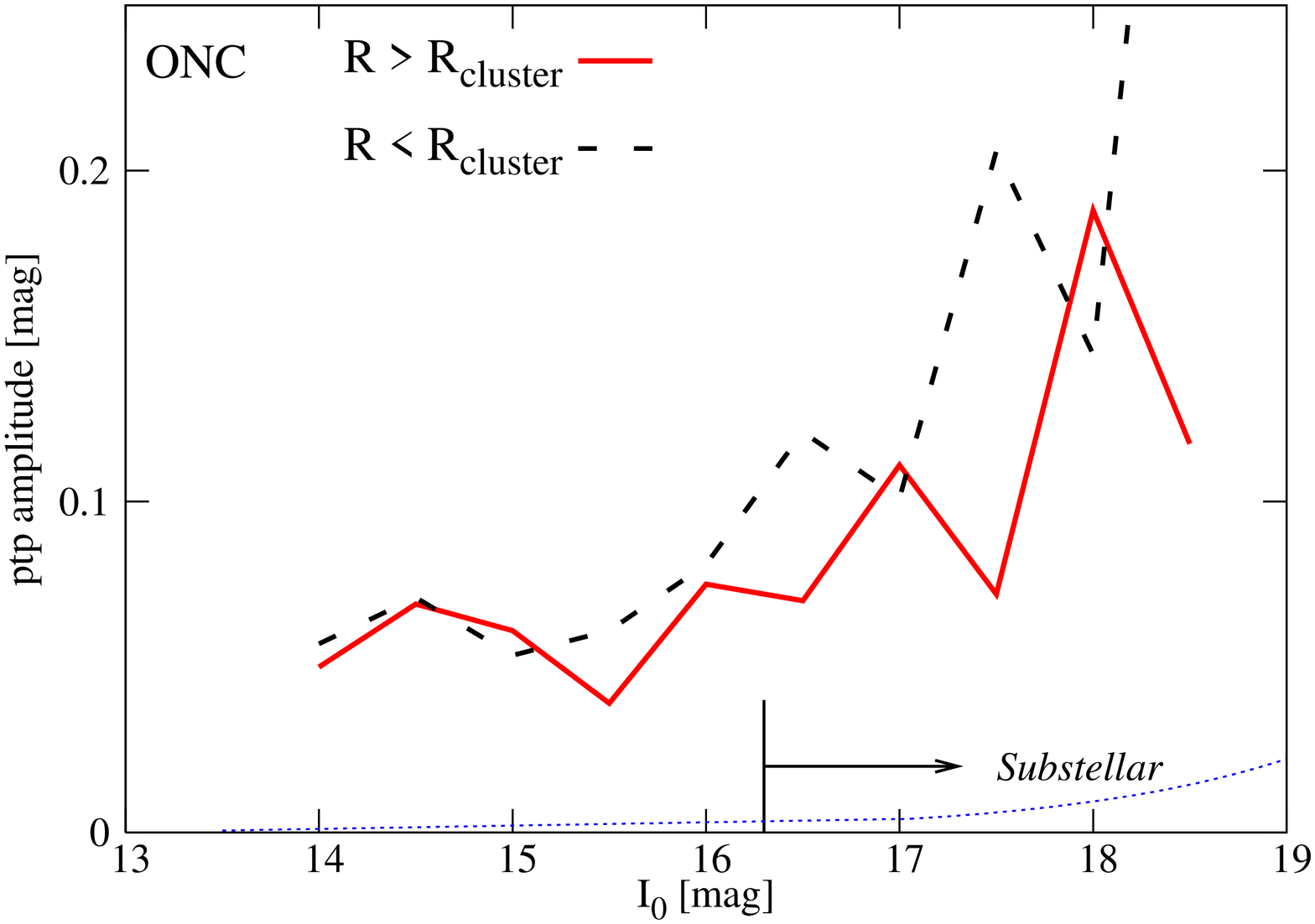}}}
\subfigure{\rotatebox{270}{\includegraphics[height=9cm]{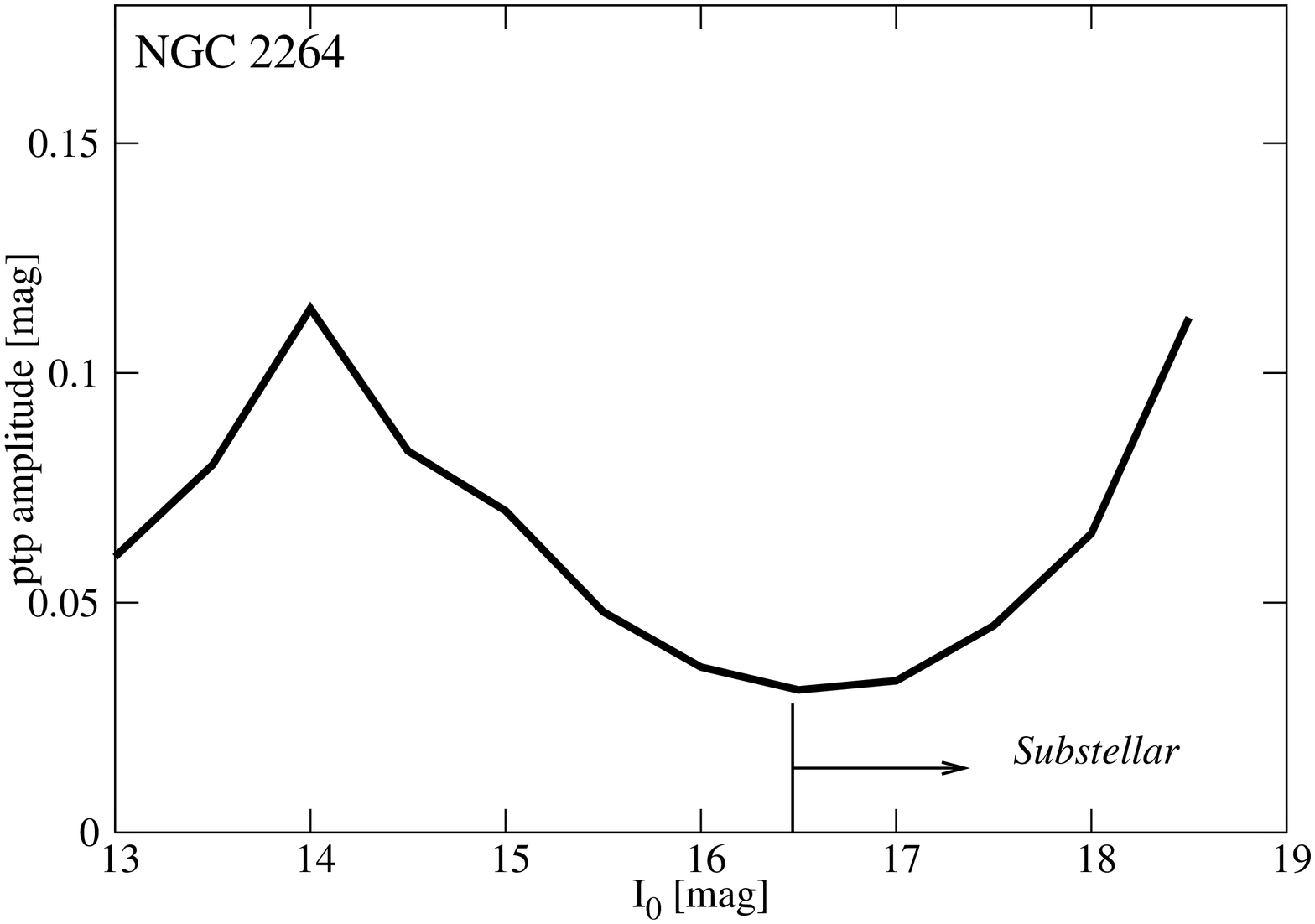}}}
\caption{Median ptp amplitude (calculated in intervals of 0.5\,mag) as a function of magnitude for 331 periodic objects in the ONC (left panel) with ptp variations that fulfil the criteria described in the text, as well as for 405 periodic variables in NGC 2264 (right panel). In the ONC the median ptp values are shown as dashed and solid lines for R\,$<$\,R$_\mathrm{cluster}$ and R\,$>$\,R$_\mathrm{cluster}$, respectively. We also indicate for comparison the error curve from Fig.\,3 as dotted lines and the adopted substellar limit in the ONC and NGC 2264. To allow a proper comparison of NGC 2264 with the ONC, the magnitudes were shifted to the distance of the ONC (450\,pc). Note that the small difference in the substellar limit (from 16.3\,mag for the ONC and 16.47\,mag for NGC 2264) is due to the higher age of the objects in NGC 2264 and the different average extinction adopted for both fields.}.
         \label{FigAvsI}
   \end{figure*}
Since H2002 did not report any ptp amplitudes, we only performed this analysis with the periodic variables measured in this study (487 objects). The small number of objects in the higher mass bin does not allow any meaningful analysis and therefore we only analyse the results for stars with I$_\mathrm{0}$\,$>$\,13.6\,mag (\textit{M}\,$<$\,0.4\,$M_{\odot}$).\\
In order to avoid any observational bias due to the higher photometric errors measured in objects located in the regions with strong nebular background, we first restricted our sample to objects with ptp amplitudes that are at least 2.5 times higher that the mean photometric error in the light-curve of each single star. For the remaining objects (470), we selected those with ptp amplitudes located above the mean error curve (see Figure 3\textit{b}) plus 1 sigma rms. We finally used 346 objects to perform the analysis of the ptp amplitude distribution.  \\
Fig. 15 shows the ptp amplitude distribution for stars located inside and outside R$_\mathrm{cluster}$ and considering only the middle and lowest mass bins previously defined in the period distribution analysis. This analysis is restricted to stars with ptp amplitudes $<$\,0.5\,mag (95.7\% of the sample) and we regard the 4.3\% of objects with ptp variations $\geq$\,0.5\,mag as stars not having cool spots but rather, with hot spots (H2002).
The comparison of the ptp variations for objects in and outside R$_\mathrm{cluster}$ exhibits a slight tendency for the objects inside R$_\mathrm{cluster}$ to have higher ptp amplitudes than the stars located outside R$_\mathrm{cluster}$. As for the period distribution we performed a two-sided K-S test resulting in a 23\% and a 74\% probability that the two distributions come from the same population for the middle and lowest mass regimes, respectively. Note that the size of the sample in the substellar regime is small.\\
Besides the small sample size, it is evident that in the substellar regime the ptp amplitudes are on average higher than for stars with masses between 0.4\,\textit{M$\mathrm{_{\odot}}$} and 0.075\,\textit{M$\mathrm{_{\odot}}$}. We found that the ptp amplitudes have a tendency to increase towards lower masses (i.e. fainter I magnitudes), as can be seen in Fig.\,16\textit{a} where the median amplitude values were computed in bins of 0.5 magnitudes, and the error curve is shown as blue dotted lines for comparison. Again we distinguish between objects inside and outside R$_\mathrm{cluster}$. Objects with I$_\mathrm{0}$\,$>$\,16 tend to have larger variations than stars with I$_\mathrm{0}$\,$<$\,16 (i.e. more massive stars) in the cluster, independently of their location inside or outside R$_\mathrm{cluster}$. This tendency is stronger for the faintest objects. Due to our conservative sample selection, we regard the increasing ptp amplitudes of the faint objects as an intrinsic phenomenon. Moreover, as can be seen in Fig.\,16, the mean photometric error curve shows a clear increase above 16\,mag but still is one order of magnitude lower than the ptp amplitudes in the plotted magnitude range. The typical mean photometric error of an I$_\mathrm{0}$\,=\,18\,mag periodic object is about 0.01\,mag while the median ptp amplitude of an object with the same brightness is about 0.15\,mag. In addition, 32\% of the lowest mass objects present low amplitude modulations below 0.1\,mag (see Fig.\,15). Since we indeed detect very low variation among the substellar candidates, it is likely that an intrinsic physical phenomenon is at least in part responsable for the increasing ptp amplitudes of the faintest objects. The same analysis was done for the NGC 2264 data set published by L2004 as shown in Fig.\,16. Quite in contrast to the ONC, the median ptp amplitude curve shows a maximum around 14\,mag and a minimum at around 16.5\,mag, which coincides with the substellar limit. The only similarity between the ONC and NGC 2264 is the increase in the ptp amplitudes at fainter magnitudes. We have no explanation for the differences shown in Fig.\,16. Presumably, the VLM objects in NGC 2264 develop smaller or more symmetrically distributed spots on their surface, as was argued by Herbst et al. (2007). This probably requires different magnetic field topologies for objects belonging to the two clusters. 
\section{Rotational period and ptp amplitude correlation in the ONC and NGC 2264}
\begin{figure*}
\centering
\subfigure{\rotatebox{270}{\resizebox{!}{9cm}{%
\includegraphics[width=9cm]{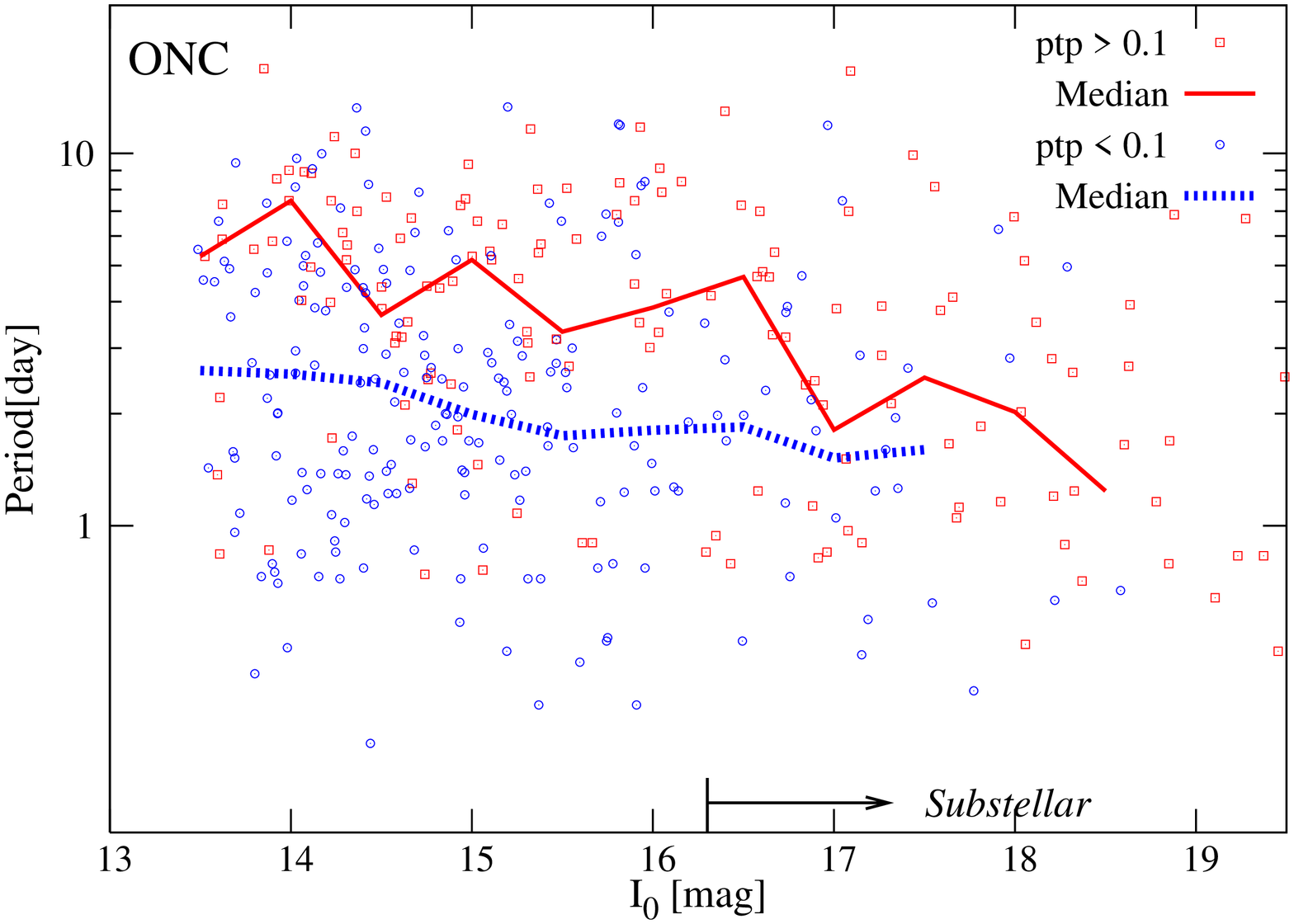}}}}
\subfigure{\rotatebox{270}{\resizebox{!}{9cm}{%
\includegraphics[width=9cm]{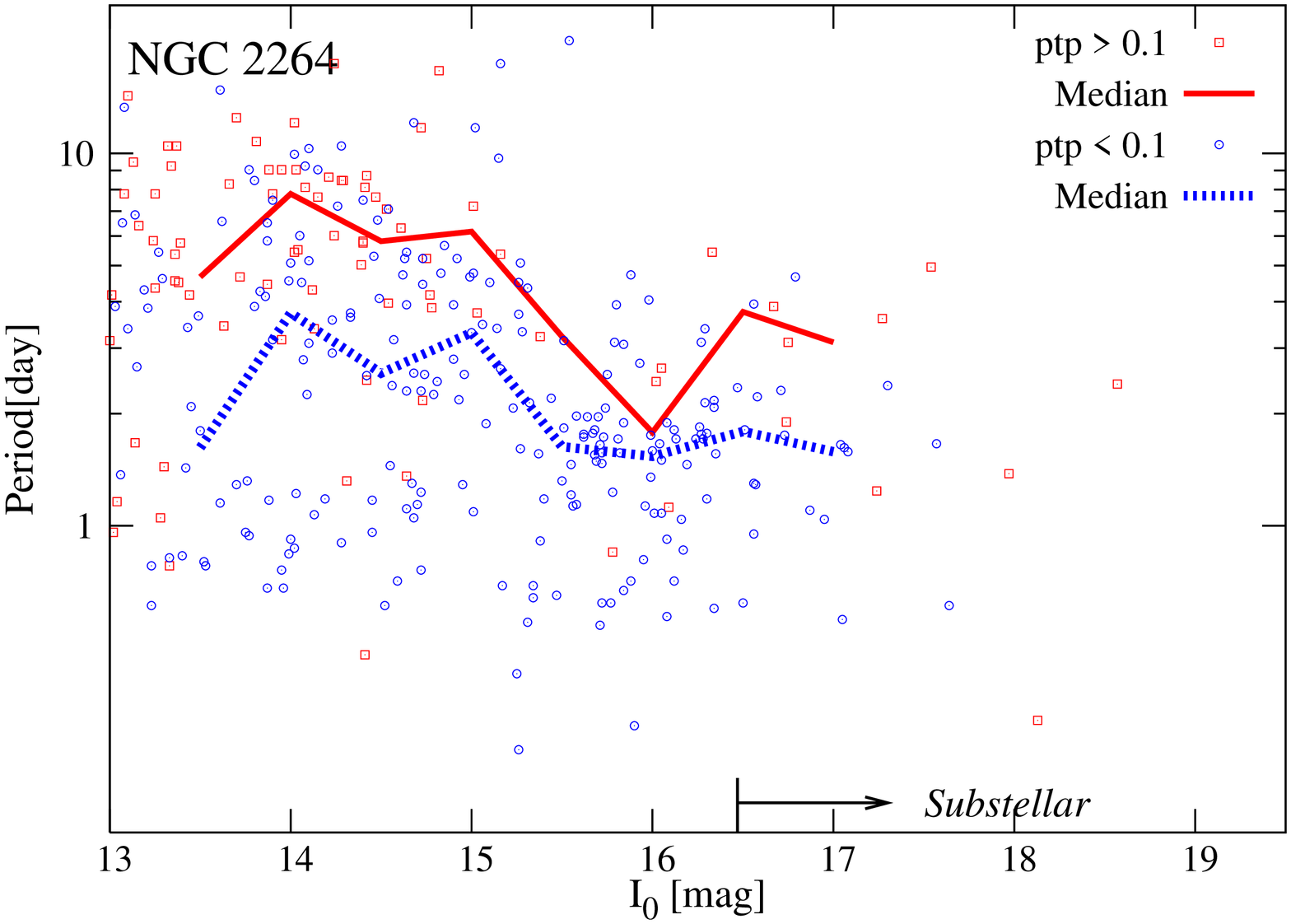}}}}
\caption{Period as a function of magnitude (mass) for two different variation amplitudes, namely for ptp amplitudes (ptp) below 0.1mag (blue open circles) and between 0.1 and 0.5 magnitude (red open squares), respectively for the ONC (left) and NGC 2264 (right). The medians of the rotational periods are calculated within 0.5 magnitudes bins for both cases and are indicated as blue dashed and red solid lines for ptp amplitudes below 0.1mag and between 0.1 and 0.5 magnitudes, respectively. In the case of NGC 2264 the magnitudes were shifted to the distance of ONC (450\,pc). }
\label{PvsI01-ONC}
\end{figure*}
We analysed the dependence of the period on the ptp amplitudes. For this purpose, we considered only objects showing ptp variations $\leq$\,0.5\,mag and divided them into two subsamples, namely one with amplitudes $\leq$\,0.1\,mag and the other with amplitudes from 0.1 to 0.5\,mag. The left panel in Fig.\,17 shows the rotational period as a function of magnitude for these two samples in the ONC. The medians for each sample were calculated in bins of 0.5 magnitudes and appear in Fig.\,17 as a dashed (blue) and a solid (red) line for the objects with ptp amplitudes below 0.1 mag and above this value, respectively. It is evident from this figure that the stars with higher amplitudes have on average a significantly longer period. This is true for all masses, although the median rotation periods for both subgroups seem to approach each other in the very low mass regime.\\
Furthermore, the left hand part of Fig.\,17 is rather interesting with respect to the already discussed decrease in period with decreasing mass (see Fig.\,14). This trend is only evident for the objects with ptp amplitudes between 0.1 and 0.5\,mag. This decrease in periods towards lower masses is not so clear for objects showing the lowest level variations (i.e. ptp amplitudes $<$\,0.1\,mag). In the last group, independently of their I values (i.e. for all masses) all objects are on average fast rotators with a median period close to 2\,days. \\
The right panel in Fig.\,17 shows the same correlation for NGC 2264 (i.e. we have shifted the I magnitudes to the distance of the ONC, 450\,pc). As in the ONC, objects with the smallest ptp amplitudes rotate on average faster than the highly variable objects. The differences between the two subgroups in NGC 2264 are roughly constant (and smaller than in the ONC) over the whole mass range, but due to the small number of objects fainter than 16 mag with ptp amplitudes above 0.1\,mag, it is difficult to perform a reliable statistical analysis. \\
It is intriguing that a correlation between rotational period and level of variability is found for periodic variables in two different clusters. Irwin et al. (\cite{Irwin2008}) did not find such a correlation between rotational period and amplitude of the modulation in the 5\,Myr old cluster NGC 2362. It is worth to note that their sample only extends down to about 0.1\,\textit{M$\mathrm{_{\odot}}$}, with a high fraction of objects even more masive than 0.4\,\textit{M$\mathrm{_{\odot}}$}.\\
This finding can probably be explained by differences in the spot coverage and/or differences in the magnetic field topologies. Recent studies of T-Tauri magnetic field structures (Jardine et al., 2006) argue that a variety of field topologies, from dipole to complex configurations, are required to explain the whole range of X-ray emission measurements for COUP\footnote{Chandra Orion Ultradeep Project} stars. In complex field configurations, coronal gas is confined within compact loops and covers a large fraction of the surface. These multipolar fields produce small and almost uniformly distributed spots which result, on average, in a low level periodic variability. In addition, this kind of magnetic configuration hardly allows for any magnetic braking through coupling with the disk and therefore these objects are more likely to be fast rotators. On the other hand, an extended dipolar field more likely produces large spot groups near the magnetic poles which could result in a high level periodic variability if the magnetic and rotational axes are inclined. In this scenario, due to the slower decline in magnetic field strength with distance from the star, the magnetic coupling to the disk is much stronger, and therefore these objects are more likely to be slow rotators.\\
\section{Summary and conclusions}
We have monitored about 3000 objects in the ONC in the I-band over 19 nights. Relative photometry resulted in 2910 objects with good photometric precision ranging from 13 to 21\,mag, which means three magnitudes deeper than previous studies in the ONC (H2002). In the following we summarise the main results: 
\begin{enumerate}
\item We found 487 periodic variables in the ONC, 377 of which are new detections relative to H2002 and 124 of which are BD candidates on the basis of the I magnitudes and from comparison with theoretical models. The combined data-set resulted in 746 periodic variables, 139 of which are BD candidates.
\item We detected 808 irregular variables (i.e. variable objects for which it was not possible to assign a rotational period) with a 99.9\% probability of being variables according to the $\chi^{2}$ test.
\item We analysed the spatial distribution of the variable objects. We found that both irregular variables and periodic variables are highly clumped in the inner region of the ONC, however, their locations are not the same. A K-S test resulted in a 99\% probability that the distributions of declinations for irregular and periodic variables are different and 43\% for the right ascension values. Irregular variables are most likely found south-west of the central Trapezium stars, while most periodic variables are clumped in a region to the north-west.
\item The analysis of the rotational period distributions for different magnitude bins (translated into estimated mass bins) confirms the already known trend of fast rotators lying towards lower masses. This trend extends well into the substellar mass regime and was found for the whole ONC field as well as for the regions inside and outside R$_\mathrm{cluster}$.
\item The analysis of the rotational period distributions for different magnitude bins for objects located inside and outside the R$_\mathrm{cluster}$ suggests differences between the two regions, where objects inside R$_\mathrm{cluster}$ tend to rotate on average slower than the outer ones. This trend seems to be true for all mass bins but at different significance levels. It can be interpreted as an age difference between the inner and outer population. 
\item In comparing the ONC with the $\sim$\,2 times older cluster NGC 2264, we used the L2004 data but adopted the models of Baraffe et al. (1998) and Chabrier et al. (2000) for mass estimates instead of D'Antona \& Mazzitelli (\cite{Dantona07}). This results in a larger number of low mass objects and even substellar objects than published by L2004. The period distributions of NGC 2264 were compared, for all mass bins, with the period distributions of objects located inside and outside R$_\mathrm{cluster}$ in the ONC. We found a factor $\sim$\,1.5 faster rotators in NGC 2264 relative to objects located inside R$_\mathrm{cluster}$ in the ONC for each considered mass bin. However, the median values of the period distributions for objects outside R$_\mathrm{cluster}$ in the ONC are almost identical to the ones in NGC 2264 for all three mass regimes. This finding favours the hypothesis of an age spread in the ONC, in which the older outer population is presumably as old as NGC 2264, according to the median rotational periods found.
\item We studied the ptp amplitude distribution of a reduced sample of periodic variables (331 objects) in order to avoid observational biases due to the higher photometric errors measured in the central part of the ONC due to the strong nebular background. Using the whole sample of periodic variables would have biased our results towards smaller ptp variations in the outer region (i.e. outside R$_\mathrm{cluster}$) with a stronger effect for the faintest objects. We did not find statistically significant differences between inside and outside R$_\mathrm{cluster}$.
\item We investigated the period dependence on variability level in both the ONC and NGC 2264. We found that low level variables (i.e. ptp amplitudes below 0.1\,mag) rotate on average faster than the high level variables in both clusters. This is true for the whole magnitude range studied, although in the substellar regime there is a tendency to have mainly fast rotators. Interestingly, very low level variables in the ONC are fast rotators, with periods of about 2\,days, quasi-independently of their mass. The more highly variable objects in the ONC, as well as the two groups in NGC 2264, show a continuous decrease in rotational period towards the lowest masses. \\
This intriguing correlation found in the two clusters probably indicates that different magnetic field topologies act on these objects, resulting in large and small amplitude variations for slow and fast rotators, respectively. Complex magnetic field configurations hardly allow magnetic coupling with the circumstellar disk. In addition, they pressumably produce a more uniform surface spot coverage with smaller spot groups resulting in small ptp amplitudes for the fast rotators, while extended dipolar fields allow strong disk coupling and therefore much stronger magnetic braking. This will produce large spot groups near the magnetic poles, resulting in large ptp amplitudes for the slow rotators. 
\item We believe that the best explanation for the results mentioned in 5 and 6 is an age spread in the ONC region, in which objects located inside R$_\mathrm{cluster}$ are on average younger than objects outside. This can explain why the younger inner population rotates slower than the somewhat older objects in the outer regions, since it has not had enough time to spin up. Moreover, since the comparison with NGC 2264 reveals that the median values of the rotational periods in the outer regions of the ONC are almost identical to the ones in NGC 2264 for the whole magnitude range, we estimate the age of this older population surrounding the ONC to be roughly twice the age of the inner region close to the Trapezium cluster.\\
The age spread in star forming regions has been discussed in detail by several authors (Huff \& Stahler 2006, 2007, Mayne \& Naylor 2008). Whether star formation acts in a rapid single episode or slowly, with different star formation rates or in several episodes, is still under debate, but we cannot exclude the latter scenario as a possible interpretation of our observed age spread. Particularly in the ONC, an age spread was suggested by several authors (e.g. Hillenbrand (1998), Hillenbrand \& Hartmann (1998) and H2000) to account for the correlations found for low mass stars (above 0.25\,M$\mathrm{_{\odot}}$) between rotational periods and projected radius and also due to the higher fraction of objects with disks (by means of IR-excess) found in the central region. We investigate this correlation in a follow-up paper (Rodriguez-Ledesma et al., 2009, in prep). 
\end{enumerate}
\begin{acknowledgements}
The authors thank William Herbst for detailed comments on this work and for critically reading the manuscript. M.V.R.L. kindly thanks Coryn Bailer-Jones for his help during the early stages of the data reduction. M.V.R.L. acknowledges support of the International Max Planck Reseach School for Astronomy and Cosmic Physics of the University of Heidelberg.

\end{acknowledgements}

\end{document}